\documentclass[12pt,twoside]{article}

\def\TheAuthor{H. Schoeller}                                       % Author name
\def\TheTitle{Dynamics of open quantum systems \footnotemark}      % Title of contribution
\def\TheHeading{Open quantum systems}                              % Short running title
\def\TheInstitute{Institut f\"ur Theorie der Statistischen Physik} % Institute
\def\TheAddress{RWTH Aachen}                                       % Place e.g. University ...
\def\ThePart{B}                                                    % Part of book
\def\TheChapter{5}                                                 % Chapter of contribution

\usepackage{ferienschule_14}                                  % Style file for layout

\makeindex
\begin{document}

\MakeTitel           %%% Displays title, author name, etc.
\tableofcontents     %%% Displays table of contents

\footnotetext{Lecture Notes of the $45^{{\rm th}}$ IFF Spring
School ``Computing Solids - Models, ab initio methods and supercomputing''
(Forschungszentrum J{\"{u}}lich, 2014). All rights reserved. }

\newpage

%%%%%%%%%%%%%%%%%%%%%%%%%%%%%%%%%% Add text here ... %%%%%%%%%%%%%%%%%%%%%%%%%%%%%%%%%%%%%%%%%%%%%

%iffindex{general relativity}
%creates an index entry in the .idx file
%
%\newpage
%\begin{figure}
%    \centering
%     \includegraphics[width=0.8\hsize]{emc.eps} \caption{This is a figure caption.}
%\end{figure}

\begin{figure}
  \centering
  \includegraphics[width=0.5\hsize]{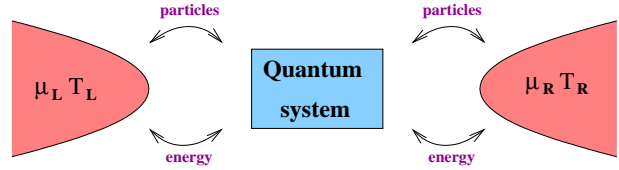}
%  \centerline{\psfig{figure=sketch.eps,scale=0.5}}
  \caption{A small quantum system coupled to several infinitely large
    reservoirs via energy and/or particle exchange. The reservoirs are
    characterized by temperatures $T_\alpha$ and chemical potentials $\mu_\alpha$.}
\label{fig:sketch}
\end{figure}

\section{Introduction}

An open quantum system \iffindex{open quantum system}
consists of a local quantummechanical system of fixed size coupled 
to infinitely large reservoirs in statistical equilibrium via a well-defined interaction, 
see Fig.~\ref{fig:sketch} for a sketch of the system. The analysis of the time
evolution \iffindex{time evolution} of the reduced density matrix of the local system
is of fundamental importance for nonequilibrium statistical mechanics. Of particular interest
is the interplay of quantum coherence in the local quantum system and dissipation generated 
by the reservoirs, which is the reason why this field is called {\it dissipative quantum mechancis}
\iffindex{dissipative quantum mechanics}. Many topics are here of current interest: 
(1) The development of a microscopic theory for irreversible
time evolution of the local density matrix; (2) The characterization of the reduced dynamics,
in particular by generic features independent of the microscopic details of high-energy
processes; (3) The calculation of typical relaxation and decoherence rates; (4) The analysis of
quantum fluctuations induced by the system-reservoir interaction beyond perturbation theory; 
(5) The analysis of the influence of strong correlations in the local quantum system, induced by 
charging energies; (6) The analysis of the influence of inhomogeneous boundary conditions, induced by different
temperatures and/or chemical potentials of several reservoirs, leading to energy, particle,
and spin currents; (7) The analysis of deviations of the stationary local density matrix from a
grandcanonical one, induced by quantum fluctuations from the system-reservoir coupling or by the
presence of several reservoirs; (8) The development of nonequilibrium renormalization group (RG)
\iffindex{renormalization group}
methods capable of resumming logarithmic divergencies occuring in higher-order perturbation
theory in the system-reservoir coupling, either at high energies w.r.t. the band width
of the reservoirs or at low energies w.r.t. the inverse time in the long-time limit; 
(9) The study of non-Markovian dynamics \iffindex{non-Markovian dynamics}
leading to additional terms in the time evolution with
unexpected oscillation frequencies and decay rates together with non-exponential time evolution;
(10) The crossover between coherent and incoherent dynamics induced by the sign and size of the
system-reservoir coupling and other tunable parameters. It is the purpose of this
tutorial introduction to present a microscopic theory for the time evolution of open quantum
systems, to discuss some of the above aspects from a generic point of view, and to characterize 
explicitly the reduced dynamics for elementary 2-level quantum systems coupled via energy, 
particle or spin exchange to external reservoirs. \newline

Although the field of dissipative quantum mechanics has a long history, the field has regained an
enormous interest in the last decades due to its relevance in quantum transport phenomena in 
nanoelectronic systems and quantum information processing, and due to the controlled realization of 
low-dimensional quantum systems in cold atom gases. To describe the time evolution of the reduced
density matrix of the local quantum system microscopically, one starts from the von Neumann equation 
for the total density matrix $\rho_{\text{tot}}(t)$ of the full system (we set $\hbar=e=k=1$)
\begin{equation}
\label{eq:von_Neumann}
i\,\dot{\rho}_{\text{tot}}(t)\,=\,[H_{\text{tot}}(t),\rho_{\text{tot}}(t)]\,=\,
L_{\text{tot}}(t)\,\rho_{\text{tot}}(t)\quad,
\end{equation}
where $L_{\text{tot}}(t)$ is the so-called 
Liouville operator, a superoperator \iffindex{superoperator} which acts on
an arbitrary operator via $L_{\text{tot}}(t)A=[H_{\text{tot}}(t),A]$. The central idea is always 
to integrate out the reservoir degrees of freedom and to set up a formally exact kinetic equation
\iffindex{kinetic equation} for the local density matrix 
$\rho(t)=\text{Tr}_{\text{res}}\rho_{\text{tot}}(t)$, defined by the
trace $\text{Tr}_{\text{res}}$ over the reservoir degrees of freedom of the total density matrix.
This kinetic equation has the form
\begin{equation}
\label{eq:kinetic_equation}
i\,\dot{\rho}(t)\,=\,\int_{t_0}^t\,dt'\,L(t,t')\,\rho(t')\quad,
\end{equation}
where $t_0$ is the initial time and $L(t,t')$ is an effective Liouville operator acting only on 
operators of the local quantum 
system. This superoperator contains all the information of the reservoir degrees of freedom and the 
system-reservoir interaction. For a time-translational invariant Hamiltonian, $L(t,t')=L(t-t')$ 
depends only on the relative time difference. The effective Liouvillian \iffindex{effective Liouvillian}
$L(t,t')$ is only defined for $t>t'$,
i.e. it acts as a response function relating the density matrix at time $t'$ to the one at the later
time $t$. This acounts for memory effects
and leads to non-Markovian dynamics. The only assumption needed to derive the kinetic equation 
\eqref{eq:kinetic_equation} is the factorization of the total density matrix at the initial time 
$t_0$ in an arbitrary local part $\rho(t_0)$ and an equilibrium part for the reservoirs
\begin{equation}
\label{eq:initial_dm}
\rho_{\text{tot}}(t_0)\,=\,\rho(t_0)\,\rho_{\text{res}}^{\text{eq}}\quad,\quad
\rho_{\text{res}}^{\text{eq}}\,=\,\prod_\alpha\rho_\alpha^{\text{eq}}\quad,\quad 
\rho_\alpha^{\text{eq}}\,=\,
{1\over Z_\alpha}\,e^{-(H_\alpha-\mu_\alpha N_\alpha)/T_\alpha}\quad,
\end{equation}
where $T_\alpha$, $\mu_\alpha$, $H_\alpha$, $N_\alpha$, and $Z_\alpha$ are the temperature, the chemical 
potential, the Hamiltonian, the particle number, and the partition function of reservoir $\alpha$, respectively.
However, by changing the Hamiltonian at a certain quench time $t_q>t_0$ abruptly, other initial
conditions can be realized where system and reservoirs are correlated. \newline

Various techniques have been developed to calculate the effective Liouvillian $L(t,t')$. The 
traditional ones are projection operator techniques \cite{projection_operator} and functional 
integrals \cite{functional_integral}. Recently, a quantum field theoretical approach has been
developed, which allows for a systematic classification of all processes in all orders of
perturbation theory in the system-reservoir coupling \cite{schoeller_EPJ09}. With this method,
it is possible to go beyond bare perturbation theory which is necessary at low temperatures
due to various logarithmic divergencies at high and low energies. The method is capable of  
identifying these logarithmic divergencies very effectively and an RG method in nonequilibrium 
has been set up to resum them. This allows a systematic weak-coupling expansion in
the renormalized coupling constants to be formulated with which the time evolution on all
time scales even when the reservoirs have different chemical potentials or temperatures
can be discussed. This technique has been applied successfully to the Kondo model \iffindex{Kondo model} 
\cite{schoeller_PRB09,pletyukhov_PRL10,kondo_E-RTRG}, the interacting
resonant level model (IRLM) \iffindex{interacting resonant level model}
\cite{RTRG_irlm,kennes_PRL13,kashuba_kennes_PRB13}, 
and the ohmic spin boson model \iffindex{spin boson model} \cite{kashuba_PRB13}. 
In particular, it has turned out that the RG formulation is most effective for the calculation of the time
evolution if the Fourier variable $E$ conjugate to the time $t$ is used as flow parameter, i.e. as
the paramater w.r.t. which derivatives of the various quantities of interest are taken to obtain
differential equations (the so-called RG equations). This technique is called
the E-RTRG \iffindex{E-RTRG} method \cite{kondo_E-RTRG,kashuba_PRB13}. The models treated so far fall
into the special class where the density of states in the reservoirs and the frequency dependence
of the system-reservoir couplings is weak and varies on the scale of the high-energy cutoff $D$. 
Physically, the high-energy cutoff can either be the band width of the reservoirs or it is some 
internal high-energy scale of the local quantum system, like e.g. charging energies, arising when 
effective models are used by integrating out high-energy processes (e.g. quantum dots in the 
Coulomb blockade regime where charge degrees of freedom can be eliminated, see the lecture B3 by 
T. Costi). For such models it is often possible to find universal physics where the special form of the 
high-energy cutoff function is not important and influences only the value of certain low-energy
scales (e.g. the Kondo temperature for the Kondo model). In such a case the high-energy cutoff $D$
does no longer occur explicitly. Furthermore, for a wide class of time-translational invariant models 
it turns out that the effective Liouvillian has the form
\begin{equation}
\label{eq:L_general_form}
L(E)\,=\,L_\Delta(E)\,+\,E\,L'(E)\quad,
\end{equation}
where $L(E)=\int_0^\infty dt e^{iEt}L(t)$ is the Fourier-transform of the response function
$L(t-t')=L(t,t')\theta(t-t')$. In this decomposition $L_\Delta(E)$ and $L'(E)$ are slowly varying
logarithmic functions, where $L_\Delta(E)$ is proportional to some energy scale $\Delta$ of the model 
which can be anything except for the Fourier variable $E$. This form will be shown by the RG analysis in 
Section~\ref{sec:RG} for the concrete models under consideration but it remains an interesting 
question for the future how generic this form is. A large part of this tutorial deals with
the technical details of calculating the appearing functions $L_\Delta$ and $L'$. Before we
do that, we will first investigate the physical consequences for the time evolution in
Section~\ref{sec:time_evolution}. We will see that when $L(E)$ has the form \eqref{eq:L_general_form} 
the time evolution can generically be decomposed as
\begin{equation}
\label{eq:time_evolution_generic}
\rho(t)\,=\,\sum_n\,F_n(t)\,e^{-iz_n t}\,\rho_{t=0}\quad,
\end{equation}
where $z_n=\pm\Omega_n-i\Gamma_n$, with $\Omega_n,\Gamma_n\ge 0$, determine the oscillation
frequencies and decay rates of exponential decay, and $F_n(t)$ are pre-exponential functions,
which typically consist of power-laws $\sim 1/t^k$ ($k=1,2,\dots$) and logarithmic corrections
in the long-time limit $t\gg 1/|z_n|$. At least one of the exponential scales is zero $z_{\text{st}}=0$, which
determines the stationary state. \newline

It is the purpose of the present article to first discuss the generic physics of the time evolution
on the basis of the form \eqref{eq:L_general_form} of the effective Liouvillian, and with this 
motivation discuss the E-RTRG method for the calculation of $L(E)$ and its decomposition into
\eqref{eq:L_general_form}. Then we will summarize the results for the time evolution of the 
Kondo model, the ohmic spin boson model, and the IRLM. We note that other RG methods 
have been developed recently to discuss the 
time evolution of open quantum systems. The most important ones are the flow-equation method 
\cite{flow_equation} and the functional RG \cite{functional_RG}. The latter will be introduced in
the lecture B7 by V. Meden and is a method where one expands systematically in the short-ranged
renormalized interaction parameter present in the local system but not in the system-reservoir coupling, 
i.e. it is complementary to the RTRG technique where arbitrary local interactions can be treated but an
expansion in the renormalized system-reservoir coupling is needed. Besides the analytical RG methods,
there is also an extensive research going on to develop numerical methods to describe the time evolution,
like e.g. time-dependent numerical renormalization group \cite{TD-NRG}, time-dependent density matrix
renormalization group \cite{TD-DMRG}, iterative stochastic path integrals \cite{ISPI}, and quantum 
Monte Carlo \cite{QMC}. Furthermore, for special models, field-theoretical methods have been used to
find exact results \cite{exact}.

\section{Basic models}
\label{sec:models}

We start with the description of the basic models under consideration, where the quantum system 
consists of $2$ states coupled via spin (Kondo model), charge and potential (IRLM), or energy (spin boson)
fluctuations to a noninteracting environment. The total Hamiltonian is assumed to be time-translational
invariant and consists of three parts
\begin{equation}
\label{eq:hamiltonian}
H_{\text{tot}}\,=\,H\,+\,H_{\text{res}}\,+\,V\quad,\quad H_{\text{res}}\,=\,\sum_\alpha H_\alpha\quad,\quad
H_\alpha\,=\,\sum_{k\sigma}\,\epsilon_{\alpha\sigma k}\,a^\dagger_{\alpha\sigma k}a_{\alpha\sigma k}
\end{equation}
where $H$ is the Hamiltonian of the local quantum system, $V$ is the system-reservoir interaction, and
$H_{\text{res}}$ describes the noninteracting (fermionic or bosonic) reservoirs with field operators
$a_{\alpha\sigma k}$. $\alpha$ is the reservoir index, $\sigma$ the channel index (e.g. spin), and
the quantum number $k$ characterizes the energy. For convenience, for given $\alpha$ and $\sigma$, we 
will denote by $\omega=\epsilon_{\alpha\sigma k}-\mu_\alpha$ the energy of the reservoir states 
measured relative to the chemical potential, and we assume that the relation between $\omega$ and $k$ is
unique. As a consequence, the field operators of the reservoirs can be characterized by the multi-index
$1\equiv\eta\alpha\sigma\omega$, where $\eta=\pm$ distinguishes between creation ($\eta=+$) and 
annihilation operators ($\eta=-$). Depending on the model under consideration, we will define below
convenient forms of the field operators $a_1\equiv a_{\eta\alpha\sigma}(\omega)$ in continuum notation,
such that the commutation relations read (the upper/lower case refers always to bosons/fermions)
\begin{equation}
\label{eq:commutation_relation_1}
[a_{\alpha\sigma}(\omega),a^\dagger_{\alpha'\sigma'}(\omega')]_\mp\,=\,
\delta_{\alpha\alpha'}\,\delta_{\sigma\sigma'}\,\delta(\omega-\omega')\,\rho_{\alpha\sigma}(\omega)\quad,
\end{equation}
where $[\cdot,\cdot]_\mp$ denotes the commutator/anticommutator for bosons/fermions.
As defined below the spectral function $\rho_{\alpha\sigma}(\omega)$ contains the d.o.s. of the 
reservoirs and possibly frequency-dependencies of the system-reservoir couplings. Together with
the commutation relations 
\begin{equation}
\label{eq:commutation_relation_2}
[a_1,H_\alpha]\,=\,-\eta\,(\omega\,+\,\mu_\alpha)\,a_1 \quad,\quad 
[a_1,N_\alpha]\,=\,-\eta \,a_1\quad,
\end{equation}
it follows that the contraction of two reservoir field operators w.r.t. the equilibrium distribution
is given by
\begin{equation}
\label{eq:contraction}
{a_1\,a_{1'}
  \begin{picture}(-20,11) 
    \put(-22,8){\line(0,1){3}} 
    \put(-22,11){\line(1,0){12}} 
    \put(-10,8){\line(0,1){3}}
  \end{picture}
  \begin{picture}(20,11) 
  \end{picture}
}
\,=\,
\text{Tr}_{\text{res}}\,a_1\,a_{1'}\,\rho_{\text{res}}^{\text{eq}}
\,=\,
\delta_{1\bar{1}'}\,\rho_{\alpha\sigma}(\omega)\,f^\eta_\alpha(\omega)
\,=\,
\delta_{1\bar{1}'}\,
\left\{
\begin{array}{cl}
\eta \\ 1
\end{array}
\right\}\,\rho_{\alpha\sigma}(\omega)\,
f_\alpha(\eta\omega)\quad,
\end{equation}
where $\bar{1}\equiv -\eta\alpha\sigma\omega$ is defined by reversing the sign of $\eta$,
$\delta_{12}=\delta_{\eta_1\eta_2}\delta_{\alpha_1\alpha_2}\delta_{\sigma_1\sigma_2}\delta(\omega_1-\omega_2)$,
$f^+_\alpha(\omega)=f_\alpha(\omega)$, $f^-_\alpha(\omega)=1\pm f_\alpha(\omega)$, and
$f_\alpha(\omega)=(e^{\omega/T_\alpha}\mp 1)^{-1}$ is the Bose/Fermi distribution. \newline

In terms of the continuum field operators, the system-reservoir interaction $V$ is 
generically written as a sum of terms of the form 
\begin{equation}
\label{eq:V_generic}
V\,=\,{1\over n!}\,
\left\{
\begin{array}{cl}
1 \\ \eta_1\eta_2\dots\eta_n
\end{array}
\right\}
\, :a_n a_{n-1}\dots a_1: \,g_{12\dots n}
\,\rightarrow\,{1\over n!}\,g_{12\dots n}\,:a_1a_2\dots a_n: \quad,
\end{equation}
where $n=1,2,\dots$ is any integer, imlicit summation/integration is assumed over the multi-indices
$i\equiv \eta_i\alpha_i\sigma_i\omega_i$, the operator $g_{12\dots n}$ acts only on the 
local system, and $:\dots :$ denotes normal-ordering w.r.t. to the equilibrium distribution 
\eqref{eq:initial_dm} of the reservoirs (i.e. in any Wick-decomposition contractions are
not allowed within the normal-ordered expression). We call the operators $g_{1\dots n}$ $n$-point vertex
operators since, together with the corresponding superoperators \eqref{eq:G_vertex_liouville},
they will appear in the diagrammatic technique as vertices with $n$ reservoir lines, see 
Section~\ref{sec:diagrammatic_expansion}. For bosons the two forms for $V$ shown in \eqref{eq:V_generic}
are the same. For fermions, the first form is needed for the definition of the vertex operators and, for 
$n$ odd, $g_{1\dots n}$ is of fermionic nature and anticommutes with the reservoir field operators. However,
it can be shown \cite{schoeller_EPJ09} that, for the calculation of any local observables, the second
form for $V$ can be used and local and reservoir operators can be taken as if they commute.
The vertex operators have the properties 
\begin{equation}
\label{eq:prop_vertex}
g_{1\dots i\dots j\dots n}\,=\,\pm g_{1\dots j\dots i\dots n}\quad,\quad
g^\dagger_{1\dots n}\,=\,g_{\bar{n}\dots\bar{1}}\quad.
\end{equation}
The first relation can always be achieved by (anti-)symmetrization of the reservoir field operators
within the normal-ordering in \eqref{eq:V_generic}, whereas the second one is necessary for 
the property $V=V^\dagger$. In the following we will specify the definition of the continuum 
reservoir field operators $a_1$, the spectral density $\rho_{\alpha\sigma}(\omega)$ and the 
vertex operators $g_{1\dots n}$ for the various models. \newline

\begin{figure}
  \centering
  \includegraphics[width=0.5\hsize]{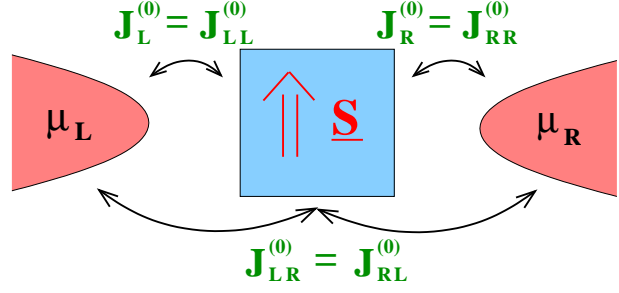}
%  \centerline{\psfig{figure=sketch.eps,scale=0.5}}
  \caption{A sketch of the nonequilibrium Kondo model. A local spin is coupled 
    via isotropic exchange couplings $J^{(0)}_{\alpha\alpha'}$ to the reservoir spins. 
    The two reservoirs are characterized by the same temperature $T$ but the chemical 
    potentials $\mu_L$ and $\mu_R$ can be different defining the voltage $V=\mu_L-\mu_R$ 
    across the system. The nondiagonal exchange couplings $J^{(0)}_{LR}=J^{(0)}_{RL}$ describe 
    spin exchange processes where a particle is transferred between the reservoirs, giving rise to a current.}
\label{fig:kondo}
\end{figure}
{\bf The Kondo model.} In its most basic form the Kondo model describes a local spin-${1\over 2}$ system coupled via
short-ranged and isotropic exchange couplings to fermionic reservoir spins, see Fig.~\ref{fig:kondo} for a 
sketch of the system. It is a model system to describe local spin fluctuations. \iffindex{spin fluctuations}
For the case of a single channel the Hamiltonian reads
\begin{equation}
\label{eq:kondo_ham}
H\,=\,h^{(0)}\,S_z \quad,\quad 
V\,=\,\sum_{\alpha\alpha'}\,
{J_{\alpha\alpha'}^{(0)}\over \sqrt{\rho_\alpha^{(0)}\rho_{\alpha'}^{(0)}}}\,
\sum_{\sigma\sigma'}\,{1\over 2}\underline{\sigma}_{\sigma\sigma'}\,
:\sum_k a^{\dagger}_{\alpha\sigma k}\sum_{k'}a_{\alpha'\sigma' k'}:\,\underline{S}\quad,
\end{equation}
where the isotropic exchange couplings $J^{(0)}_{\alpha\alpha'}=J^{(0)}_{\alpha'\alpha}$ are real and
dimensionless, $\underline{\sigma}=(\sigma_x,\sigma_y,\sigma_z)$ are the Pauli matrices, 
$\underline{S}$ is the local spin, and $h^{(0)}$ is the local bare magnetic field. $\rho^{(0)}_\alpha$ denotes the
d.o.s. of the reservoirs at the Fermi level. The Kondo model can be derived via a Schrieffer-Wolff 
transformation from the single-impurity Anderson model (see the lecture B3 by T. Costi),
in which case the exchange couplings fulfil the relation 
\begin{equation}
\label{eq:J_property}
J^{(0)}_{\alpha\alpha'}\,=\,2\,\sqrt{x_\alpha x_{\alpha'}}\,J^{(0)}\quad,\quad \sum_\alpha x_\alpha \,=\,1
\quad,\quad 0<x_\alpha<1\quad,
\end{equation}
where $x_\alpha$ are asymmetry factors weighting the energy broadening of the local level
from reservoir $\alpha$. Defining the continuum field operator by 
$a_1={1\over\sqrt{\rho_\alpha^{(0)}}}
\sum_k \delta(\omega-\epsilon_{\alpha\sigma k}+\mu_\alpha)a_{\eta\alpha\sigma k}$, 
with $1\equiv\eta\alpha\sigma\omega$, and
assuming a flat d.o.s. in the reservoirs, we obtain for the spectral function and the vertex operator
\begin{align}
\label{eq:kondo_spectral_function}
\rho_{\alpha\sigma}(\omega)\,&=\,\rho(\omega)\,=\,
{1\over \rho_\alpha^{(0)}}\sum_k \delta(\omega-\epsilon_{\alpha\sigma k}+\mu_\alpha)\,=\,{D^2\over D^2+\omega^2}
\quad,\\
\label{eq:kondo_vertex_operator}
g_{11'}\,&=\,{1\over 2}\,J_{\alpha\alpha'}^{(0)}\,\,\underline{\sigma}_{\sigma\sigma'}\cdot\underline{S}
\quad,\quad \text{for}\quad \eta=-\eta'=+\quad,
\end{align}
where $D$ is the band width of the reservoirs and, for convenience, we have chosen a Lorentzian for
the high-energy cutoff function. The case $\eta=-\eta'=-$ is obtained from $g_{11'}=-g_{1'1}$. \newline

{\bf The IRLM.} The IRLM is a basic model to describe charge and potential fluctuations.
\iffindex{potential fluctuations} It consists
of a single fermionic level, which is coupled to fermionic reservoirs via tunneling and
a local Coulomb interaction, see Fig.~\ref{fig:irlm} for a sketch of the system. Disregarding
the spin, the Hamiltonian is defined by 
\begin{equation}
\label{eq:irlm_ham}
H\,=\,\epsilon\,c^\dagger c \quad,\quad
V\,=\,\sum_\alpha\,{t_\alpha\over \sqrt{\rho_\alpha^{(0)}}}\,\sum_k\,
\left(a^{\dagger}_{\alpha k}c\,+\,\text{h.c.}\right)
\,+\,\sum_\alpha\,{U_\alpha\over\rho_\alpha^{(0)}}\,(c^\dagger c - {1\over 2})\,
\sum_{kk'}\,:a^\dagger_{\alpha k}a_{\alpha k'}:\quad,
\end{equation}
where $c$ is the field operator annihilating a particle on the local system, 
$t_\alpha$ are the tunneling matrix elements (in units of ${1\over\sqrt{\rho_\alpha^{(0)}}}$),
$U_\alpha$ denote the dimensionless Coulomb couplings, and $\epsilon$ is the bare energy of
the local level. At $\epsilon=0$ the model fulfils particle-hole symmetry. Defining the 
continuum field operators as for the Kondo model, with $1\equiv\eta\alpha\omega$ (i.e. omitting the
spin index), we find the same result \eqref{eq:kondo_spectral_function}
for the spectral function, and the vertex operators are given by
\begin{equation}
\label{eq:irlm_vertex_operator}
g_1\,=\,t_\alpha\,
\left\{\begin{array}{ll}
c& \textrm{for $\;\eta=+$}\\
c^{\dagger}& \textrm{for $\;\eta=-$}
\end{array}\right.
\quad,\quad
g_{11'}\,=\,\delta_{\eta,-\eta'}\,\delta_{\alpha\alpha'}\,\eta \,U_\alpha\,(c^\dagger c -{1\over 2})\quad.
\end{equation}
\begin{figure}
  \centering
  \includegraphics[width=0.5\hsize]{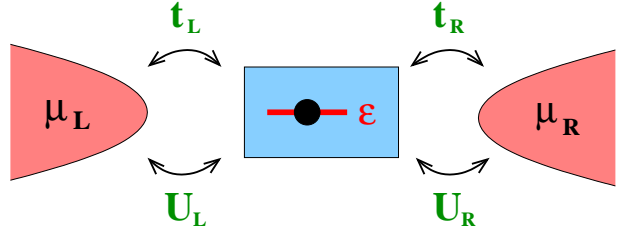}
%  \centerline{\psfig{figure=sketch.eps,scale=0.5}}
  \caption{A sketch of the IRLM. A local level without spin is coupled via
  tunneling and Coulomb interaction to several reservoirs.} 
\label{fig:irlm}
\end{figure}
\\

{\bf The spin boson model.} The spin boson model describes energy fluctuations, \iffindex{energy fluctuations} 
where a $2$-level system is
coupled linearly to a phonon bath, see Fig.~\ref{fig:spin_boson} for a sketch of the system.
The Hamiltonian is given by
\begin{equation}
\label{eq:spinboson_ham}
H_{\text{res}}\,=\,\sum_k\,\omega_k\,a^\dagger_k a_k\quad,\quad
H\,=\,{1\over 2}\,\epsilon\,\sigma_z\,-\,{1\over 2}\,\Delta\,\sigma_x \quad,\quad
V\,=\,{1\over 2}\,\sigma_z\,\sum_k\,\alpha_k\,(a_k\,+\,a_k^\dagger)\quad,
\end{equation}
where $\epsilon$ and $\Delta$ denote the bias and the tunneling of the local
$2$-level system, respectively. The phonon frequencies $\omega_k>0$ are positive, and the 
equilibrium phonon distribution is characterized by temperature $T$. The $k$-dependence of
the real coupling constants $\alpha_k$ and the phonon frequencies $\omega_k$ is considered by defining the 
continuum field operators by $a_1=\sum_k \alpha_k \delta(\omega-\omega_k)a_{\eta k}$ with $1\equiv\eta\omega$. 
This leads to the following spectral function and vertex operator
\begin{equation}
\label{eq:spinboson_sf_vo}
\rho(\omega)\,=\,\sum_k\,\alpha_k^2\,\delta(\omega-\omega_k)\,=\,
2\,\alpha\,\omega\,\left({\omega\over D}\right)^{s-1}\theta(\omega)\,{D^2\over D^2+\omega^2}
\quad,\quad
g_1\,=\,{1\over 2}\,\sigma_z \quad,
\end{equation}
where $\alpha$ is a dimensionless coupling constant, and we have again chosen a Lorentzian
high-energy cutoff function with band width $D$. The special form chosen for $\rho(\omega)$ 
describes the ohmic case for $s=1$ considered in this article, whereas $s<1$ ($s>1$) define 
the sub-ohmic (super-ohmic) cases. For the special case $\Delta=0$ the spin boson model can
be solved exactly \cite{functional_integral} with the result
\begin{equation}
\label{eq:sb_exact}
\langle\sigma_{x,y}\rangle(t)\,=\,e^{-h(t)}\,\langle\sigma_{x,y}\rangle_{t=0}
\quad,\quad
\langle\sigma_z\rangle(t)\,=\,\langle\sigma_z\rangle_{t=0}\quad,
\end{equation}
with $h(t)=-\int d\omega(\rho(\omega)/\omega)(1-\cos(\omega t))(1+2f(\omega))$, where
$f(\omega)=(exp(\omega/T)-1)^{-1}$ is the Bose function.
\newline

\begin{figure}
  \centering
  \includegraphics[width=0.5\hsize]{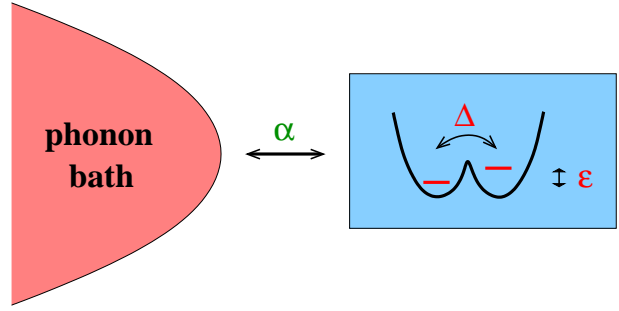}
%  \centerline{\psfig{figure=sketch.eps,scale=0.5}}
  \caption{A sketch of the spin boson model. A $2$-level system, characterized by tunneling $\Delta$
  and bias $\epsilon$ is coupled linearly via the dimensionless coupling constant $\alpha$ to a
  phonon bath of harmonic oscillators.}
\label{fig:spin_boson}
\end{figure}
For the special case of $\alpha$ close to ${1\over 2}$, the ohmic spin boson model can be mapped
on the IRLM with a single reservoir (with $\mu=0$) \cite{functional_integral}. The parameters
$U$ and $t$ of the IRLM are related to $\alpha$ and $\Delta$ of the ohmic spin boson
model in the following way
\begin{equation}
\label{eq:irlm_sb_parameters}
U\,=\,1\,-\,\sqrt{2\alpha}\quad,\quad
\Gamma^{(0)}\,=\,2\,\pi\,t^2\,=\,{\Delta^2\over D}\quad.
\end{equation}
The local occupation $\langle n\rangle(t)=\langle c^\dagger c\rangle(t)$ of the IRLM is 
related to the expectation value $\langle \sigma_z\rangle(t)$ of the ohmic spin boson model via
\begin{equation}
\label{eq:irlm_sb_occupation}
2\,\langle n \rangle(t)\,-\,1\,=\,\langle \sigma_z\rangle(t)\quad,
\end{equation}
whereas the expectation value $\langle\sigma_{x,y}\rangle(t)$ of the spin boson model 
is related to expectation values of highly nonlinear operators involving reservoir degrees of
freedom in the IRLM. The value $\alpha={1\over 2}$ is of special importance since, at this point,
the time evolution of $\langle\sigma_z\rangle(t)$ changes from an oscillating one 
(for $\alpha<{1\over 2}$) to a purely decaying one (for $\alpha>{1\over 2}$) 
\cite{functional_integral,kennes_PRL13,kashuba_kennes_PRB13}. 
Correspondingly, for the IRLM, this crossover
occurs when the sign of the Coulomb interaction $U$ is changed.

\section{Kinetic equation and time evolution}
\label{sec:time_evolution}

In this section we aim at discussing the time evolution from a generic point of view
based on the general form \eqref{eq:kinetic_equation} of the kinetic equation and the
form \eqref{eq:L_general_form} of the effective Liouvillian for the case of a time-translational
invariant Hamiltonian. Using $L(t,t')=L(t-t')$ the kinetic equation reads
\begin{equation}
\label{eq:kinetic_equation_special}
i\dot{\rho}(t)\,=\,\int_{0}^t\,dt'\,L(t-t')\,\rho(t')\quad,
\end{equation}
where, for convenience, we have set the initial time $t_0=0$.
The reduced density matrix $\rho(t)$ acts only in local space, i.e. has matrix elements
$\rho(t)_{ss'}=\langle s|\rho(t)|s'\rangle$, where $s$ and $s'$ are states of the local quantum system.
In contrast, the superoperator $L(t)$ acts on local operators A, i.e. the matrix elements can be written
as $L_{s_1s_2,s_1^\prime s_2^\prime}=\langle s_1 s_2|L(t)|s_1^\prime s_2^\prime\rangle$, where
$|ss'\rangle=|s\rangle\langle s'|$ are the basis elements ($=$ operators) in Liouville space 
\iffindex{Liouville space} and
$\langle ss'|A=\langle s|A|s'\rangle$ are the corresponding dual vectors. The density matrix fulfils
the property of conservation of probability $\text{Tr}\rho(t)=1$ and is self-adjoint $\rho(t)=\rho(t)^\dagger$.
It is straightforward to show that the kinetic equation respects these properties if and only if the 
effective Liouville operator fulfils the properties
\begin{equation}
\label{eq:L(t)_properties}
\text{Tr}\,L(t)\,=\,\sum_s\,L(t)_{ss,\cdot\cdot}\,=\,0 \quad,\quad
L(t)^c\,=\,-\,L(t) \quad,
\end{equation}
where the $c$-transform is defined by 
$L(t)^c_{s_1s_2,s_1^\prime s_2^\prime}=L(t)^*_{s_2s_1,s_2^\prime s_1^\prime}$ and fulfils the useful property
$(L(t)A)^\dagger = L(t)^c A^\dagger$. In Fourier space $L(E)=\int_{0}^\infty dt e^{iEt}L(t)$ this 
means
\begin{equation}
\label{eq:L(E)_properties}
\text{Tr}\,L(E)\,=\,0 \quad,\quad
L(E)^c\,=\,-\,L(-E^*) \quad,
\end{equation}
or for the quantities $L_\Delta(E)$ and $L'(E)$ appearing in the decomposition \eqref{eq:L_general_form}
\begin{equation}
\label{eq:properties_L_Delta_Z}
\text{Tr}\,L_\Delta(E)\,=\,\text{Tr}\,L'(E)\,=\,0\quad,\quad
L_\Delta(E)^c\,=\,-L_\Delta(-E^*)\quad,\quad
L'(E)^c\,=\,L'(-E^*)\quad.
\end{equation}
\\

With $\rho(E)=\int_{0}^\infty dt e^{iEt}\rho(t)$, the kinetic equation reads in Fourier space
$E\rho(E)-i\rho_{t=0}=L(E)\rho(E)$ leading to the formal solution 
\begin{equation}
\label{eq:solution}
\rho(E)\,=\,i\,R(E)\,\rho_{t=0} \quad,\quad R(E)\,=\,{1\over E\,-\,L(E)} \quad.
\end{equation}
We now investigate the consequences of the generic form \eqref{eq:L_general_form} of the 
effective Liouvillian $L(E)$. Using inverse Fourier transform, the time evolution can be 
calculated for $t>0$ from
\begin{equation}
\label{eq:time_evolution}
\rho(t)\,=\,{i\over 2\pi}\,\int_{-\infty+i0^+}^{\infty+i0^+}\hspace{-0.5cm}dE\,
e^{-iEt}\,R(E)\,\rho_{t=0}\,=\,
{i\over 2\pi}\,\int_{-\infty+i0^+}^{\infty+i0^+}\hspace{-0.5cm}dE\,
e^{-iEt}\,\tilde{R}(E)\,Z'(E)\,\rho_{t=0}\quad,
\end{equation}
where we have defined 
\begin{equation}
\label{eq:tilde_L_Z}
\tilde{R}(E)\,=\,{1\over E\,-\,\tilde{L}_\Delta(E)}\quad,\quad
\tilde{L}_\Delta(E)\,=\,Z'(E)\,L_\Delta(E) \quad,\quad
Z'(E)\,=\,{1\over 1\,-\,L'(E)} \quad.
\end{equation}
By convention, $Z'(E)$ is called the $Z'$-factor operator. 
\begin{figure}
  \centering
  \includegraphics[width=0.5\hsize]{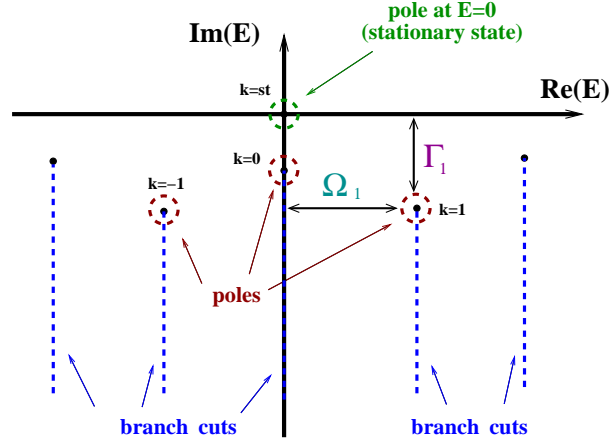}
%  \centerline{\psfig{figure=sketch.eps,scale=0.5}}
  \caption{The analytic structure of the resolvent $R(E)=1/(E-L(E))$. 
    Generically, there is a pole at $E=z_{\text{st}}^p=0$ corresponding to the stationary state. Branch cuts
    can occur in the lower half starting either at a branching point or at a pole.
    This will be demonstrated in Section~\ref{sec:diagrammatic_expansion}, see Eq.~\eqref{eq:branching_point}.
    The analytic structure is symmetric w.r.t. the imaginary axis. The poles are
    denoted by $z_k^p=\pm\Omega_k-i\Gamma_k$. A pole lying on the imaginary axis gets 
    the index $k=0$, the others are labelled by $\pm k$ with $k=1,2,\dots$.}
\label{fig:analytic_structure}
\end{figure}
The last form of (\ref{eq:time_evolution}) is very helpful for the evaluation of the
energy integral because it explicitly exhibits the slowly varying logarithmic functions
$\tilde{L}_\Delta(E)$ and $Z'(E)$. 
The energy integral $\int dE$ is calculated by closing
the integration contour in the lower half of the complex plane and deforming the contour
such that the poles and branch cuts of the integrand are enclosed, see Fig.~\ref{fig:analytic_structure}.
To identify the singularities of the integrand we use the spectral decomposition of the 
Liouvillian $\tilde{L}_\Delta(E)$ in terms of its eigenvalues $\lambda_k(E)$ and 
corresponding projectors $P_k(E)$ 
\begin{equation}
\label{eq:spectral}
\tilde{L}_\Delta(E)\,=\,\sum_k\,\lambda_k(E)\,P_k(E)\quad.
\end{equation}
Since we deal with a non-hermitian superoperator, we have to distinguish the left and
right eigenvectors, which we denote in Dirac notation by $|x_k(E)\rangle$ and $\langle \bar{x}_k(E)|$
\begin{equation}
\label{eq:eigenvectors}
\tilde{L}_\Delta(E)\,|x_k(E)\rangle\,=\,\lambda_k(E)\,|x_k(E)\rangle\quad,\quad
\langle \bar{x}_k(E)|\,\tilde{L}_\Delta(E)\,=\,\langle \bar{x}_k(E)|\,\lambda_k(E)\quad.\quad
\end{equation}
The eigenvectors fulfil the orthonormalization condition 
$\langle \bar{x}_k(E)|x_{k'}(E)\rangle=\delta_{kk'}$ and the projectors are given by 
$P_k(E)=|x_k(E)\rangle\langle \bar{x}_k(E)|$ with $\sum_k P_k(E)=1$. \newline

Due to the condition $\text{Tr}\tilde{L}_\Delta(E)=0$, we obtain either $\lambda_k(E)=0$
or $\text{Tr}\,|x_k(E)\rangle=0$. Therefore, the Liouvillian has always an eigenvalue zero,
which we characterize by the index $k=\text{\text{st}}$ since it corresponds to the stationary state (see below).
The other eigenvalues are numerated by $k=0,\pm 1,\pm 2, \dots$. Normalizing the eigenvector with 
$k=\text{st}$ according to $\text{Tr}|x_{\text{st}}(E)\rangle=1$ and using 
$\langle \bar{x}_{\text{st}}(E)|=\text{Tr}$, we get
\begin{eqnarray}
\label{eq:zero_eigenvectors}
\quad\text{Tr}\,|x_{\text{st}}(E)\rangle\,&=&\,\sum_s\,\langle ss|x_{\text{st}}(E)\rangle\,=\,1\quad,\quad
\langle \bar{x}_{\text{st}}(E)|ss\rangle\,=\,1\\
\label{eq:tr_eigenvectors}
\text{Tr}\,|x_k(E)\rangle\,&=&\,\sum_s\,\langle ss|x_k(E)\rangle\,=\,0\quad,\quad
\text{for}\quad k=0,\pm 1,\pm 2,\dots\quad.
\end{eqnarray}
As a consequence we get 
\begin{equation}
\label{eq:P_st}
P_{\text{st}}(E)\,=\,|x_{\text{st}}(E)\rangle\,\text{Tr}\quad,
\end{equation}
and the property
$\text{Tr}L_\Delta(E)=\text{Tr}L'(E)=0$ can also be written as
\begin{equation}
\label{eq:P_st_property}
P_{\text{st}}(E)\,Z'(E)\,=\,P_{\text{st}}(E)\quad,\quad
P_{\text{st}}(E)\,L_\Delta(E)\,=\,0\quad.
\end{equation}
Due to the condition $\tilde{L}_\Delta(E)^c=-\tilde{L}_\Delta(-E^*)$, the eigenvalues and projectors 
occur always in pairs (except for $k=0,\text{st}$ where we define $k\equiv -k$) with 
\begin{equation}
\label{eq:pairs}
\lambda_{-k}(E)\,=\,-\lambda_k(-E^*)\quad,\quad
P_{-k}(E)\,=\,P_k(-E^*)^c\quad.
\end{equation}
\\

Using the spectral representation, the time evolution can be written as
\begin{equation}
\label{eq:time_evolution_spectral}
\rho(t)\,=\,
{i\over 2\pi}\,\sum_k\,\int_\gamma dE\,e^{-iEt}\,
{1\over E-\lambda_k(E)}\,P_k(E)\,Z'(E)\,\rho_{t=0}\quad,
\end{equation}
where $\gamma$ is an integration contour which encloses the lower half of the
complex plane including the real axis. Poles are located at $E=z_k^p=\lambda_k(z_k^p)=\pm\Omega_k-i\Gamma_k$,
with $\Omega_k,\Gamma_k\ge 0$, where $z_{\text{st}}^p=0$ is a pole at the origin, see 
Fig.~\ref{fig:analytic_structure}.
At {\bf zero temperature}, which we consider from now on, additional nonanalytic features occur from
branch cuts since $\lambda_k(E)$, $P_k(E)$ and $Z'(E)$ depend
logarithmically via terms $\sim \ln({D\over E-z_n})$ generated by the ultraviolet
divergencies from the high-energy cutoff $D$ (at finite temperature the branch cuts turn
into an infinite number of discrete poles separated by $2\pi T$). From the structure of the
perturbation theory (see below) it can be seen that the singularities $z_n$ are associated
with poles of the resolvents $\tilde{R}(E_{1\dots n})$, where 
\begin{equation}
\label{eq:E_1...n}
E_{1\dots n}\,=\,E\,+\,\bar{\mu}_{1\dots n}\quad,\quad
\bar{\mu}_{1\dots n}\,=\,\bar{\mu}_1\,+\dots +\,\bar{\mu}_n\quad,\quad
\bar{\mu}_1\,=\,\eta_1\,\mu_{\alpha_1}\quad,
\end{equation}
i.e. are located at $z_n=E$ with $E_{1\dots n}=E+\bar{\mu}_{1\dots n}=z_k^p$.
Therefore, the singularities $z_n=z_k^p-\bar{\mu}_{1\dots n}$ are generically given by the 
poles shifted by some linear combination of the chemical potentials of the reservoirs. \newline

In Section~\ref{sec:RG} we will see how $\tilde{L}_\Delta(E)$ and $Z'(E)$ can be determined from 
differential equations, see Eq.~\eqref{eq:L_Delta_L'_rg},
where we differentiate w.r.t the Fourier variable $E$. These differential 
equations are defined in the whole complex plane and will be the RG equations of the E-RTRG method. 
$E$ is called the flow paramater and a solution of the RG equations along a certain path is called
the RG flow. The particular advantage
is that these RG equations can be solved along the paths $E=z_n+i\Lambda \pm O^+$, with $\Lambda$
real, starting at some high value $\Lambda\sim D$ down to $\Lambda=-\infty$. Since no singularities
are present on these paths, it can even be numerically enforced that the branch cuts start at $z_n$
and point into the direction of the negative imaginary axis. Furthermore, the jump of the Liouvillian 
at the branch cuts can be determined from the difference of the two solutions and the integrals
around the branch cuts can be calculated. The choice that the branch cuts point into the direction
of the negative imaginary axis is very convenient since $e^{-iEt}=e^{-iz_n t}e^{-xt}$ 
is exponentially decaying in $xt$, which allows an analytical discussion of the long-time limit
(see below). Using $E=z_n-ix\pm O^+$, the integration around a particular branch cut (including the
case when the branching point is a pole) gives the contribution $\rho_n(t)=F_n(t)e^{-iz_n t}\rho_{t=0}$
to the time evolution with
\begin{equation}
\label{eq:F_contribution}
F_n(t)\,=\,{1\over 2\pi}\,\int_0^\infty\,dx\,e^{-xt}\,\left\{R(z_n-ix+0^+)-R(z_n-ix-0^+)\right\}
\end{equation}
such that the total time evolution can be written in the form \eqref{eq:time_evolution_generic}
\begin{equation}
\label{eq:total_evolution}
\rho(t)\,=\,\sum_n\,\rho_n(t)\,=\,\sum_n\,F_n(t)\,e^{-iz_n t}\,\rho_{t=0}\quad.
\end{equation}
For the further evaluation of $F_n(t)$ it is important to distinguish between the cases
when the branching point is a pole or not. We label the contributions from branching poles
$z_k^p$ by $F_k^p(t)$ and $\rho_k^p(t)$ and the others by $F_n^b(t)$ and $\rho_n^b(t)$, such
that \eqref{eq:total_evolution} reads
\begin{align}
\nonumber
\rho(t)\,&=\,\sum_k\,\rho_k^p(t)\,+\,\sum_n\,\rho_n^b(t)\\
\label{eq:total_evolution_pb}
\,&=\,\sum_k\,F_k^p(t)\,e^{-iz_k^p t}\,\rho_{t=0}\,+\,\sum_n\,F_n^b(t)\,e^{-iz_n^b t}\,\rho_{t=0}\quad.
\end{align}
Thereby we note that the same singularity $z_k^p=z_n^b$ can appear as a branching pole and as a branching point, 
since a certain term involving $\lambda_k(E)$ in \eqref{eq:time_evolution_spectral} can have a branch
cut at $z_n^b=z_{k'}^p$ with $k'\ne k$.
Generically, for weakly coupled system-reservoir systems, the contributions $\rho_n^b(t)$ are smaller 
since they are proportional to the system-reservoir coupling (see below Eq.~\eqref{eq:branching_point}). 
However, if
the decay rates occuring in $z_n^b$ are smaller than those ones of $z_k^p$, the relative order of the
various terms can change as function of time, as discussed e.g. in detail in 
Refs.~\cite{kennes_PRL13,kashuba_kennes_PRB13}
for the IRLM with positive Coulomb interaction or the ohmic spin boson model for $\alpha$ close but
slightly below the value $\alpha={1\over 2}$. In the Markovian approximation, only the contributions
$\rho_k^p(t)$ remain and the pre-exponential functions are approximated by constants of $O(1)$. \newline

{\bf Time-evolution regimes.}
Using the general expressions (\ref{eq:time_evolution}) and (\ref{eq:time_evolution_spectral}),
one can discuss the qualitative form of the time evolution in different time regimes. For
{\bf short times} $t\ll 1/|z_n|$, only high frequencies $E\sim 1/t\gg |z_n|$ matter in 
Eq.~\eqref{eq:time_evolution}, i.e. the cutoff scales $z_n$ 
in the logarithmic terms are unimportant and can be neglected. Furthermore, to leading order,
we can replace $E\rightarrow 1/t$ in the logarithmic parts, and we obtain from 
(\ref{eq:time_evolution})
\begin{equation}
\label{eq:short_times}
\rho(t)\,=\,{i\over 2\pi}\,\int_\gamma dE\,e^{-iEt}\,
{1\over E-\tilde{L}_\Delta(1/t)}\,Z'(1/t)\,\rho_{t=0}
\,=\,e^{-i\tilde{L}_\Delta(1/t)t}\,Z'(1/t)\,\rho_{t=0}\quad.
\end{equation}
Expanding the exponential one finds in leading order that the logarithmic dependence of 
$Z'(1/t)$ and $\tilde{L}_\Delta(1/t)$ at high energies determine the short time behavior. 
This means that the RG equations are cut off at the large 
energy scale $E=1/t$, which is the poor man scaling regime, where all the cutoff scales $z_n$
are unimportant. In this regime the time evolution is determined by the scaling of $Z'(1/t)$
and $\tilde{L}_\Delta(1/t)$. If, in addition, $t\gg 1/D$, where $D$ is the high-energy 
cutoff, one obtains universal time evolution in the short-time regime. It means that all 
leading logarithmic divergencies $\sim (\alpha\ln(Dt))^n$ have been resummed in the 
functions $Z'(1/t)$ and $\tilde{L}_\Delta(1/t)$, where $\alpha\ll 1$ is some 
small dimensionless coupling parameter. Based on this unified picture the universal short-time 
behaviour has been derived in 
Refs.~\cite{pletyukhov_PRL10,kashuba_kennes_PRB13,kashuba_PRB13} for
the Kondo model, the IRLM, and the ohmic spin boson model, in accordance with similiar results
of previous literature. \\

For {\bf intermediate and long times} $t\gtrsim 1/|z_n|$, we have to study the contributions from
the poles and branch cuts in detail, based on the decomposition \eqref{eq:total_evolution_pb}.
We start with the contributions from the {\bf branch cuts starting at a pole} $z_k^p$, which we evaluate by using the form
(\ref{eq:time_evolution_spectral}). For the branch cut integral
we set $E=z_k^p-ix\pm 0^+$ and replace in leading order $\lambda_k(E)\rightarrow z_k^p$ 
and the logarithmic function $P_k(E)Z'(E)$ by its average
$\bar{P}_k(z_k^p-ix)\bar{Z}'(z_k^p-ix)$ over the branch cut, where 
$\bar{A}(E)={1\over 2}(A(E+0^+)+A(E-0^+))$. Furthermore, in leading order, we can 
use $x\rightarrow 1/t$ in the logarithmic functions. This gives the result
\begin{equation}
\label{eq:branching_pole_zw}
F_k^p(t)\,\approx\,{1\over 2\pi}\,\int_{0^-}^\infty dx\,e^{-xt}\,
\left({1\over -ix+0^+}-{1\over -ix-0^+}\right)\,\bar{P}_k(z_k^p-i/t)\,\bar{Z}'(z_k^p-i/t)
\quad.
\end{equation}
Using ${1\over -ix+0^+}-{1\over -ix-0^+}=2\pi\delta(x)$, we obtain the following 
contribution to the total time evolution \eqref{eq:total_evolution}
\begin{equation}
\label{eq:branching_pole}
\rho_k^p(t)\,\approx\,\bar{P}_k(z_k^p-i/t)\,\bar{Z}'(z_k^p-i/t)\,e^{-iz_k^p t}\,\rho_{t=0}
\quad,
\end{equation}
i.e., for $z_k^p=\pm\Omega_k-i\Gamma_k$, an exponential one with oscillation $\Omega_k$
and decay rate $\Gamma_k$, modulated by a logarithmic function. 
For the special term $k=\text{st}$, where $z^p_{\text{st}}=0$, 
$P_{\text{st}}(E)=|x_{\text{st}}(E)\rangle\text{Tr}$ and $P_{\text{st}}(E)Z'(E)=P_{\text{st}}(E)$, 
we get the following contribution to the time evolution
\begin{equation}
\label{eq:branching_pole_zero}
\rho_{\text{st}}^p(t)\,\approx\,|\overline{x_{\text{st}}}(-i/t)\rangle
\,\xrightarrow{t\rightarrow\infty}\,\rho_{\text{st}}\,=\,|\overline{x_{\text{st}}}(0)\rangle
\quad,
\end{equation}
i.e. we see that for $t\rightarrow\infty$ one always gets the stationary distribution
$\rho_{\text{st}}$ but, if $z_{\text{st}}^p$ is a branching pole, logarithmic corrections 
can occur for the time evolution which do not decay exponentially. 
We note that for the models discussed here, there is no logarithmic term in the 
diagrammatic series involving the pole $z^p_{\text{st}}$. In addition, there is no accidental 
pole $z_{k\ne\text{st}}^{p}=0$, and therefore the pole at $E=0$ is isolated and has no attached branch cuts. \newline

The evaluation of a {\bf branch cut starting at a branching point} $z_n^b$ which is not a pole is more 
subtle since both $\lambda_k(E)$ and $P_k(E)Z'(E)$ can be discontinuous and cancellations can occur 
between the two contributions. Therefore, it is more convenient to start from the first expression
of (\ref{eq:time_evolution}) involving the resolvent $R(E)$. Denoting by 
$\delta A=A_+-A_-$ the jump across the branch and by $\bar{A}={1\over 2}(A_++A_-)$ the
average value, with $A_\pm=A(E\pm 0^+)=\bar{A}\pm{1\over 2}\delta A$, one finds for the 
jump of the resolvent expanding in the small quantity $\delta L$ (leading to higher orders in the
renormalized coupling constants)
\begin{equation}
\label{eq:jump_propagator}
\delta R(E)\,=\,R_+\,\delta L\,R_-\,=\,{1\over E-\bar{L}}\,\delta L\,{1\over E-\bar{L}}
\,+\,O(\delta L^3)\quad.
\end{equation}
Using $\overline{AB}-\bar{A}\bar{B}={1\over 4}\delta A\delta B$, we get
\begin{equation}
\label{eq:average_propagator}
{1\over E-\bar{L}}\,=\,\overline{{1\over E-L}}+O(\delta L^2)\,=\,
\sum_k\overline{{1\over E-\lambda_k}P_k Z'}+O(\delta L^2)\,=\,
\sum_k{1\over E-\bar{\lambda}_k}\bar{P}_k\bar{Z}'+O(\delta L^2)
\end{equation}
Inserting this in (\ref{eq:jump_propagator}), neglecting $O(\delta L^3)$, and approximating
$E=z_n^b-ix\rightarrow z_n^b-i/t$ in the
logarithmic functions $\bar{\lambda}_k$, $\bar{P}_k$ and $\bar{Z}'$, we get  
the following result for the branch cut integral
\begin{equation}
\label{eq:F_branching_point}
F_n^b(t)\,\approx\,{1\over 2\pi}\,\sum_{\stackrel{k k'}{z_k^p,z_{k'}^p\ne z_n^b}}\,\int_0^\infty dx\,e^{-xt}\,
{1\over z_n^b-ix-\bar{\lambda}^n_k}\,\bar{P}^n_k\,\bar{Z}^{\prime n}\,\delta L(z_n^b-ix)\,
{1\over z_n^b-ix-\bar{\lambda}^n_{k'}}\,\bar{P}^n_{k'}\,\bar{Z}^{\prime n}
\,,
\end{equation}
where $\bar{\lambda}_k^n=\bar{\lambda}_k(z_n^b-i/t)$, $\bar{P}_k^n=\bar{P}_k(z_n^b-i/t)$ and
$\bar{Z}^{\prime n}=\bar{Z}'(z_n^b-i/t)$. We have omitted the cases $z_k^p=z_n^b$ or $z_{k'}^p=z_n^b$ since
we consider a branching point and not a branching pole. Since
$\bar{\lambda}_k^n\approx z_k^p$, we can neglect $x$ in the denominators of the resolvents
for times $t\sim 1/x \gg 1/|z_n^b-z_{k,k'}^p|$. In this case, the long-time scaling 
is determined by the scaling of $\delta L(z_n^b-ix)$ for small $x$. Besides additional 
logarithmic corrections (which again can be treated by replacing $x\rightarrow 1/t$), 
we will show in Section~\ref{sec:RG} that 
\begin{equation}
\label{eq:delta_L_charge_fluctuations}
\delta L(z_n^b-ix)\,\sim\, \theta(x)
\end{equation}
for models with charge fluctuations \iffindex{charge fluctuations} (like the IRLM) and 
\begin{equation}
\label{eq:delta_L_spin/orbital_fluctuations}
\delta L(z_n^b-ix)\,\sim \, x\,\theta(x)
\end{equation}
for models with spin/orbital or energy fluctuations (like the Kondo and the ohmic spin boson model),
see Eq.~\eqref{eq:delta_L_rg_explicit}.
Therefore, if $x$ can be neglected in the resolvents of the integrand of (\ref{eq:F_branching_point}),
we obtain (up to logarithmic corrections) $\rho_n^b(t)\sim 1/t$ for charge fluctuations and
$\rho_n^b(t)\sim 1/t^2$ for spin/orbital and energy fluctuations. For special resonant cases, 
where $z_n^b$ comes close to $z_k^p$ or $z_{k'}^p$, one can also define time regimes 
$1/|z_n^b|\lesssim t\ll 1/|z_n^b-z_{k,k'}^p|$, where $x$ dominates in the denominators for certain
values of $k$ or $k'$, leading to different scaling. If $x$ is not neglected in \eqref{eq:branching_point},
the integral can also be calculated exactly, leading typically to exponential integrals from which
the whole crossover behaviour from intermediate $t\sim 1/|z_n|$ to long times $t\gg 1/|z_n|$
can be calculated. \newline

In the regime of intermediate to long times the cutoff scales $z_n$ are very important. Each term
of the series \eqref{eq:total_evolution_pb} has to be treated separately, leading to different
scaling of the individual terms (in contrast to the short-time regime, where all exponentials
can be approximated by one and only the sum of all pre-exponential functions matters). As we have
seen above, various functions $K(z_n-i/t)$ with logarithmic scaling occur in the projectors, 
the $Z$-factors, and the jump of the Liouvillian. In bare perturbation theory, the logarithmic
functions $K(E)$ will contain powers of terms $\sim \alpha\ln{D\over E-z_m}$. To get rid of
the high-energy cutoff $D$, a standard technique is to resum first all leading logarithmic divergencies
$\sim (\alpha\ln{D\over \Lambda_c})^n$, where $\Lambda_c\gtrsim|z_m|$ is some maximal physical
low energy scale. Technically, this can be achieved by cutting off the RG flow at $\Lambda_c$, 
defining renormalized coupling constants $\alpha_c$ at this point, and expanding the full
solution for $|E|\lesssim|z_m|$ in $\alpha_c$. This is possible if $\alpha_c$ is small, i.e. if
$\Lambda_c$ is much larger than the strong coupling scale $\Lambda^*$, where the coupling constants
become of $O(1)$. As a result, $K(E)$ will contain powers of 
logarithmic terms $\sim \alpha_c\ln{\Lambda_c\over E-z_m}$. For $E=z_n-i/t$ the most dangerous case
is $n=m$, leading to powers in the time-dependent parameter $\alpha_t\sim\alpha_c\ln{\Lambda_c t}$. 
Since $\alpha_c\ll 1$, this parameter is small $\alpha_t\ll 1$, unless time is exponentially large.
Therefore, it can be treated perturbatively, leading to logarithmic corrections $\sim\alpha_t$ 
in the pre-exponential functions. This strategy has been 
used in Refs.~\cite{pletyukhov_PRL10,RTRG_irlm,kennes_PRL13,kashuba_kennes_PRB13,kashuba_PRB13} 
to determine the time evolution at
intermediate and long times (but not exponenitally large times) for the Kondo model, the IRLM,
and the spin boson model. \newline

Finally, the most complicated time regime is the one at 
{\bf exponentially large times}, where $\alpha_t\sim O(1)$. In this regime, a perturbative treatment
is no longer possible and all powers of $\alpha_t$ are important. These logarithmic divergencies
at {\it low energies} are independent of those at {\it large energies} and can even arise if there is
no logarithmic divergence at high energies. Their occurence is related to the fact that, concerning
the time evolution, the final cutoff scale at low energies is set by inverse time $1/t$ and not by 
decay rates. The latter holds only for the calculation of stationary properties, see 
Refs.~\cite{schoeller_EPJ09,schoeller_PRB09,RG_noneq_D_cutoff}.
The E-RTRG method is unique in the sense that it is also capable of resumming the logarithmic 
divergencies at low energies, provided the renormalized coupling constants remain small when $E$ approaches
one of the singularities $z_n$. Recently, this has been achieved in a controlled way for the ohmic
spin boson model \cite{kashuba_PRB13}, where deviations from previously predicted scaling behaviour have 
been found. Results for the Kondo model and the IRLM are still under investigation in this regime.
In particular for the Kondo model, the problem is that the renormalized coupling constants become
of $O(1)$ when approaching one of the singularities although they might be small for the calculation
of stationary quantities at $E=0$. Thus, weak-coupling problems for stationary quantities can turn
into strong-coupling ones for the calculation of the long-time behaviour at exponentially large times.

\section{Diagrammatic expansion}
\label{sec:diagrammatic_expansion}

{\bf Effective Liouvillian.} In this section we will derive a quantum field theoretical diagrammatic representation of the 
effective Liouvillian by expanding in the system-reservoir interaction $V$,
following Refs.~\cite{schoeller_EPJ09,saptsov_PRB12}. Although this can
be done for the general case of an explicitly time-dependent Hamiltonian \cite{kashuba_kennes_PRB13}, here
we will restrict ourselves to the more simpler case of a time-translational invariant
Hamiltonian. To find a diagrammatic expansion of the effective Liouvillian $L(E)$ in Fourier space, we try to
bring the local density matrix $\rho(E)$ into the form \eqref{eq:solution}. We start from the 
formal solution of the von Neumann equation \eqref{eq:von_Neumann} for the total density matrix, use
the initial condition \eqref{eq:initial_dm}, and obtain by expanding in the system-reservoir interaction
\begin{align}
\nonumber
\rho(E)\,&=\,\int_0^\infty\,dt\,e^{iEt}\,\text{Tr}_{\text{res}}\,\rho_{\text{tot}}(t)
\,=\,\int_0^\infty\,dt\,e^{iEt}\,\text{Tr}_{\text{res}}\,e^{-iL_{\text{tot}}t}\,\rho_{\text{tot}}(t=0)\\
\nonumber
\,&=\,\text{Tr}_{\text{res}}\,{i\over E\,-\,L_{\text{tot}}}\,\rho_{t=0}\,\rho_{\text{res}}^{(\text{eq})}
\,=\,\text{Tr}_{\text{res}}\,{i\over E\,-\,L^{(0)}\,-\,L_{\text{res}}\,-\,L_V}\,
\rho_{t=0}\,\rho_{\text{res}}^{(\text{eq})}\\
\label{eq:formal_solution}
\,&=\,i\,\text{Tr}_{\text{res}}\,R^{(0)}(E-L_{\text{res}})\,\sum_{k=0}^\infty\,
(L_V\,R^{(0)}(E-L_{\text{res}}))^k\,\rho_{t=0}\,\rho_{\text{res}}^{(\text{eq})}\quad,
\end{align}
where we have defined
\begin{equation}
\label{eq:def_R0_L0_Lres_LV}
R^{(0)}(E)\,=\,{1\over E\,-\,L^{(0)}}\quad,\quad
L^{(0)}\,=\,[H,\cdot]\quad,\quad
L_{\text{res}}\,=\,[H_{\text{res}},\cdot]\quad,\quad
L_V\,=\,[V,\cdot]\quad.
\end{equation}
Using the form \eqref{eq:V_generic} of the system-reservoir interaction, a similiar form can
be derived for the Liouville superoperator $L_V$
\begin{equation}
\label{eq:L_V}
L_V\,=\,{1\over n!}\,\sum_{p=\pm}\,G^{(0)p\dots p}_{1\dots n}\,
:A^{p}_1\dots A^{p}_n:\quad.
\end{equation}
Here, $p=\pm$ is the so-called Keldysh index, which indicates whether the interaction $V$ arises
from the first or the second part of the commutator $L_V b=Vb-bV$ ($b$ is an arbitrary operator).
$A^p_1$ are reservoir field superoperators in Liouville space defined by 
\begin{equation}
\label{eq:liouville_field_operators}
A_1^p\,b \,=\,\sigma^p_{\text{res}}\,
\left\{
\begin{array}{cl}
a_1\,b\, &\mbox{for }p=+ \\
b\,a_1\, &\mbox{for }p=-
\end{array}
\right.\quad,
\end{equation}
and $G^{(0)p\dots p}_{1\dots n}$ is
a superoperator acting in Liouville space of the local quantum system defined by 
\begin{equation}
\label{eq:G_vertex_liouville}
G^{(0)p\dots p}_{1\dots n}\,b\,=\,
\left\{
\begin{array}{cl}
1\, &\mbox{for }n\mbox{ even} \\
\sigma^p\, &\mbox{for }n\mbox{ odd}
\end{array}
\right\}\,
\left\{
\begin{array}{cl}
g_{1\dots n}\,b\, &\mbox{for }p=+ \\
-b \,g_{1\dots n}\, &\mbox{for }p=-
\end{array}
\right.\quad.
\end{equation}
$\sigma^p$ and $\sigma_{\text{res}}^p$ are convenient sign superoperators which account for
fermionic signs and measure the parity of the fermionic particle number difference $N_s-N_{s'}$ 
of intermediate states $|ss'\rangle$ in Liouville space via the definition 
($N_s$ denotes the particle number of state $s$ and $\pm$ refers to bosons/fermions)
\begin{equation}
\label{eq:sign_superoperator}
\sigma^+\,=\,1 \quad,\quad
\sigma^-_{s_1 s_2,s_1^\prime s_2^\prime}\,=\,\delta_{s_1 s_1^\prime}\,\delta_{s_2 s_2^\prime}\,
(\pm)^{N_s-N_{s'}}\quad,
\end{equation}
and a corresponding definition for $\sigma^p_{\text{res}}$ by replacing local states $s$
by reservoir states. Since the total parity (local system plus reservoirs)
of all intermediate states must be even in Liouville space for fermions (note that it is 
impossible to prepare a nondiagonal matrix element of the total density matrix where the 
total fermionic particle number difference is odd, see Refs.~\cite{saptsov_PRB12,saptsov_arXiv}
for a detailed discussion and the consequences of this point), we obtain the important property
\begin{equation}
\label{eq:total_parity}
\sigma^p\,\sigma^p_{\text{res}}\,=\,1 \quad.
\end{equation}
From the definition of the reservoir field superoperators one can straightforwardly derive
how the product $:A_1^p\dots A_n^p:$ occuring in Eq.~\eqref{eq:L_V} acts in Liouville space
\begin{equation}
\label{eq:product_A_liouville}
:A_1^p\dots A_n^p:\,b\,=\,
\left\{
\begin{array}{cl}
1\, &\mbox{for }n\mbox{ even} \\
\sigma^p_{\text{res}}\, &\mbox{for }n\mbox{ odd}
\end{array}
\right\}\,
\left\{
\begin{array}{cl}
:a_1\dots a_n:\,b\, &\mbox{for }p=+ \\
b \,:a_1\dots a_n:\, &\mbox{for }p=-
\end{array}
\right.\quad,
\end{equation}
i.e. similiar to $G_{1\dots n}^{(0)p\dots p}$ but a minus sign is missing for $p=-$. Taking this
equation together with \eqref{eq:G_vertex_liouville} and using the property \eqref{eq:total_parity},
one can easily prove the representation \eqref{eq:L_V} for $L_V$. \newline

Most importantly, the reservoir field superoperators are defined such that the usual Wick theorem
can be applied (see Ref.~\cite{saptsov_PRB12} for an elegant proof), i.e. the average 
$\text{Tr}_{\text{res}} A_1^{p_1} \dots A_n^{p_n} \rho_{\text{res}}$
decomposes into a product of pair contractions and the sum has to be taken over all combinations,
with the usual definition of a fermionic sign to disentangle the various contractions. Using 
\eqref{eq:contraction} a single contraction is given by the expression
\begin{align}
\nonumber
\gamma_{11'}^{pp'}\,&=\,{A^p_1\,A^{p'}_{1'}
  \begin{picture}(-20,11) 
    \put(-26,10){\line(0,1){3}} 
    \put(-26,13){\line(1,0){16}} 
    \put(-10,10){\line(0,1){3}}
  \end{picture}
  \begin{picture}(20,11) 
  \end{picture}
}
\,=\,
\text{Tr}_{\text{res}}\,A^p_1\,A^{p'}_{1'}\,\rho_{\text{res}}^{\text{eq}} \\
\label{eq:contraction_liouville}
&=\,
\delta_{1\bar{1}'}\,
\left\{
\begin{array}{cl}
1 \\ p'
\end{array}
\right\}\,
\,\rho_{\alpha\sigma}(\omega)\,f^{p'\eta}_\alpha(\omega)
\,=\,
\delta_{1\bar{1}'}\,p'\,
\left\{
\begin{array}{cl}
\eta \\ 1
\end{array}
\right\}\,
\rho_{\alpha\sigma}(\omega)\,
f_\alpha(p'\eta\omega)\quad.
\end{align}

Using the form \eqref{eq:L_V} in \eqref{eq:formal_solution} one can shift all
reservoir field superoperators $A_i^{p_i}$ to the right by using the
analog of the commutation relation \eqref{eq:commutation_relation_1} in Liouville space
\begin{equation}
\label{eq:commutation_relation_liouville}
A^p_1\,L_{\text{res}}\,=\,(L_{\text{res}}\,-\,\eta(\omega+\mu_\alpha))\,A^p_1 \quad.
\end{equation}
This means that by shifting a certain field superoperator $A_1^p$ through all 
resolvents to the right, we shift all reservoir Liouville operators $L_{\text{res}}$ standing
right to $A_1^p$ by $-\eta(\omega+\mu_\alpha)$, where $1\equiv\eta\alpha\sigma\omega$. We note that, with
the second form \eqref{eq:V_generic} of the interaction, there is no fermionic sign when commuting local and 
reservoir operators. Shifting all reservoir field superoperators to the right and using the notation
\begin{align}
\nonumber
X_{1\dots n}\,=\,\bar{\omega}_{1\dots n}\,+\,\bar{\mu}_{1\dots n}\quad,\quad
\bar{\omega}_{1\dots n}\,&=\,\bar{\omega}_1\,+\,\dots\,\bar{\omega}_n\quad,\quad
\bar{\omega}_1\,=\,\eta_1\,\omega_1\quad,\\
\label{eq:X_notation}
\bar{\mu}_{1\dots n}\,&=\,\bar{\mu}_1\,+\,\dots\,\bar{\mu}_n\quad,\quad
\bar{\mu}_1\,=\,\eta_1\,\mu_1\quad,
\end{align}
we obtain for \eqref{eq:formal_solution} the form
\begin{align}
\nonumber
\rho(E)\,&=\,i\,\sum_{k=0}^\infty\,\text{Tr}_{\text{res}}\,R^{(0)}(E-L_{\text{res}})\,
({1\over n!}G^{(0)})\,R^{(0)}(E+X_{M_1}-L_{\text{res}})\,\cdot\\
\nonumber
&\hspace{1cm}
\cdot\,({1\over n!}G^{(0)})\,R^{(0)}(E+X_{M_2}-L_{\text{res}})\,\dots\,
({1\over n!}G^{(0)})\,R^{(0)}(E+X_{M_k}-L_{\text{res}})\,\rho_{t=0}\,\cdot \\
\nonumber
&\hspace{1cm}
\cdot\,(:A\dots A:)\,(:A\dots A:)\,\dots\,(:A\dots A:)\,\rho_{\text{res}}^{(\text{eq})}\\
\nonumber
&=\,i\,\sum_{k=0}^\infty\,R^{(0)}(E)\,({1\over n!}G^{(0)})\,R^{(0)}(E+X_{M_1})
\,\dots\,({1\over n!}G^{(0)})\,R^{(0)}(E+X_{M_k})\,\rho_{t=0}\,\cdot\\
\label{eq:perturbative_expansion_zw}
&\hspace{1cm}
\cdot\,\text{Tr}_{\text{res}}\,\left\{(:A\dots A:)\,\dots\,(:A\dots A:)\right\}
\,\rho_{\text{res}}^{(\text{eq})}\quad,
\end{align}
where we have used $\text{Tr}_{\text{res}}L_{\text{res}}=0$ in the last step. Thereby, the set
$M_i$ includes those indices of reservoir field superoperators which were standing
left to the corresponding resolvent in the original expression. As a result
the local and reservoir degrees of freedom have been decoupled and the trace over the reservoir
degrees of freedom can be performed by the application of Wick's theorem in Liouville space.
Since all diagrams give the same contribution when the indices of a particular vertex 
$G^{(0)p\dots p}_{1\dots n}$
are permuted, the factor ${1\over n!}$ is cancelled, except for the case when two vertices are
connected by $m$ contractions, leaving a symmetry factor ${1\over m!}$.
This leads to a sum of diagrams which symbolically are translated by the rule
\begin{equation}
\label{eq:diagram_rho}
\rho(E)\,\rightarrow\,i\,{(\pm)^{N_p}\over S}\,\left(\prod \gamma\right)\,
R^{(0)}(E)\,G^{(0)}\,R^{(0)}(E+X_{M_1})\,\dots\,G^{(0)}\,R^{(0)}(E+X_{M_k})\,\rho_{t=0}\quad,
\end{equation}
where $\prod \gamma$ denotes the product over all contractions \eqref{eq:contraction_liouville},
$N_p$ is the number of permutations of reservoir field superoperators to disentangle the
fermionic contractions, and $S=\prod_i m_i!$ is a symmetry factor arising for the case when
pairs of vertices are connected by $m_i$ contractions. \newline

The determination of the shift variables $X_{M_i}$ is simplified by noting that,
according to \eqref{eq:contraction_liouville}, a single contraction $\gamma_{12}^{p_1 p_2}$
between $A_1^{p_1}$ and $A_2^{p_2}$ is only possible for $\eta_1=-\eta_2$, 
$\alpha_1=\alpha_2$ and $\omega_1=\omega_2$. This gives $\bar{\omega}_{12}=\bar{\mu}_{12}=0$, 
i.e. if the two indices fall both into the same set $M_i$, there is no contribution 
to the shift $X_{M_i}$. As a consequence, the left index $1$ of a contraction $\gamma_{12}^{p_1 p_2}$
will contribute only to those resolvents, which stand between the two field operators $A_1^{p_1}$
and $A_2^{p_2}$ in the original series. For this reason, the last resolvent in \eqref{eq:diagram_rho}
has no shift $X_{M_k}=0$ and is given by $R^{(0)}(E)$. \newline

With the diagrammatic rules it is straightforward to translate a particular diagram, 
which we visualize as follows:
\begin{align}
\nonumber
\rho(E)\,
&
\rightarrow\,i\,
\begin{picture}(10,10)
\put(5,-20){\includegraphics[height=1.5cm]{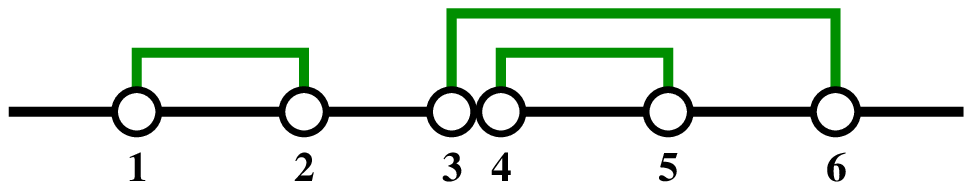}}
\end{picture}
\hspace{8cm}\rho_{t=0}\\ \nonumber \\
\nonumber
&=\,i\,\gamma_{12}^{p_1 p_2}\,\gamma_{36}^{p_3 p_6}\,\gamma_{45}^{p_4 p_5}\,
R^{(0)}(E)\,G^{(0)p_1}_1\,R^{(0)}(E_1+\bar{\omega}_1)\,G^{(0)p_2}_2\,
R^{(0)}(E)
\\
\label{eq:diagram_example}
&\hspace{1cm}
G^{(0)p_3 p_4}_{34}\,R^{(0)}(E_{34}+\bar{\omega}_{34})\,G^{(0)p_5}_5\,
R^{(0)}(E_3+\bar{\omega}_3)\,G^{(0)p_6}_6\,R^{(0)}(E)\,\rho_{t=0}\quad,
\end{align}
where we used the notation $E_{1\dots n}=E+\bar{\mu}_{1\dots n}$, see \eqref{eq:E_1...n}.
In the diagrams, the green lines are the contractions, the circles denote the vertices, and the 
black lines connecting the vertices represent the resolvents $R^{(0)}$ describing the dot propagation in 
Fourier space. The indices of the shift variables of a particular resolvent can
be determined by drawing a vertical line at the position of that resolvent and taking the left
indices of all contractions which cut through this line. We note that we do not distinguish
between diagrams which differ only by a permutation of the contractions connected to a certain
vertex, i.e. the permutation of the two green lines connected to the indices $3$ and $4$ in
the above example does not lead to a new diagram. \newline

To bring the density matrix $\rho(E)$ into the form \eqref{eq:solution} and to identify the
effective Liouvillian $L(E)$, we note that each diagram consists of a sequence of connected
blocks, defined by the property that each vertical line will at least hit one contraction,
connected by resolvents $R^{(0)}(E)$. E.g., the diagram \eqref{eq:diagram_example} consists
of a sequence of two blocks. Denoting the sum of all connected diagrams by $\Sigma(E)$, the
diagrammatic series can be written as
\begin{align}
\nonumber
\rho(E)\,&=\,i\,\left\{R^{(0)}(E)\,+\,R^{(0)}(E)\,\Sigma(E)\,R^{(0)}(E)\,+\,\right.\\
\label{eq:diagram_series}
&\left.+\,R^{(0)}(E)\,\Sigma(E)\,R^{(0)}(E)\,\Sigma(E)\,R^{(0)}(E)\,+\,\dots\right\}\,
\rho_{t=0}\,=\,{i\over E\,-\,L^{(0)}\,-\,\Sigma(E)}\,\rho_{t=0}\quad.
\end{align}
Comparing to \eqref{eq:solution}, we see that the effective Liouvillian is given by
\begin{equation}
\label{eq:effective_Liouvillian}
L(E)\,=\,L^{(0)}\,+\,\Sigma(E)\quad,
\end{equation}
and $\Sigma(E)$ consists of the sum of all connected diagrams with translation rule
\begin{equation}
\label{eq:diagram_Sigma}
\Sigma(E)\,\rightarrow\,{(\pm)^{N_p}\over S}\,\left(\prod \gamma\right)_{\text{\it con}}\,
G^{(0)}\,R^{(0)}(E_{M_1}+\bar{\omega}_{M_1})\,\dots\,
G^{(0)}\,R^{(0)}(E_{M_k}+\bar{\omega}_{M_k})\,G^{(0)}\,\quad,
\end{equation}
where $\left(\prod \gamma\right)_{\text{\it con}}$ means that only connected diagrams are considered.
E.g. some of the lowest order diagrams of $\Sigma(E)$ are given by
\begin{align}
\nonumber
&\\
\label{eq:diagrams_Sigma_example}
\Sigma(E)\,&=\,
\begin{picture}(10,10)
\put(5,-5){\includegraphics[height=0.7cm]{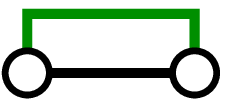}}
\end{picture}
\hspace{1.5cm}
\,+\,
\begin{picture}(10,10)
\put(5,-5){\includegraphics[height=1cm]{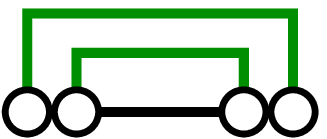}}
\end{picture}
\hspace{2.5cm}
\,+\,
\begin{picture}(10,10)
\put(5,-5){\includegraphics[height=1cm]{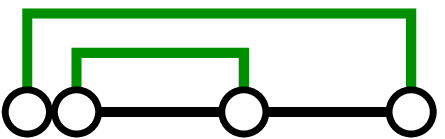}}
\end{picture}
\hspace{3.5cm}
\,+\,\dots
\end{align}
$\Sigma(E)$ is the dissipative part of the effective Liouvillian, which contains the whole 
information of the coupling to the reservoirs and leads to irreversible time evolution.
In time space we obtain $L(t)=L^{(0)}\delta(t-0^+)+\Sigma(t)$, such that the kinetic equation
\eqref{eq:kinetic_equation_special} reads
\begin{equation}
\label{eq:kinetic_equation_sigma}
i\dot{\rho}(t)\,=\,L^{(0)}\,\rho(t)\,+\,\int_{0}^t\,dt'\,\Sigma(t-t')\,\rho(t')\quad.
\end{equation}
The first term describes the von Neumann equation in the absence of the reservoirs, whereas
the second one is the dissipative part. The two terms are the analog of the
\lq\lq flow'' and the \lq\lq collision'' term of quantum Boltzmann equations.\newline

{\bf Local observables.} From the density matrix $\rho(E)$ the time evolution of all averages of local observables 
can be calculated. The diagrammatic expansion can also be formulated for the calculation of
arbitrary observables containing reservoir degrees of freedom or correlation functions. E.g., if
an observable $I$ of the generic form \eqref{eq:V_generic} is taken
\begin{equation}
\label{eq:I_generic}
I\,=\,{1\over n!}\,
\left\{
\begin{array}{cl}
1 \\ \eta_1\eta_2\dots\eta_n
\end{array}
\right\}
\, :a_n a_{n-1}\dots a_1: \,i_{12\dots n}
\,\rightarrow\,{1\over n!}\,i_{12\dots n}\,:a_1a_2\dots a_n: \quad,
\end{equation}
we define a corresponding superoperator $L_I$ by the anticommutator
\begin{equation}
\label{eq:L_I}
L_I\,=\,{i\over 2}\,[I,\cdot]_+\,=\,
{1\over n!}\,\sum_{p=\pm}\,I^{(0)p\dots p}_{1\dots n}\,
:A^{p}_1\dots A^{p}_n:\quad,
\end{equation}
with 
\begin{equation}
\label{eq:I_vertex_liouville}
I^{(0)p\dots p}_{1\dots n}\,b\,=\,{i\over 2}\,
\left\{
\begin{array}{cl}
1\, &\mbox{for }n\mbox{ even} \\
\sigma^p\, &\mbox{for }n\mbox{ odd}
\end{array}
\right\}\,
\left\{
\begin{array}{cl}
i_{1\dots n}\,b\, &\mbox{for }p=+ \\
b \,i_{1\dots n}\, &\mbox{for }p=-
\end{array}
\right.\quad,
\end{equation}
such that the average can be written as
\begin{equation}
\label{eq:I_average}
\langle\,I\,\rangle(t)\,=\,\text{Tr}_{\text{tot}}\,I\,\rho_{\text{tot}}(t)\,=\,
-i\,\text{Tr}\,\text{Tr}_{\text{res}}\,L_I\,e^{-i L_{\text{tot}}t}\,
\rho_{t=0}\,\rho_{\text{res}}^{\text{eq}}\quad.
\end{equation}
This expression has a formal similiarity to 
\begin{align}
\nonumber
\dot{\rho}(t)\,&=\,-i\,\text{Tr}_{\text{res}}\,L_{\text{tot}}\,e^{-iL_{\text{tot}}t}\,
\rho_{\text{tot}}(t=0)
\,=\,-i\,\text{Tr}_{\text{res}}(L^{(0)}+L_{\text{res}}+L_V)\,e^{-iL_{\text{tot}}t}\,
\rho_{t=0}\,\rho_{\text{res}}^{\text{eq}}\\
\label{eq:rho_dot}
&=\,-i\,L^{(0)}\,\rho(t)\,-\,i\,\text{Tr}_{\text{res}}L_V\,e^{-iL_{\text{tot}}t}\,
\rho_{t=0}\,\rho_{\text{res}}^{\text{eq}}\quad,
\end{align}
where $\text{Tr}_{\text{res}}L_{\text{res}}=0$ has been used in the last line. Comparing
to the kinetic equation \eqref{eq:kinetic_equation_sigma}, we find
\begin{equation}
\label{eq:analog_rho_I}
-\,i\,\text{Tr}_{\text{res}}L_V\,e^{-iL_{\text{tot}}t}\,\rho_{t=0}\,\rho_{\text{res}}^{\text{eq}}
\,=\,-i\,\int_{0}^t\,dt'\,\Sigma(t-t')\,\rho(t')\quad.
\end{equation}
Therefore, when applying the same perturbative expansion to \eqref{eq:I_average}, we obtain
the result
\begin{equation}
\label{eq:I_result}
\langle\,I\,\rangle(t)\,=\,-i\,\int_{0}^t\,dt'\,\text{Tr}\,\Sigma_I(t-t')\,\rho(t')
\quad,\quad
\langle\,I\,\rangle(E)\,=\,-i\,\text{Tr}\,\Sigma_I(E)\,\rho(E)\quad,
\end{equation}
with the only difference that the first vertex of the kernel $\Sigma_I(E)$ has to be the
vertex $I^{(0)}$ instead of $G^{(0)}$, i.e. the diagrammatic rule \eqref{eq:diagram_Sigma} changes to 
\begin{align}
\nonumber
\Sigma_I(E)\,&\rightarrow\,{(\pm)^{N_p}\over S}\,\left(\prod \gamma\right)_{\text{\it con}}\\
\label{eq:diagram_Sigma_I}
&\hspace{1cm}
I^{(0)}\,R^{(0)}(E_{M_1}+\bar{\omega}_{M_1})\,
G^{(0)}\,R^{(0)}(E_{M_2}+\bar{\omega}_{M_2})\,\dots\,
G^{(0)}\,R^{(0)}(E_{M_k}+\bar{\omega}_{M_k})\,G^{(0)}\,\quad.
\end{align}
A prominent example for an observable is the particle current operator flowing from reservoir 
$\alpha$ into the local system defined by
\begin{equation}
I_\alpha\,=\,-{d\over dt}\,N_\alpha\,=\,-i\,[H_{\text{tot}},N_\alpha]
\,=\,-i\,[V,N_\alpha]\quad.
\end{equation}
Inserting the form \eqref{eq:V_generic} of $V$, one finds after some straightforward manipulations 
\begin{equation}
\label{eq:current_vertex}
(i_\alpha)_{1\dots n}\,=\,i\,\sum_{k=1}^n\,\eta_k\,\delta_{\alpha_k\alpha}\,g_{1\dots n}
\quad,\quad
(I_\alpha)^{(0)p\dots p}_{1\dots n}\,=\,-{1\over 2}\,\sum_{k=1}^n\,\eta_k\,\delta_{\alpha_k\alpha}\,
p\,G^{(0)p\dots p}_{1\dots n}\quad.
\end{equation}
\newline

Using the diagrammatic expansion \eqref{eq:diagram_Sigma} and \eqref{eq:diagram_Sigma_I}, 
the kernels $\Sigma(E)$ and $\Sigma_I(E)$ can be calculated in perturbation
theory w.r.t. the bare vertices $G^{(0)}$, and the local density matrix $\rho(E)$ and the 
average $\langle I \rangle(E)$ of any observable $I$ follow from \eqref{eq:diagram_series} 
and \eqref{eq:I_result} in Fourier space. Using inverse Fourier transform the time evolution 
can finally be discussed following the strategy described in Section \ref{sec:time_evolution}.
Stationary quantities are obtained from 
\begin{align}
\rho_{\text{st}}\,&=\,\lim_{t\rightarrow\infty}\,\rho(t)
\,=\,\lim_{E\rightarrow 0^+}\,(-iE)\,\rho(E)
\quad \Leftrightarrow \quad L(E=0^+)\,\rho_{\text{st}}\,=\,0\\
\label{eq:stationary_I}
\langle I \rangle_{\text{st}}\,&=\,\lim_{t\rightarrow\infty}\,\langle I \rangle(t)
\,=\,\lim_{E\rightarrow 0^+}\,(-iE)\,\langle I \rangle(E)
\,=\,-i\,\text{Tr}\,\Sigma_I(E=0^+)\,\rho_{\text{st}}\quad.
\end{align}
Applications of these perturbative schemes for the calculation of transport properties will be discussed  
in the lecture C7 by M. Wegewijs. Similiar schemes have also been developed to
calculate correlation functions \cite{correlation_function} and to consider explicitly
time-dependent Hamiltonians \cite{kashuba_kennes_PRB13,kashuba_EPL12}. Concerning the latter first 
applications have considered adiabatic response \cite{kashuba_EPL12} and quantum quenches 
\cite{kashuba_kennes_PRB13} for the IRLM.\newline

{\bf Analytic properties.} From the perturbative expansion one finds that the effective Liouvillian $L(E)$ has
a branch cut on the real axis and is analytic in the upper and lower half of the complex plane. This can
be seen from the resolvents $R^{(0)}(E_M+\bar{\omega}_M)$ since $L^{(0)}=[H^{(0)},\cdot]$ is a self-adjoint
superoperator with real eigenvalues and all frequency variables $\bar{\omega}$ are integrated over the real axis. 
The same analytic property holds for the resolvent 
$R(E)={1\over E-L(E)}$, since, due to \eqref{eq:solution}, we get for any initial density matrix $\rho_{t=0}$
\begin{align}
\nonumber
R(E)\,\rho_{t=0}\,&=\,-i\,\rho(E)\,=\,-i\,\int_0^\infty\,dt\,e^{iEt}\,\rho(t)\,=\,
-i\,\int_0^\infty\,dt\,e^{iEt}\,\text{Tr}_{\text{res}}\,\rho_{\text{tot}}(t)\\
\label{eq:resolvent_analytic_prop}
\,&=\,-i\,\int_0^\infty\,dt\,e^{iEt}\,\text{Tr}_{\text{res}}\,
e^{-iL_{\text{tot}}t}\,\rho_{\text{tot}}(t=0)\,=\,
\text{Tr}_{\text{res}}\,{1\over E\,-\,L_{\text{tot}}}\,\rho_{\text{tot}}(t=0)\quad.
\end{align}
This function can only have a branch cut on the real axis since 
$L_{\text{tot}}=[H_{\text{tot}},\cdot]$ is a self-adjoint superoperator with real
eigenvalues. To calculate the time evolution we have seen from \eqref{eq:time_evolution} that the 
integration $\int dE$ is slightly above the real axis and has to be closed in the lower half of 
the complex plane (due to $t>0$). It is very inconvenient to calculate this integral by enclosing 
the branch cut of the integrand on the real
axis due to the rapidly oscillating function $e^{-iEt}$ in the integrand on the scale $1/t$. Therefore,
in analogy to the standard procedure for response functions, one tries to find an appropriate analytic
continuation of the functions $L(E)$ and $R(E)$ into the lower half of the complex plane such that all branch cuts
point into the direction of the negative imaginary axis starting at certain singularities $z_n$ with
$\text{Im}z_n\le 0$. We achieve this in two steps. First, we will transform the perturbative series for $L(E)$
into a self-consistent equation by resumming all blocks of connected
diagrams on the propagators connecting the vertices. The diagrammatic representation allows this to
be done in a unique way and, as a result, the bare resolvents $R^{(0)}(E_M+\bar{\omega}_M)$ are replaced 
by the full ones $R(E_M+\bar{\omega}_M)$ and no diagrams are allowed with connected sub-blocks without any
free lines on the propagators which we indicate by $\left(\prod \gamma\right)_{\text{\it irr}}$
\begin{equation}
\label{eq:diagram_Sigma_selfconsistent}
\Sigma(E)\,\rightarrow\,{(\pm)^{N_p}\over S}\,\left(\prod \gamma\right)_{\text{\it irr}}\,
G^{(0)}\,R(E_{M_1}+\bar{\omega}_{M_1})\,\dots\,
G^{(0)}\,R(E_{M_k}+\bar{\omega}_{M_k})\,G^{(0)}\,\quad.
\end{equation}
For $\text{Im}E>0$, all resolvents are analytic functions w.r.t. the integration variables $\bar{\omega}_i$
in the upper half of the complex plane. Therefore, in the second step, we can close all 
integration contours in the upper half and have to enclose only the nonanalytic features arising 
from the spectral function $\rho_{\alpha\sigma}(\omega)$ and the Bose/Fermi distribution 
$f_\alpha(p'\eta\omega)$ in the contraction $\gamma_{11'}^{pp'}$, defined in 
\eqref{eq:contraction_liouville}. Thereby we assume that the frequency dependence of the vertices  
$G^{(0)p\dots p}_{1\dots n}$ can be neglected. Decomposing the Bose/Fermi distribution
in symmetric and antisymmetric parts and using the representation in terms of the Masubara frequencies
$\omega_n^\alpha=2n\pi T_\alpha$ ($\omega_n^\alpha=(2n+1)\pi T_\alpha$) for bosons (fermions), 
we can write the contraction in the form 
\begin{align}
\label{eq:contraction_decomposition}
\gamma_{11'}^{pp'}\,&=\,\delta_{1\bar{1}'}\,p'\,
\left\{
\begin{array}{cl}
\eta \\ 1
\end{array}
\right\}\,
\bar{\rho}_{\alpha\sigma}(\bar{\omega})
\,\left\{\mp\,{1\over 2}\,+\,p'\,(f_\alpha(\bar{\omega})\,\pm\,{1\over 2})\right\}
\,=\,
\delta_{1\bar{1}'}\,(p'\,\gamma_1^s\,+\,\gamma_1^a)\quad,\\
\label{eq:contraction_symmetric}
\gamma_1^s\,&=\,\mp\,{1\over 2}
\left\{
\begin{array}{cl}
\eta \\ 1
\end{array}
\right\}\,
\bar{\rho}_{\alpha\sigma}(\bar{\omega})\quad,\\
\nonumber
\gamma_1^a\,&=\,
\left\{
\begin{array}{cl}
\eta \\ 1
\end{array}
\right\}\,
\bar{\rho}_{\alpha\sigma}(\bar{\omega})
\,\left\{f_\alpha(\bar{\omega})\,\pm\,{1\over 2}\right\}\\
\label{eq:contraction_antisymmetric}
&=\,
\left\{
\begin{array}{cl}
\eta \\ 1
\end{array}
\right\}\,
\bar{\rho}_{\alpha\sigma}(\bar{\omega})\,
\,T_\alpha\,{1\over 2}\,\sum_n\,\left({1\over \bar{\omega}-i\omega_n^\alpha}\,+\,
{1\over \bar{\omega}+i\omega_n^\alpha}\right)\quad,
\end{align}
where $\bar{\rho}_{\alpha\sigma}(\bar{\omega})=\rho_{\alpha\sigma}(\omega)$. Thus, after performing
all integrations $\int d\bar{\omega}_i$, and assuming for the moment that the spectral 
function $\bar{\rho}_{\alpha\sigma}(\bar{\omega})$ is an analytic function in the upper half, 
the quantities $\bar{\omega}_M$ occuring in the resolvents $R(E_M+\bar{\omega}_M)$ will consist of 
a sum of positive Matsubara frequencies $i\sum_{j\in M}|\omega_{n_j}^{\alpha_j}|$. As a consequence, 
the analytic continuation w.r.t. $E$ of this result for $L(E)$ into the lower half of the 
complex plane will lead to nonanalytic features at $E_M+i\sum_{j\in M}|\omega_{n_j}^{\alpha_j}|=z_k^p$, 
where $z_k^p$ are the poles of the
resolvent $R(E)$ after the analytic continuation into the lower half of the complex plane. Since $R(E)$ is
analytic in the upper half, the poles $z_k^p$ have to lie in the lower half, and we find that $L(E)$ has
an infinite series of poles in the lower half located at
\begin{equation}
\label{eq:pole_positions}
E\,=\,z_k^p\,-\,\bar{\mu}_M\,-\,i\sum_{j\in M}\,|\omega_{n_j}^{\alpha_j}|\quad,
\end{equation}
which, at zero temperature, turn into a series of branch cuts in the direction of the negative imaginary
axis with branching points $z_n$ located at the poles $z_k^p$ shifted by any combinations of the
chemical potentials of the reservoirs 
\begin{equation}
\label{eq:branching_point}
z_n\,=\,z_k^p-\,\bar{\mu}_M\quad.
\end{equation}
This result has formed the basis for the generic discussion of the time evolution in 
Section~\ref{sec:time_evolution}.\newline

{\bf Influence of spectral function.}
We note that the spectral function 
$\bar{\rho}_{\alpha\sigma}(\bar{\omega})$ of the models introduced in Section~\ref{sec:models}
does not change this picture. For quantum dots coupled to Fermi liquid leads, like the Kondo 
model or the IRLM, the spectral function 
$\bar{\rho}_{\alpha\sigma}(\bar{\omega})={D^2\over D^2+\bar{\omega}^2}$ defines just a high-energy
cutoff function with pole at $\bar{\omega}=iD$ in the upper half and residuum $-iD/2$. The contribution
of this pole to the frequency integration leads for $D\rightarrow\infty$ either to a vanishing or to
a regular contribution in $E$. For the ohmic spin boson model we get from \eqref{eq:spinboson_sf_vo} that 
$\bar{\rho}(\bar{\omega})=2\alpha |\bar{\omega}|\theta(\eta\bar{\omega}){D^2\over D^2+\bar{\omega}^2}$,
which has a branch cut on the whole imaginary axis. However, since the vertex $g_1={1\over 2}\sigma_z$ 
is independent of $\eta$, we can sum the contraction $\gamma_{11'}^{pp'}$ over $\eta$ and $\eta'$ at
fixed $\bar{\omega}$ and $\bar{\omega}'$ (which are the integration variables) and get from
\eqref{eq:contraction_liouville} for the case of bosons the effective contraction
\begin{equation}
\label{eq:effective_contraction_spinboson} 
\gamma_{11'}^{pp'}\,=\,\delta(\bar{\omega}+\bar{\omega}')\,p'\,2\,\alpha\,
\bar{\omega}\,{D^2\over D^2+\bar{\omega}^2}\,f(p'\bar{\omega})
\,=\,\delta(\bar{\omega}+\bar{\omega}')\,(p'\gamma_1^s\,+\,\gamma_1^a)
\quad,
\end{equation}
with
\begin{equation}
\label{eq:gamma_sa_spinboson}
\gamma_1^s\,=\,-\alpha\,\bar{\omega}\,{D^2\over D^2+\bar{\omega}^2}\quad,\quad
\gamma_1^s\,=\,\alpha\,\bar{\omega}\,{D^2\over D^2+\bar{\omega}^2}\,(2f(\bar{\omega})\,+\,1)\quad,
\end{equation}
where the index $1\equiv\bar{\omega}$ contains only the frequency variable. Since $\bar{\omega}$ is
an analytic function, there is no change of the analytic structure of $L(E)$ for the ohmic spin boson model.
For a generic frequency dependence of the spectral function, the analytic structure might change. If
$\bar{\rho}(\bar{\omega})$ has a branch cut in the upper half in the direction of the positive 
imaginary axis starting at $\Delta_\rho + i\gamma_\rho$, with $\gamma_\rho\ge 0$, the position 
\eqref{eq:branching_point} of the branching points of $L(E)$ can be shifted by multiples of
$-\Delta_\rho-i\gamma_\rho$. This can e.g. happen for superconducting leads, where $\Delta_\rho$
corresponds to the superconducting gap and $\gamma_\rho=0$. For sub- or super-ohmic
spin boson models there is no change of the analytic properties since the branch cuts of the
spectral function start at the origin. \newline

{\bf Symmetric part of the contraction.} 
We note that the part $\gamma_1^s$ of the contraction \eqref{eq:contraction_decomposition} involving
the symmetric part of the Bose/Fermi distribution plays a special role. It is the only
part of the contraction $\gamma_{11'}^{pp'}$ which depends on the Keldysh indices via $p'$ and it depends on 
the frequency only via the spectral function. In particular for a spectral function of the form 
$\bar{\rho}(\bar{\omega}_1)={D^2\over D^2+\bar{\omega}_1^2}$, i.e. if it just acts as a high-energy cutoff 
function but has no other special form, the frequency integration $\int d\bar{\omega}_1$ will involve only
the pole of the spectral function at $\bar{\omega}_1=iD$ when closed in the upper half. In the limit
$D\rightarrow\infty$ this means that this integration gives either zero (if more than one resolvent 
involves $\bar{\omega}_1$) or a constant if this contraction connects two consecutive vertices
\begin{equation}
\label{eq:st_freq_integration} 
\int d\bar{\omega}_1\,{D^2\over D^2+\bar{\omega}_1^2}\,R(E_{1\dots n}+\bar{\omega}_{1\dots n})\,=\,
\pi\,D\,R(E_{1\dots n}+\bar{\omega}_{2\dots n}+iD)\,\xrightarrow{D\rightarrow\infty}\,-i\,\pi\quad.
\end{equation}
As a result, the symmetric part of the contraction can be integrated out analytically and can be 
incorporated in an effective vertex by taking the two consecutive vertices together to a single one. 
The same can be shown for the ohmic spin boson model \cite{kashuba_PRB13} due to its special algebra,
whereas for more general spectral functions with nonanalytic features in the upper half this is not 
the case. If it holds, one important consequence of this property is that the special pole 
at $E_M+\bar{\omega}_M=z_{\text{st}}^p=0$ of the resolvents $R(E_M+\bar{\omega}_M)$ leads to regular contributions 
in the limit $D\rightarrow\infty$ and does not contribute to the branch cuts of $L(E)$.
The reason is the special form $P_{\text{st}}(E)=|x_{\text{st}}(E)\rangle\,\text{Tr}$ for the projector
of the mode $k=st$ (see Eq.~\eqref{eq:P_st}) together with the property
\begin{equation}
\label{eq:G0_trace}
\sum_p\,\text{Tr}\,G^{(0)p\dots p}_{1\dots n}\,=\,0,
\end{equation}
which follows straightforwardly from the definition \eqref{eq:G_vertex_liouville}. Since the contractions
$\gamma_{11'}^{pp'}$ are independent of the first Keldysh index $p$ (see Eq.~\eqref{eq:contraction_liouville}),
this means that if the projector $P_{\text{st}}(E_M+\bar{\omega}_M)$ is inserted between two 
consecutive vertices, at least one 
of the contractions $\gamma_{11'}^{pp'}$ associated with the right vertex must point into the left direction and 
only its $p'$-dependent symmetric part $p'\gamma_1^s$ will contribute
\begin{align}
\nonumber &\\
\label{eq:st_picture}
&\begin{picture}(10,10)
\put(-80,-20){\includegraphics[height=2cm]{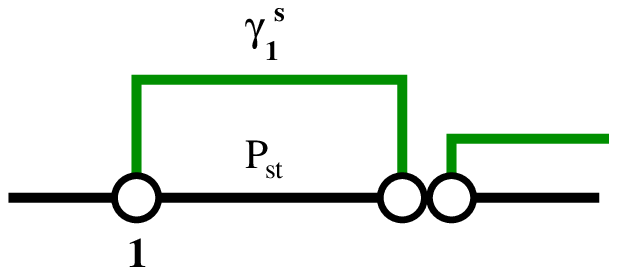}}
\end{picture}
\\
\nonumber &
\end{align}
As shown above this means that this contraction has to connect the two consecutive vertices and the 
frequency integration gives a constant. Thus, the case $k=\text{st}$ does not contribute to the
positions \eqref{eq:branching_point} of branch cuts for $L(E)$ or $R(E)$. There might be other 
accidental poles $z_k^p=0$ for $k\ne st$, like e.g. for multi-channel Kondo models with non-Fermi
liquid behaviour \cite{kondo_E-RTRG}, 
but for the models discussed in Section~\ref{sec:models}, this is not the case. Therefore, for these
models, the pole at $z_{\text{st}}=0$ is isolated, as already stated in Section~\ref{sec:time_evolution}
after Eq.~\eqref{eq:branching_pole_zero}.
\newline

{\bf Breakdown of perturbation theory for time-evolution problems.}
Finally we note that it is very important to use the perturbative expansion of $L(E)$ in the self-consistent 
form \eqref{eq:diagram_Sigma_selfconsistent} in order to find the right position of the branching
points $z_n$ of the branch cuts of $R(E)$. For the original series \eqref{eq:diagram_Sigma} involving
the bare resolvent $R^{(0)}(E_M+\bar{\omega}_M)$, the same considerations as above lead to
branch cuts of $L(E)$ starting on the real axis at the value $z_n=\lambda^{(0)}-\bar{\mu}_M$,
where $\lambda^{(0)}$ is a real eigenvalue of the bare Liouvillian $L^{(0)}$. This would have a dramatic
effect on the long-time evolution because it leads to non-exponential decay. However, this result
is not correct since perturbation theory is very dangerous in the regime $|E-z_n|\sim \Gamma$, where
$\Gamma$ is a typical decay rate. At low frequencies, the resolvents can then become very large of the 
order of the inverse coupling constant, raising serious questions about convergence. In particular 
for the original series \eqref{eq:diagram_Sigma} involving $R^{(0)}$ this effect is most dramatic 
since a series of 
connected sub-blocks contains an arbitrary number of resolvents with exactly the same argument, i.e.
the singularity at low frequency appears to an arbitrary power. Such a series is certainly not 
convergent and it is necessary to resum it first to the self-consistent version 
\eqref{eq:diagram_Sigma_selfconsistent} before determining the position of the branching points. E.g.
consider a contribution to the effective Liouvillian of the form ($\alpha$ is some small dimensionless
coupling constant and $\Delta$ denotes a typical low-energy scale)
\begin{equation}
\label{eq:convergence}
\alpha\,(-iE\,+\,\alpha\Delta)\,\ln{D\over -iE\,+\,\alpha\Delta}\,=\,
\alpha\,(-iE\,+\,\alpha\Delta)\,\ln{D\over -iE}
\,-\,\alpha^2\,\Delta\,-\,{1\over 2}\,i\,\alpha^3\,{\Delta^2\over E}\,+\,O(\alpha^4)\quad.
\end{equation}
The logarithm on the l.h.s. has a branching point at $E=-i\alpha\Delta$ but the expanded form gives in 
$O(\alpha)$ and $O(\alpha^2)$ a branching point at $E=0$. The mistake can only be seen by 
considering higher orders in $\alpha$, where an infinite series of terms 
with a pole at $E=0$ is obtained. Due to the factor
$-iE+\alpha\Delta$ in front of the logarithm, this artifact is even not visible in $O(\alpha^2)$ but starts
in $O(\alpha^3)$ or higher. The form \eqref{eq:convergence} arises e.g. for the Kondo model and for
the ohmic spin boson model, where we will show in Section~\ref{sec:RG} that 
${\partial\over\partial E}L(E)$ must be a slowly varying logarithmic function leading to 
typical terms of the form \eqref{eq:convergence}, see also Eq.~\eqref{eq:L_generic_structure}. 
For the ohmic spin boson model previous calculations \cite{functional_integral} have 
predicted terms with non-exponential decay which have been corrected recently 
\cite{kashuba_PRB13,slutskin_EPL11,egger_PRE97}.

\section{Renormalization group}
\label{sec:RG}

{\bf General remarks.} At low temperatures the perturbative calculation of the effective Liouvillian can 
break down even at small reservoir-system coupling for two reasons. First, at high energies
(the so-called ultraviolett regime), the frequency integrals are typically logarithmic leading 
to logarithmic contributions 
$\sim\alpha^k(\ln{D\over E-z_n})^l$ in higher-order perturbation theory, with $k\ge l$, 
where $z_n$ are the branching points \eqref{eq:branching_point} of the resolvent $R(E)$ and $\alpha$ is an
appropriate dimensionless coupling constant. Secondly, even if perturbation theory does
not contain ultraviolett logarithmic divergencies in the limit $D\rightarrow\infty$, it may contain
logarithmic terms $\sim\alpha\ln{E-z_n\over E-z_m}$, which, for $E\rightarrow z_n$, turn into 
the form $\sim \alpha\ln{E-z_n\over z_n-z_m}$, which can lead to a breakdown
of perturbation theory at low energies (the so-called infrared regime). Therefore, a method is
needed capable of reorganizing perturbation theory such that all ultravioltett and infrared 
logarithmic divergencies are resummed. Concerning high energies, resumming all logarithmic 
contributions $\sim\alpha^k(\ln{D\over E-z_n})^l$ with $l=k,k-1,k-2,\dots$ is called 
leading order, sub-leading order, sub-sub leading order, etc. approximation (sometimes also referred to 
as $1$-loop, $2$-loop, $3$-loop, etc.). In traditional (so-called poor man
scaling) RG methods \cite{pms}, one tries to perform this resummation by integrating out 
high-energy scales, i.e. the band width $D$ is successively reduced in infinitesimal steps and the 
physical quantity of interest is kept invariant by renormalizing the coupling constants and other 
energy scales. Provided that the renormalized coupling constants remain small (the so-called 
weak-coupling regime), a well-controlled truncation scheme can be set
up by neglecting higher-order terms in the renormalized couplings. This strategy has also been used 
for calculating stationary quantities of nonequilibrium problems \cite{RG_noneq_D_cutoff}, but it 
turns out that for the calculation of the effective Liouvillian $L(E)$ it is very hard to set up a 
systematic truncation scheme. This has been improved by using a high-energy cutoff on the imaginary
axis by cutting off the Matsubara frequencies of the Bose/Fermi distribution function \cite{schoeller_EPJ09}
and applied to various models 
\cite{schoeller_EPJ09,pletyukhov_PRL10,RTRG_irlm,saptsov_PRB12,correlation_function}.
Here we will follow another route by describing the E-RTRG method \cite{kondo_E-RTRG}, which is 
unique in the sense that it is capable of dealing with all logarithmic divergencies at high 
{\it and} low energies. Technically, 
this is achieved by considering the perturbation theory not for $L(E)$ but for its first or second 
derivative w.r.t. the Fourier variable $E$ together with a proper resummation in terms of effective vertices. 
Whether a first or a second derivative is needed depends on the model under consideration. 
This leads to a series where all frequency integrals converge at high energies and 
the limit $D\rightarrow\infty$ can be performed in all orders. As a consequence one obtains a {\it universal} 
differential equation (called RG equation) for $L(E)$ independent of the specific 
choice of the high-energy cutoff function. Furthermore, the RG equation for $L(E)$ turns out
to be such that the divergence at low energies for $E\rightarrow z_n$ is at most $\sim {1\over E-z_n}$
multiplied with a perturbative series in terms of effective vertices which exists in the limit $E\rightarrow z_n$.
This allows for a systematic solution at low energies as well. Besides the effective Liouvillian 
also effective vertices will appear in the RG equation due to the resummation procedure, for which 
similiar universal RG equations can be derived. Provided that the effective vertices stay small 
(so-called weak coupling problems) the RG equations can be systematically truncated and well-controlled 
universal properties can be determined at high as well as at low energies. According to the discussion
in Section~\ref{sec:time_evolution} this allows a well-controlled discussion of the time evolution
at short and long times together with the crossover behaviour. The high-energy cutoff $D$ 
will only appear in the initial condition for the various quantities which are calculated by a 
well-controlled perturbation theory in the bare couplings at $E=iD$. This procedure has the 
advantage that by construction only the universal properties of the model are obtained, although 
it is also possible to keep $D$ fixed and solve the RG equations for a given high-energy cutoff 
function. Furthermore, the use of a physical scale $E$ as flow parameter of the RG equations has 
the advantage that at each stage of the flow the solution $L(E)$ provides a result for a
physical quantity. Moreover, since $E$ is a complex flow parameter, the flow can be solved on any path in 
the complex plane which is very helpful to find
appropriate analytic continuations of retarded functions into the lower half of the complex plane, even by 
using numerical methods, see also the discussion in Section~\ref{sec:time_evolution} after
Eq.~\eqref{eq:E_1...n}.\newline

{\bf Derivation of the E-RTRG equations.}
To illustrate the general strategy for the derivation of the RG equations within the E-RTRG method we consider 
here, for simplicity, a spectral function of the form 
$\bar{\rho}(\bar{\omega})={D^2\over D^2+\bar{\omega}^2}$, which arises typically for fermionic metallic
reservoirs where the d.o.s. is approximately a constant in the physically relevant energy regime.
Therefore, we consider only the fermionic case in the following.
Furthermore, we assume that only $1$- and $2$-point vertices occur in the original model and that
the frequency dependence of the bare vertices $g_{1}$ and $g_{12}$ can be neglected.
This applies to the Kondo model and the IRLM introduced in Section~\ref{sec:models}. For the ohmic spin
boson model, a similiar procedure can be used to derive the RG equations, see Ref.~\cite{kashuba_PRB13}.
For such models, one obtains a problem with convergence at high energies if the number of frequency 
integrations is larger or equal to the number of resolvents where the frequencies occur. For models with 
$1$- and $2$-point vertices this means that diagrammatic sub-elements of the form
\begin{align}
\nonumber
&\\
\nonumber
&
\begin{picture}(10,10)
\put(-50,-0){\includegraphics[height=0.5cm]{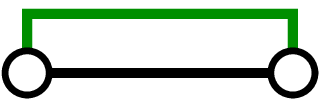}}
\end{picture}
\hspace{4cm}
 \begin{picture}(10,10)
\put(-50,-0){\includegraphics[height=0.7cm]{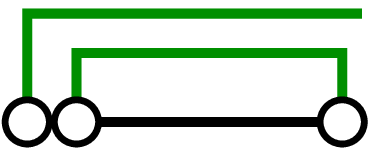}}
\end{picture}
\hspace{4cm}
 \begin{picture}(10,10)
\put(-50,-0){\includegraphics[height=0.5cm]{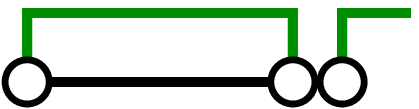}}
\end{picture}
\\
\nonumber
\\
\label{eq:diagrams_divergent}
&
 \begin{picture}(10,10)
\put(-50,-0){\includegraphics[height=0.7cm]{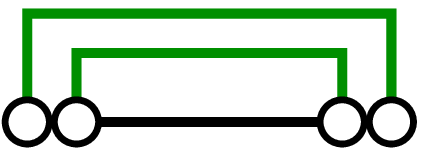}}
\end{picture}
\hspace{4cm}
 \begin{picture}(10,10)
\put(-50,-0){\includegraphics[height=0.7cm]{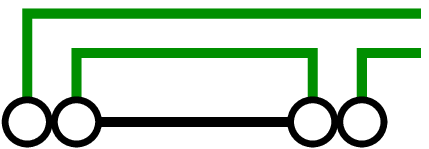}}
\end{picture}
\hspace{4cm}
 \begin{picture}(10,10)
\put(-50,-0){\includegraphics[height=1cm]{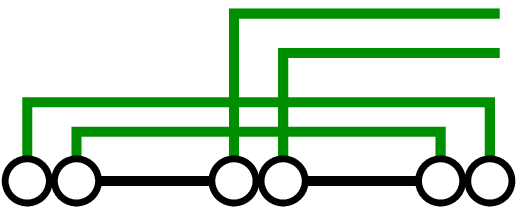}}
\end{picture}
%\\
%\nonumber
\end{align}
lead to problems at high energies and have to be avoided. This can be achieved by taking a single 
or a double derivative w.r.t. $E$ of the resolvents occuring in these diagrams.
Therefore, the idea is to consider a perturbative expansion
for the derivatives ${\partial\over\partial E}L(E)$ or ${\partial^2\over\partial E^2}L(E)$ and to
resum the series such that no subelements of the form \eqref{eq:diagrams_divergent} remain. 
The procedure is quite straightforward and we illustrate it for 
the case of a model where only $2$-point vertices occur, like e.g. the Kondo model. Here, to 
guarantee convergence we consider two derivatives w.r.t. $E$ of the diagrammatic series 
\eqref{eq:diagram_Sigma_selfconsistent} of $L(E)$ in the self-consistent form. Since the 
$E$-dependence occurs only in the resovents $R(E_M+\bar{\omega}_M)$, we can either take two
derivatives of a single resolvent or two single derivatives of different resolvents. Fixing the
positions of the resolvents we can then resum all remaining diagrams in a unique way such that
the bare $2$-point vertices $G^{(0)pp}_{12}$ are replaced by full effective $2$-point 
vertices $G^{p_1 p_2}_{12}(E)$, which are defined as the sum of all connected diagrams 
with $2$ external reservoir lines. With the convention that these two external lines are directed
to the right, it turns out that the energy argument of an effective vertex is identical to
the one of the preceding resolvent, i.e. only the combination 
$R(E_M+\bar{\omega}_M)G^{p_1 p_2}_{12}(E_M+\bar{\omega}_M)$ can occur in the diagrammatic expansion.
Furthermore, for all diagrams
contributing to the effective vertex $G_{12}^{p_1 p_2}(E)$, where the two external lines have the 
sequence $21$, a fermionic sign has to be added. After this resummation, the diagrammatic series
for ${\partial^2\over\partial E^2}L(E)$ up to third order in the effective vertices reads
\begin{equation}
\label{eq:L_rg}
\frac{1}{2}\frac{\partial^2}{\partial E^2}L(E)\,=\,
\frac{1}{2}\,\frac{1}{2}\,
\raisebox{-0.5em}{
\includegraphics[scale=0.45]{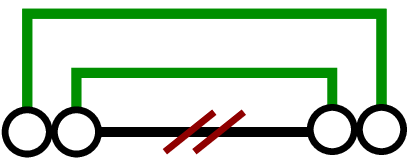}
}
\,+\,
\raisebox{-0.5em}{
\includegraphics[scale=0.45]{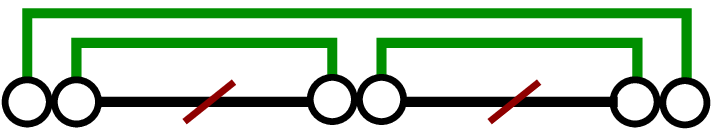}
}
\,+\,\mathcal{O}\left(G^4\right)\quad,
\end{equation}
where the red slash indicates a derivative ${\partial\over\partial E}$ of the corresponding resolvent
(two slashes indicate the second derivative ${\partial^2\over\partial E^2}$). This is one of the
central equations in the E-RTRG approach.
Prefactors arsing from the symmetry factor ${1\over S}$ have explicitly been indicated and all
vertices are full effective $2$-point vertices from now on. All frequency integrations are 
convergent even if one neglects the frequency-dependence of the effective vertices (those can
only enhance convergence). Therefore on the r.h.s. of this differential equation we can take
the limit $D\rightarrow\infty$. This property holds in all orders since, by construction, all
diagrammatic subelements \eqref{eq:diagrams_divergent} leading to a divergence in the infinite-$D$
limit have been eliminated by the resummation procedure. To close the equation one can also derive
in the same way a differential equation for the effective $2$-point vertex 
\begin{align}
\nonumber
&\frac{\partial}{\partial E}G^{p_1p_2}_{12}(E)\,=\,
\Bigg[
\raisebox{-0.75em}{
\includegraphics[scale=0.45]{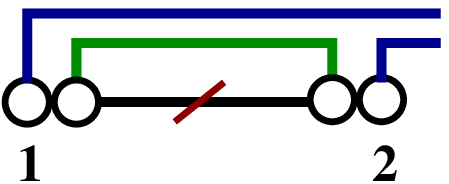}
}
\,-\,(1\leftrightarrow 2)\Bigg]
\\
\nonumber
&\hspace{1.5cm}
\,+\,\frac{1}{2}\,
\raisebox{-1em}{
\includegraphics[scale=0.45]{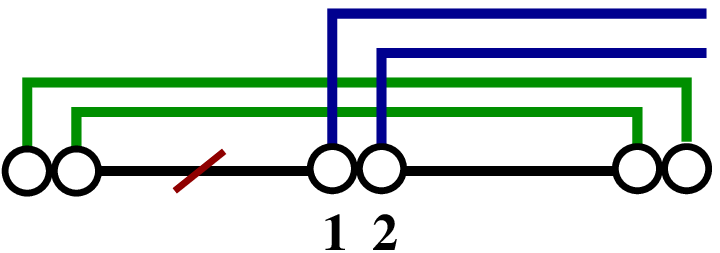}
}
\,+\,\frac{1}{2}\,
\raisebox{-1em}{
\includegraphics[scale=0.45]{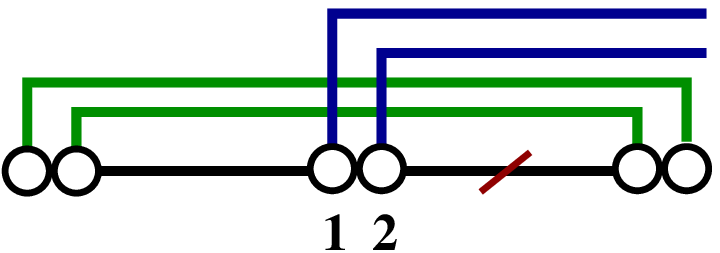}
}
\\
\label{eq:G_rg}
&\hspace{1.5cm}
\,+\,\Bigg[
\raisebox{-1em}{
\includegraphics[scale=0.45]{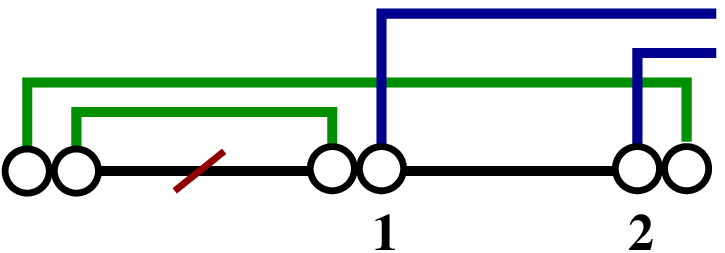}
}
\,+\,
\raisebox{-1em}{
\includegraphics[scale=0.45]{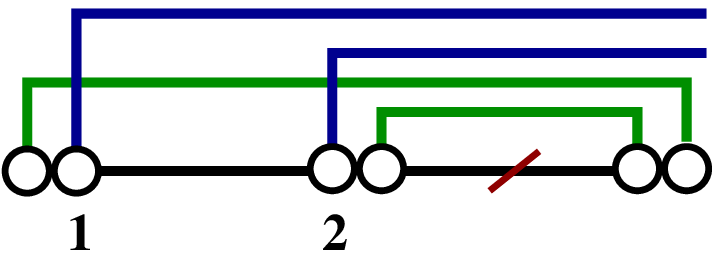}
}
\,-\,(1\leftrightarrow 2)\Bigg]
\,+\,
O(G^4)\,\,.
\end{align}
After the limit $D\rightarrow\infty$ has been taken, the symmetric part \eqref{eq:contraction_symmetric} of 
the contraction becomes an analytic function and does not contribute to the frequency integration when closing
the integration in the upper half of the complex plane. This means that the Keldysh indices no longer appear
explicitly in the RG equations, i.e. only the effective $2$-point vertices averaged over the Keldysh 
indices are needed, which we denote by $G_{12}(E)=\sum_{p_1 p_2}G_{12}^{p_1 p_2}(E)$. This simplifies the
analysis considerably. As a result all contractions can be replaced by the antisymmetric part given by
\eqref{eq:contraction_antisymmetric}
\begin{equation}
\label{eq:RG_contraction}
\gamma_{11'}^{pp'}\,\rightarrow\,\gamma_1^a\,=\,f_\alpha(\bar{\omega})\,-\,{1\over 2}
\quad,
\end{equation}
where we have already taken the limit $D\rightarrow\infty$ and integrated out the trivial
part $\delta_{1\bar{1}}$ of all contractions in the RG diagrams. By convention, $\bar{\omega}$ is
always the frequency variable of the left vertex. \newline

{\bf Frequency dependence.} 
To calculate the integrals over the internal frequencies in the RG diagrams, it is necessary to
know the frequency dependence of the effective vertices and the Liouvillian. This can be treated
systematically by the formalism. Provided that the bare vertices are frequency-independent, one
finds for the vertices that the diagrammatic series for the difference 
$G_{12}(E)-G_{12}(E)_{\bar{\omega}_1=\bar{\omega}_2=0}$ can be resummed by a similiar procedure in terms
of effective $2$-point vertices such that the limit $D\rightarrow\infty$ is well-defined. The reason is 
that at least one resolvent in the original perturbative series must involve the difference 
$R(E_M+\bar{\omega}_M+\bar{\omega}_{M_{\text{ex}}})-R(E_M+\bar{\omega}_M)$, where $M_{\text{ex}}$ contains 
some of the external indices $\{1,2\}$. Fixing this resolvent and resumming the rest of the diagram in
terms of effective $2$-point vertices yields in lowest order the equation 
\begin{align}
\label{eq:G_omega}
\raisebox{-1.em}{
\includegraphics[scale=0.45]{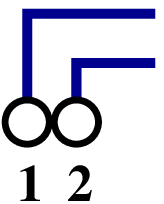}
}
&\,=\,
\raisebox{-1.em}{
\includegraphics[scale=0.45]{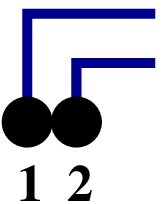}
}
\,+\,
\raisebox{-1.em}{
\includegraphics[scale=0.45]{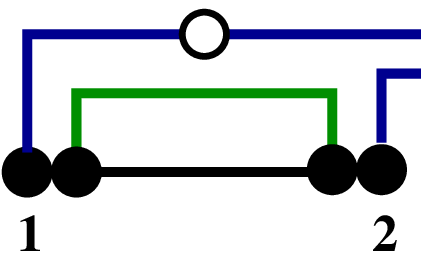}
}
\,-\,
\raisebox{-1.em}{
\includegraphics[scale=0.45]{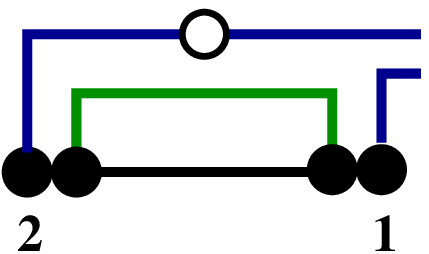}
}
\,+\,\mathcal{O}\left(G^3\right)\quad,
\end{align}
where the filled double dots represent the effective vertices at zero frequency.
This is the second key equation in the E-RTRG approach.
A contraction with an open circle and external frequency $\bar{\omega}_i$ indicates that the
resolvent corresponding to the vertical cut at the position of that circle has to be replaced 
by the difference $R(E_M+\bar{\omega}_M +\bar{\omega}_i)-R(E_M+\bar{\omega}_M)$. This difference falls off
$\sim(\bar{\omega}_M)^2$ w.r.t. the internal frequency integration variables $\bar{\omega}_M$ and, 
therefore, all frequency integrations are convergent in the limit $D\rightarrow\infty$. For 
the frequency dependence of the Liouvillian $L(E_M+\bar{\omega})$ it turns out that the similiar diagrammatic 
series for the difference $L(E+\bar{\omega})-L(E)$ does not exist in the 
limit $D\rightarrow\infty$, similiar to the fact that two derivatives are needed for convergence
(see above). Therefore, one defines a discrete version of the second derivative
$\Delta^2_{\bar{\omega}}L(E)$ via
\begin{equation}
\label{eq:L_omega}
L(E+\bar{\omega})\,=\,L(E)\,+\,{\partial\over\partial E}L(E)\,\bar{\omega}
\,+\,\Delta^2_{\bar{\omega}} L(E)
\,=\,L(E)\,+\,{\partial\over\partial E}L(E)\,\bar{\omega}\,+\,O(G^2)\quad,
\end{equation}
and finds that $\Delta^2_{\bar{\omega}}L(E)$ exists in the limit $D\rightarrow\infty$ and is 
at least of $O(G^2)$ since it involves second and higher-order derivatives of the Liouvillian.
Neglecting $O(G^2)$ (note that this contributes $O(G^4)$ to the RG equations
\eqref{eq:L_rg} and \eqref{eq:G_rg}), the resolvents occuring in the RG diagrams 
of \eqref{eq:L_rg} and \eqref{eq:G_rg} can be written as
\begin{equation}
\label{eq:resolvent_omega}
R(E_M+\bar{\omega}_M)\,=\,{1\over \bar{\omega}_M\,+\,\chi(E_M)}\,Z(E_M)\,+\,O(G^2)\quad,
\end{equation}
with 
\begin{equation}
\label{eq:chi_Z}
\chi(E)\,=\,Z(E)\,(E\,-\,L(E))\quad,\quad
Z(E)\,=\,{1\over 1\,-\,{\partial\over\partial E}L(E)}\quad,
\end{equation} 
where the RG equation for $Z(E)$ follows from the one for ${\partial^2\over\partial E^2}L(E)$ by
\begin{equation}
\label{eq:Z_rg}
{\partial\over\partial E}Z(E)\,=\,Z(E) \left\{{\partial^2\over\partial E^2}L(E)\right\} Z(E)
\,\sim\,O(G^2)\quad.
\end{equation}
Inserting \eqref{eq:G_omega} and \eqref{eq:resolvent_omega} into the RG equations \eqref{eq:L_rg} and
\eqref{eq:G_rg}, calculating all frequency integrations and neglecting all terms of $O(G^4)$, 
one obtains a closed set of RG equations for $G_{12}(E)|_{\bar{\omega}_1=\bar{\omega}_2=0}$ and $L(E)$,
which can be easily solved numerically. These constitute the basic equations of the E-RTRG approach.
The crucial step in the formalism is the parametrization of the frequency dependence,
otherwise a numerical solution would be very
time consuming. Truncating the RG equations at $O(G^2)$ provides the solution up to leading order,
whereas a truncation at $O(G^3)$ includes in addition all sub-leading terms. An important check for the
reliability of the solution is whether these two truncation schemes lead approximately to the same universal
solution. For the nonequilibrium Kondo model at zero magnetic field, the equations have been solved 
in Ref.~\cite{kondo_E-RTRG} to calculate the stationary conductance with reliable results even
in the strong coupling regime. Similiar RG equations can be 
set up for the IRLM and the spin boson model which have been studied in Refs.~\cite{RTRG_irlm,kashuba_PRB13}.
\newline

{\bf RG equations for the slowly varying parts of the Liouvillian.}
We are now ready to show how the decomposition \eqref{eq:L_general_form} can be derived together with 
RG equations for the slowly varying functions $L_\Delta(E)$ and $L'(E)$. First of all, one can see from
the RG equations \eqref{eq:L_rg} and \eqref{eq:G_rg} that ${\partial\over\partial E}L(E)$ and
$G_{12}(E)$ are slowly varying logarithmic functions. At large $E$ we find from dimensional arguments
that 
\begin{equation}
\label{eq:large_E}
{\partial^2\over\partial E^2}L(E)\,,\,{\partial\over \partial E}G_{12}(E)\,\sim\,
{1\over E}\,\left(1\,+\,O({\Delta\over E})\right)\quad,
\end{equation}
where $\Delta$ is some physical scale except $E$. For large $E$, we can neglect the higher orders 
$\sim O({\Delta\over E})$ and we see that, due to the factor ${1\over E}$, logarithmic functions are generated 
by integrating over $E$. For $E$ close to some branching point $z_n$, we find, that even in the worst case
when all resolvents contain the same branching point, that 
${\partial^2\over\partial E^2}L(E)$ and ${\partial\over \partial E}G_{12}(E)$ can at most diverge 
$\sim{1\over E-z_n}$ for $E\rightarrow z_n$. As a result, also for $E\rightarrow z_n$, 
${\partial\over\partial E}L(E)$ and $G_{12}(E)$ are slowly varying logarithmic functions of $E-z_n$. 
This can only be the case if $L(E)$ consists of terms 
\begin{equation}
\label{eq:L_generic_structure}
L(E)\,\sim\, (E-z_n)\,K_n(E-z_n)\,=\,-z_n\,K_n(E-z_n)\,+\,E\,K_n(E-z_n)\quad,
\end{equation}
where $K(E)$ is a slowly varying function, or, more precisely, $z_n=z_k^p-\bar{\mu}_M$ will be replaced 
by $\lambda_k(E_M)-\bar{\mu}_M$ if $E$ is not close to one of the singularities. Therefore, we see
that $L(E)$ can be decomposed in the form \eqref{eq:L_general_form},
\begin{equation}
\label{eq:L_general_form_2}
L(E)\,=\,L_\Delta(E)\,+\,E\,L'(E)\quad,
\end{equation}
with slowly varying functions 
$L_\Delta(E)\sim -z_n K_n(E-z_n)$ and $L'(E)\sim K_n(E-z_n)$. We note that we used precisely 
this form at the end of Section~\ref{sec:diagrammatic_expansion} in Eq.~\eqref{eq:convergence}.
It shows that $L_\Delta(E)$ and $L'(E)$ have a quite similiar structure. \newline

We note that the property that ${\partial\over\partial E}L(E)$ and $G_{12}(E)$ are slowly varying 
logarithmic functions can also be seen directly from the original perturbative expansion 
\eqref{eq:diagram_Sigma_selfconsistent} since in all orders of perturbation theory the number of
frequency integrations is identical to the number of resolvents. This leads to logarithmic integrals
at large and low energies even if all resolvents contain the same cutoff scale at low energies. For
the proof it is essential that the perturbation theory is taken in the self-consistent form 
\eqref{eq:diagram_Sigma_selfconsistent} since this leads to the property that all resolvents 
involve a different combination of the frequencies. The same can be shown for the ohmic 
spin boson model where the decomposition \eqref{eq:L_general_form} holds also in all orders 
of perturbation theory. For models with $1$-point vertices and a flat spectral function 
(like e.g. quantum dot models in the charge fluctuation regime), the number of resolvents 
can be arbitrarily larger than the number of frequency integrations. Here, to show the 
logarithmic scaling at low energies in all orders of perturbation theory, it is very important that the 
resolvents do not only have different frequency combinations but many of them have also different 
cutoff scales at low energies. In contrast to models with spin/orbital fluctuations, it turns out
that already the first derivative ${\partial\over\partial E}L(E)$ exists in the limit 
$D\rightarrow\infty$, see e.g. the first diagram of \eqref{eq:diagrams_divergent}. This means that
${\partial\over\partial E}L(E)\sim {\Gamma\over E-z_n}$ multiplied with 
a well-controlled series with no divergence at high or low energies. This part influences
only the function $L_\Delta(E)$ but not $E L'(E)$. The systematic treatment of all orders
in the tunneling for models with charge fluctuations is still an issue of ongoing research.\newline

To find RG equations for $L_\Delta(E)$ and $L'(E)$, we try to bring the 
RG equation \eqref{eq:L_rg} for ${\partial^2\over\partial E^2}L(E)$ into the form
\begin{equation}
\label{eq:L_2der_special_form}
{\partial^2\over \partial E^2}L(E)\,=\,{\partial\over \partial E}L'(E)
\,+\,{\partial\over \partial E}\left\{{\partial\over \partial E}L_\Delta(E)
\,+\,E{\partial\over \partial E}L'(E)\right\}\quad,
\end{equation}
such that ${\partial\over\partial E}L_\Delta(E)$ and ${\partial\over\partial E}L'(E)$ can be 
identified and that $L_\Delta(E)$ is proportional to some physical scale $\Delta$ except $E$.
For simplicity we show the procedure only up to $O(G^2)$, for $O(G^3)$ see Ref.~\cite{goettel_preprint}.
Taking only the first term on the r.h.s. of the RG equation \eqref{eq:L_rg}, replacing the
vertices by the ones at zero frequency via \eqref{eq:G_omega}, and shifting the two derivatives of
the resolvent 
${\partial^2\over\partial E^2}R(E_{12}+\bar{\omega}_{12})={\partial\over\partial\bar{\omega}_1}
{\partial\over\partial\bar{\omega}_2}R(E_{12}+\bar{\omega}_{12})$ via two partial integrations
to the contractions, we obtain
\begin{align}
\label{eq:L_rg_special_form}
{\partial^2\over \partial E^2}L(E)\quad &= \quad
{1\over 2}
\begin{picture}(10,10)
\put(5,-7){\includegraphics[height=0.7cm]{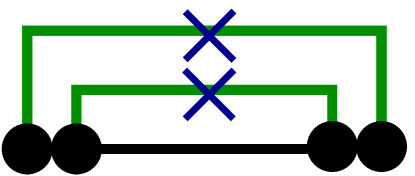}}
\end{picture}
\hspace{2cm}
+\quad{\partial\over\partial E}\left\{{1\over 2}
\begin{picture}(10,10)
\put(5,-7){\includegraphics[height=0.7cm]{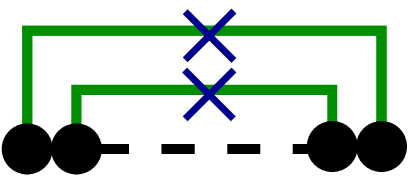}}
\end{picture}
\hspace{1.5cm}\right\}\quad+\quad O(G^3)\quad,
\end{align}
where a cross at a contraction denotes the derivative 
${\partial\over\partial\bar{\omega}}\gamma_1^a={\partial\over\partial\bar{\omega}}f_\alpha(\bar{\omega})$,
see \eqref{eq:RG_contraction}. The dashed line in the second term indicates that
the resolvent is replaced by the $Z'$-factor $R(E_M+ß\bar{\omega}_M)\rightarrow Z'(E_M+\bar{\omega})$, 
defined in \eqref{eq:tilde_L_Z}. Therefore, this term is of $O(G^3)$ and can be added without violating
the consistency of the truncation scheme up to $O(G^2)$. The term has been added in such a way that 
when identifying \eqref{eq:L_rg_special_form} with \eqref{eq:L_2der_special_form}, the derivative 
${\partial\over\partial E}L_\Delta(E)$ will become proportional to a physical scale $\Delta$. 
Together with the relation
\begin{align}
\label{eq:Z_chi_Delta}
Z'(E_M+\bar{\omega}_M)\,-\,E\,R(E_M+\bar{\omega}_M)
\,&=\,\chi_\Delta(E,\bar{\mu}_M+\bar{\omega}_M)\,R(E_M+\bar{\omega}_M)\quad,\\
\label{eq:chi_Delta}
\chi_\Delta(E,\bar{\mu}_M+\bar{\omega}_M)\,&=\,\bar{\mu}_M\,+\,\bar{\omega}_M\,-\,
\tilde{L}_\Delta(E_M+\bar{\omega}_M)\quad,
\end{align}
which follows from \eqref{eq:tilde_L_Z} with the definition $\tilde{L}_\Delta(E)=Z'(E)L_\Delta(E)$, we obtain
\begin{align}
\label{eq:L_Delta_L'_rg}
{\partial\over \partial E} L_\Delta(E)\,=\,
{1\over 2}
\begin{picture}(10,10)
\put(5,-12){\includegraphics[height=0.9cm]{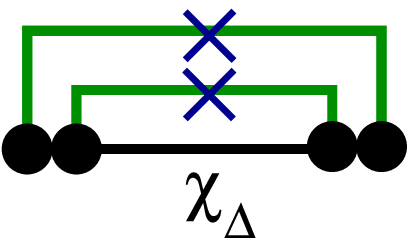}}
\end{picture}
\hspace{1.6cm}\,+\,O(G^3)
\quad,\quad
{\partial\over \partial E} L'(E)\,=\, 
{1\over 2}\,
\begin{picture}(10,10)
\put(5,-7){\includegraphics[height=0.7cm]{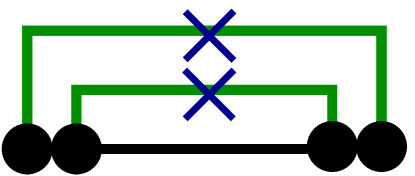}}
\end{picture}
\hspace{1.6cm}\,+\,O(G^3)\,,
\end{align}
where the symbol $\chi_\Delta$ at the resolvent means that the resolvent multiplied
with $\chi_\Delta(E,\bar{\mu}_M+\bar{\omega}_M)$ has to be taken. Obviously, $\chi_\Delta$ is
proportional to a physical scale, since $\bar{\mu}_M$, $\bar{\omega}_M$ and
$\tilde{L}_\Delta(E_M+\bar{\omega})$ have this property. For $\bar{\omega}_M$ this follows
from the fact that the RG equations contain only the derivatives 
${\partial\over\partial\bar{\omega}}\gamma_1^a={\partial\over\partial\bar{\omega}}f_\alpha(\bar{\omega})$
of the contractions, such that $|\bar{\omega}_i|\lesssim T_{\alpha_i}$. In contrast to the 
RG equation \eqref{eq:L_rg} for the full Liouvillian $L(E)$, the RG equations \eqref{eq:L_Delta_L'_rg}
for $L_\Delta(E)$ and $L'(E)$ are first order differential equations. Therefore, the 
differences $L_\Delta(E_M+\bar{\omega}_M)-L_\Delta(E_M)$ and $L'(E_M+\bar{\omega}_M)-L'(E_M)$ are
of $O(G^2)$ such that the frequency dependence of the resolvent and $\chi_\Delta$ entering the
RG equations \eqref{eq:L_Delta_L'_rg} can be approximated by
\begin{align}
\label{eq:R_omega_Delta}
R(E_M+\bar{\omega}_M)\,&=\,{1\over \bar{\omega}_M\,+\,E_M\,-\,\tilde{L}_\Delta(E_M)}\,Z'(E_M)
\,+\,O(G^2)\\
\label{eq:chi_omega_Delta}
\chi_\Delta(E,\bar{\mu}_M+\bar{\omega}_M)\,&=\,
\bar{\mu}_M\,+\,\bar{\omega}_M\,-\,\tilde{L}_\Delta(E_M)\,+\,O(G^2)\quad.
\end{align}
As a consequence, all frequency integrations can be straightforwardly performed such that the
differential-integro equations \eqref{eq:L_Delta_L'_rg} are converted into differential equations.
E.g., at zero temperature, the two frequency integrations in \eqref{eq:L_Delta_L'_rg}
are trivial leading to the explicit expression
\begin{align}
\label{eq:L_Delta_T=0_rg}
{\partial\over\partial E}L_\Delta(E)\,&=\,{1\over 2}\,G_{12}(E)\,
{\bar{\mu}_{12}\,-\,\tilde{L}_\Delta(E_{12}) \over E_{12}\,-\,\tilde{L}_\Delta(E_{12})}\,
Z'(E_{12})\,G_{\bar{2}\bar{1}}(E_{12})\quad,\\
\label{eq:L'_T=0_rg}
{\partial\over\partial E}L'(E)\,&=\,{1\over 2}\,G_{12}(E)\,
{1 \over E_{12}\,-\,\tilde{L}_\Delta(E_{12})}\,
Z'(E_{12})\,G_{\bar{2}\bar{1}}(E_{12})\quad,
\end{align}
together with the RG equation for the vertex which follows from the lowest order term of
\eqref{eq:G_rg} as
\begin{equation}
\label{eq:G_T=0_rg}
{\partial\over\partial E}G_{12}(E)\,=\,
G_{13}(E)\,{1 \over E_{13}\,-\,\tilde{L}_\Delta(E_{13})}\,Z'(E_{13})\,G_{\bar{3}2}(E_{13})
\,-\,(1\leftrightarrow 2)\quad.
\end{equation}
\newline

{\bf Solution of approximate E-RTRG equations.}
The first-order RG equations \eqref{eq:L_Delta_T=0_rg} and \eqref{eq:L'_T=0_rg} for
$L_\Delta(E)$ and $L'(E)$ provide the most convenient starting point 
for an analytical solution of the RG equations at least in that regime of the complex plane where 
the effective vertices stay small, see Refs.~\cite{kashuba_PRB13,goettel_preprint} for details. 
The strategy is to solve the RG equations approximately in three different energy regimes by expanding in the
effective vertices but keeping large logarithmic terms (either at large or low energies) to all orders, and 
matching the different solutions to fix the integration constants. Denoting the small dimensionless 
coupling constant by $\alpha$, we distinguish the following regimes: (1) The regime of high energies
$|E|\gg|z_n|$, where the RG resums all ultraviolett logarithmic terms $\sim (\alpha\ln{D\over -iE})^k$; 
(2) The regime of intermediate and small energies $|E-z_n|\lesssim O(|z_n|)$ but $E$ not too close to
the branching points such that one can expand in the small parameter 
$\alpha|\ln{|z_n|\over|E-z_n|}|\ll 1$; (3) The regime of small energies exponentially close to some 
of the branching points, i.e. $|E-z_n|\ll O(|z_n|)$ and $\alpha|\ln{|z_n|\over|E-z_n|}|\sim 1$, where
the RG resums all infrared logarithmic terms $\sim(\alpha|\ln{|z_n|\over|E-z_n|}|)^k$. In particular for
the ohmic spin boson model and the IRLM, we will see in 
Section~\ref{sec:results} that the coupling constant $\alpha$ stays
small in the whole complex plane such that a well-controlled analytical solution is possible for all $E$,
showing that the resummation of logarithmic terms for high and low energies gives very different results.
For the Kondo problem a weak-coupling solution is only possible for high, intermediate and small energies,
but not for exponentially small energies where the coupling constant $\alpha\sim O(1)$.\newline

{\bf Initial conditions.}
The initial conditions for the RG flow
at large energies are set up at the value $E=iD$, where $D\gg|z_n|$ is the high-energy cutoff.
The motivation for the choice $E=iD$ lies in the fact that, for $D\gg|E|\gg|z_n|$, the bare perturbation 
series for $L_\Delta(E)$ and $L'(E)$ contain logarithmic terms $\sim (\ln{D\over -iE})^k$ of all
powers $k$ (we have chosen $-iE$ in the argument, such that the branch cut is directed towards 
the negative imaginary axis). All other terms $\sim ({|E|\over D})^n$ are neglected since they vanish in the
limit $D\rightarrow\infty$ and thus do not contribute to the universal solution
which is independent of the cutoff details. Extrapolating
this result up to $E=iD$ has the effect that all logarithmic terms vanish, which sets the initial point
for the universal RG flow. The calculation of the initial values can be done by bare perturbation theory
for $D\gg|E|\gg|z_n|$ and omitting all logarithmic contributions. For small bare coupling constants it is
sufficient to take the lowest order term if it is universal, otherwise one takes zero for the initial 
condition. We note that this procedure works well to determine the universal initial condition for  
$L_\Delta(E)$ and $L'(E)$ at $E=iD$ but fails for the initial condition of $L(E)$ since $L(E)$ contains
terms linear in $E$ which are very large for $E=iD$. Therefore, for the RG equation \eqref{eq:L_rg} one
either has to keep the high-energy cutoff function in the RG equations and start the RG flow at 
$|E|\gg D$ (where one can take the bare values as initial condition), or one has to find a reference
point at low energies where $L(E)$ is known from exact results, see e.g. the solution of the Kondo 
model in strong coupling in Ref.~\cite{kondo_E-RTRG}.\newline

{\bf RG for the Liouvillian discontinuity jumps.}
Finally we show how RG equations can be derived for the jump $\delta L$ of the Liouvillian at a particular 
branch cut, as this jump is needed to evaluate the branch cut contributions to the
time evolution, see Eq.~\eqref{eq:F_branching_point}. As described in Section~\ref{sec:diagrammatic_expansion}
the branch cuts of $L(E)$ occur only at zero temperature and can be identified in the perturbative expansion 
\eqref{eq:diagram_Sigma_selfconsistent} by closing all frequency integrations in the upper half of the
complex plane and considering the branch cuts of the Fermi distribution functions on the positive imaginary
axis. This means that the frequencies are shifted to the positive imaginary axis
$\bar{\omega}_M\rightarrow i|\bar{\omega}_M|$. In leading order, a given branch cut at 
$E=z_n-ix\pm 0^+$ is generated by some resolvent which is resonant, i.e. the jump of this 
resolvent across the branch cut becomes a $\delta$-function. If the resolvent contains the 
eigenvalue $\lambda_k(E_M+i|\bar{\omega}_M|)$ the resonance occurs if $z_n=z_k^p-\bar{\mu}_M$.
With $E_M=z_n+\bar{\mu}_M-ix\pm 0^+=z_k^p-ix\pm 0^+$, we replace approximately 
$\lambda_k\rightarrow z_k^p$, $P_k\rightarrow \bar{P}_k(z_k^p-ix)$ and 
$Z'\rightarrow \bar{Z}'(z_k^p-ix)$, which gives for the jump of the resolvent the 
following $\delta$-function
\begin{eqnarray}
\nonumber
\delta R\,&=&\,\left({1\over -ix+i\bar{\omega}_{1\dots n}+0^+}-
{1\over -ix+i|\bar{\omega}_M|-0^+}\right)\,\bar{P}_k(z_k^p-ix)\,\bar{Z}'(z_k^p-ix)\\
\nonumber\\
\label{eq:R_jump}
&=&\,2\pi \delta(|\bar{\omega}_M|-x)\,\bar{P}_k(z_k^p-ix)\,\bar{Z}'(z_k^p-ix)\quad.
\end{eqnarray}
As expected, the frequency integrals give only a contribution for $x>0$, since this is
the region where the branch cut starts.
The RG equation for ${\partial\delta L\over \partial E}(z_n-ix)$ is obtained by fixing
the resonant resolvent together with the resolvent where the $E$-derivative is taken 
and resumming the rest of the perturbative series in terms of effective $2$-point vertices.
If both resolvent are the same, the $E$-derivative is replaced by a frequency derivative
${\partial\over\partial\bar{\omega}_i} $ and is shifted via partial integration to the 
derivative of some contraction crossing over the resolvent. Thus, we obtain in leading order
\begin{align}
\label{eq:delta_L_rg}
{\partial\delta L\over\partial E}(z_n-ix)\,=\,
i{\partial\over\partial x}\delta L(z_n-ix)\, = \,-\,{1\over 2}
\begin{picture}(10,10)
\put(5,-7){\includegraphics[height=1cm]{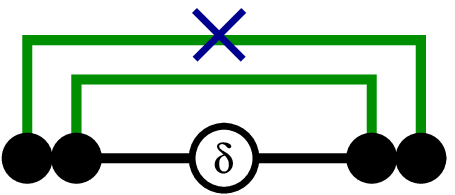}}
\end{picture}
\end{align}
where the symbol $\delta$ at the resolvent means that we replace the resolvent by its jump
$\delta R$ given by \eqref{eq:R_jump}. The initial condition for the RG equation is 
$\delta L(z_n)=0$. At zero temperature both frequency integrations are trivial. The contraction
with the cross gives 
$\int d\bar{\omega}f_\alpha^\prime(\bar{\omega})\{\dots\}=-\{\dots\}_{\bar{\omega}=0}$,
whereas the contraction without the cross leads to
\begin{equation}
\label{eq:int_omega}
\int \,d\bar{\omega}\,(f_\alpha(\bar{\omega})-{1\over 2})\,\{\dots\}\,=\,
-{1\over 2}\,\int \,d\bar{\omega}\,\text{sign}(\bar{\omega})\,\{\dots\}\,=\,
-i\,\int_0^\infty \,d\bar{\omega}\,\{\dots\}_{\bar{\omega}\rightarrow i\bar{\omega}}\quad,
\end{equation}
where we have used that the jump of the sign-function at the branch cut is given by $2$.
Therefore, \eqref{eq:delta_L_rg} gives explicitly
\begin{align}
\label{eq:delta_L_rg_explicit}
{\partial\over\partial x}\delta L(z_n-ix)\,=\,
-\pi\,\theta(x)\,\bar{G}_{12}(z_k^p-\bar{\mu}_{12}-ix)\,
\bar{P}_k(z_k^p-ix)\,\bar{Z}'(z_k^p-ix)\,\bar{G}_{\bar{2}\bar{1}}(z_k^p-ix)\quad,
\end{align}
where we have replaced all vertices that are discontinous across the branch cut by their
average value $\bar{G}$. Up to the corrections from the weak $x$-dependence of the 
logarithmic functions on the r.h.s. of this equation, we obtain 
\begin{equation}
\label{eq:delta_L_spin/orbital_fluctuations_2}
\delta L(z_n-ix)\,\sim \, x\,\theta(x)\quad,
\end{equation}
giving Eq.~\eqref{eq:delta_L_spin/orbital_fluctuations} used in Section~\ref{sec:time_evolution}.
Therefore, if $x$ can be neglected in the 
resolvents of the integrand of (\ref{eq:branching_point}),
we obtain (up to logarithmic corrections) $\rho_t^{n,b}\sim 1/t^2$ in the long-time limit
for {\it all} models with spin or orbital fluctuations, like e.g. the Kondo model. Similiar
considerations show that the same holds for the ohmic spin boson model, whereas for models
with charge fluctuations (like the IRLM), one obtains $\delta L(z_n-ix)\sim \theta(x)$.

\section{Results}
\label{sec:results}

In this section we will discuss the application of the formalism to the models introduced
in Section~\ref{sec:models}. Since the Kondo model has the simplest algebra, we will take
this model as a tutorial example to discuss the solution of the RG equations and the consequences
for the time evolution in all detail. Since the general strategy is always the same, we will
then show briefly the results for the ohmic spin boson model and the IRLM, and will
concentrate on interesting features which are different from the ones for the Kondo model.

\subsection{Kondo model}

We consider the nonequilibrium Kondo model at zero magnetic field $h^{(0)}=0$ and zero
temperature $T=0$ for the antiferromagnetic case $J^{(0)}>0$. We assume that the 
local spin-${1\over 2}$ is coupled to several 
reservoirs with chemical potentials $\mu_\alpha$. For the special case of two reservoirs
$\alpha\equiv L/R\equiv \pm$, we take $\mu_\alpha=\alpha {V\over 2}$, where $V$ denotes
the bias voltage across the system. Following Refs.~\cite{pletyukhov_PRL10,kondo_E-RTRG} 
our aim is to calculate the time evolution of the local spin $\langle \underline{S}\rangle(t)$. 
\newline

The model is spin rotational invariant and therefore the effective Liouvillian $L(E)$ should 
be an invariant under spin rotations. Defining two basis spinoperators $\underline{L}_\pm$ in 
Liouville space by ($A$ is an arbitrary local operator) $\underline{L}^+ A=\underline{S}A$ and
$\underline{L}^- A=-A\underline{S}$, the only two invariants are given by the identity and
$\underline{L}^+\cdot\underline{L}^-$. Since the Liouvillian must also fulfil $\text{Tr}L(E)=0$, we
find that the Liouvillian can be parametrized by
\begin{equation}
\label{eq:L_kondo}
L(E)\,=\,-i\,\Gamma(E)\,L^a \quad,\quad L^a\,=\,{3\over 4}\,+\,\underline{L}^+\cdot\underline{L}^- \quad.
\end{equation}
$\Gamma(E)$ is the energy dependent spin relaxation rate. The Liouvillian
has one zero eigenvalue with projector $P_{\text{st}}=1-L^a$ and three degenerate
eigenvalues at $-i\Gamma(E)$ with projector $1-P_{\text{st}}=L^a$.
Therefore, by using \eqref{eq:time_evolution}, we can write for the time evolution of the local
density matrix
\begin{equation}
\label{eq:kondo_time_evolution}
\rho(t)\,=\,\rho_{\text{st}}\,+\,{i\over 2\pi}\,\int_{-\infty+i0^+}^{\infty+i0^+}\hspace{-0.5cm}dE\,
e^{-iEt}\,{1\over E\,+\,i\Gamma(E)}\,L^a\,\rho_{t=0}\quad,
\end{equation}
where $\rho_{\text{st}}=(1-L^a)\rho_{t=0}$ is the diagonal stationary density matrix with
equal probabilities for both spin directions $(\rho_{\text{st}})_{ss}={1\over 2}$. Using
$\text{Tr}\underline{S}\rho(t)=\text{Tr}\underline{S}L^a\rho(t)$ and $(L^a)^2=L^a$, we find
that the spin relaxation rate $\Gamma(E)$ determines the spin dynamics via
\begin{equation}
\label{eq:kondo_spin_dynamics}
\langle \underline{S} \rangle(t)\,=\,{i\over 2\pi}\,\int_{-\infty+i0^+}^{\infty+i0^+}\hspace{-0.5cm}dE\,
e^{-iEt}\,{1\over E\,+\,i\Gamma(E)}\,\langle \underline{S}\rangle_{t=0}\quad.
\end{equation}
As a consequence, the operator structure of the Liouvillian is no longer important, and we can
use all formulas derived in Section~\ref{eq:time_evolution} for the time evolution with the 
replacement $L(E)\rightarrow -i\Gamma(E)$ or $R(E)\rightarrow{1\over E+i\Gamma(E)}$. 
This means that all projectors $P_k$ can be left out and $Z'(E)$ can be used from the decomposition
\begin{equation}
\label{eq:gamma_decomposition}
\Gamma(E)\,=\,\Gamma_\Delta(E)\,+\,E\,\Gamma'(E)\quad,\quad 
R(E)\,=\,{1\over E\,+\,i\Gamma(E)}\,=\,{1\over E\,+\,i\tilde{\Gamma}_\Delta(E)}\,Z'(E)\quad,
\end{equation}
with 
\begin{equation}
\label{eq:kondo_tilde_Gamma_Z'}
\tilde{\Gamma}_\Delta(E)\,=\,Z'(E)\,\Gamma_\Delta(E)\quad,\quad 
Z'(E)\,=\,{1\over 1\,+\,i\Gamma'(E)}\quad.
\end{equation}
Analysing the functions $\tilde{\Gamma}_\Delta(E)$ and $Z'(E)$ in leading order from the 
RG equations \eqref{eq:L_Delta_T=0_rg} and \eqref{eq:L'_T=0_rg}, we will show below 
that the branching poles and branching points of the resolvent $R(E)$ are given by
\begin{equation}
\label{eq:kondo_branch_cuts}
z_0^p\,=\,-i\Gamma^* \quad,\quad 
z_{\alpha\alpha'}^b\,=\,-i\Gamma^*\,+\,\mu_{\alpha}\,-\,\mu_{\alpha'}\quad\text{for}\quad \alpha\ne\alpha'
\end{equation}
where we used the notation of Eq.~\eqref{eq:total_evolution_pb} and  assumed
that all reservoirs have different chemical potentials (otherwise they can be taken together). 
$z_0^p=i\tilde{\Gamma}_{\Delta}(z_0^p)$ is the pole of the resolvent $R(E)$ and
$\Gamma^*$ is called the Korringa rate \iffindex{Korringa rate} which, as will be shown below, is given by
\begin{equation}
\label{eq:kondo_pole}
\Gamma^*\,=\,2\pi\,J_V^2\,\sum_{\alpha\ne\alpha'}\,x_\alpha\,x_{\alpha'}\,|\mu_\alpha-\mu_{\alpha'}|
\quad,\quad J_V\,=\,{1\over 2\,\ln{V\over T_K}}\quad,\quad T_K\,=\,D\,e^{-1/(2J^{(0)})}\quad,
\end{equation}
where we have used the notation of Eq.~\eqref{eq:J_property}. $J_V$ is the renormalized exchange coupling 
at the scale $V=\text{max}_{\alpha\alpha'}\{|\mu_{\alpha}-\mu_{\alpha'}|\}$ 
(which is the bias voltage for two reservoirs) and $T_K$ is called the Kondo temperature
\iffindex{Kondo temperature}. The result
for the Korringa rate holds in the weak-coupling case $V\gg T_K$ which we will consider from now on.
In this case, we get $J_V\ll 1$ and $\Gamma^*\ll V$. 
Using \eqref{eq:total_evolution_pb} and \eqref{eq:kondo_branch_cuts}, we get 
the following general form for the time evolution of the local spin 
\begin{equation}
\label{eq:kondo_total_evolution_pb}
\langle \underline{S} \rangle(t)\,=\,
F_0^p(t)\,e^{-\Gamma^* t}\,\langle \underline{S} \rangle(t)_{t=0}\,+\,
\sum_{\alpha\ne\alpha'}\,F_{\alpha\alpha'}^b(t)\,e^{-\Gamma^*t}\,e^{-i(\mu_\alpha-\mu_{\alpha'})t}\,
\langle \underline{S} \rangle(t)_{t=0}\quad.
\end{equation}
For {\bf intermediate and long times} $t\gtrsim{1\over V}$, the pre-exponential functions can be calculated from
\eqref{eq:branching_pole_zero} and \eqref{eq:F_branching_point} as
\begin{align}
\label{eq:kondo_F_p}
F^p_0(t)\,&=\,\bar{Z}'(-i\Gamma^*-i/t)\quad,\\
\label{eq:kondo_F_b}
F^b_{\alpha\alpha'}(t)\,&=\,-{i\over 2\pi}\,\int_0^\infty\,dx\,e^{-xt}\,
{\bar{Z}'(z_{\alpha\alpha'}^b-i/t)^2\,\delta\Gamma(z_{\alpha\alpha'}^b-ix)\over
(\mu_\alpha\,-\,\mu_{\alpha'}\,-\,ix)^2}\quad,
\end{align}
where we have neglected $-i\Gamma^*+i\bar{\tilde{\Gamma}}_\Delta(z_{\alpha\alpha'}^b-ix)$
in the dominator of the integrand of the last equation. This can be done for 
$|\mu_\alpha-\mu_{\alpha'}|\gg\Gamma^*$ for all $\alpha\ne\alpha'$, i.e. if the branch cuts 
are sufficiently apart from each other (other cases can be treated as well but need a special 
procedure \cite{pletyukhov_PRL10}). For {\bf short times} $t\ll {1\over V}$, we can use 
\eqref{eq:short_times} and get
\begin{equation}
\label{eq:kondo_short_times}
\langle \underline{S} \rangle(t)\,=\,
Z'(1/t)\,e^{-\tilde{\Gamma}_\Delta(1/t) t}\,\langle \underline{S} \rangle(t)_{t=0} \quad.
\end{equation}
To evaluate \eqref{eq:kondo_F_p}, \eqref{eq:kondo_F_b} and \eqref{eq:kondo_short_times} explicitly,
we need the functions $\tilde{\Gamma}_\Delta(E)$, $Z'(E)$ and $\delta\Gamma(z_{\alpha\alpha'}-ix)$,
which we will derive in the following by considering the RG equations
\eqref{eq:L_Delta_T=0_rg}, \eqref{eq:L'_T=0_rg} and \eqref{eq:delta_L_rg_explicit},
together with the RG equation \eqref{eq:G_T=0_rg} for the vertex.\newline

In leading order it can be shown that the effective vertex $G_{11'}(E)$ can be parametrized in the
same form as the initial vertex $G^{(0)}_{11'}$, defined by \eqref{eq:kondo_vertex_operator} and 
\eqref{eq:G_vertex_liouville}. This gives the form
\begin{equation}
\label{eq:G_kondo}
G_{11'}(E)\,=\,-\,J_{\alpha\alpha'}(E)\,\underline{L}^2\cdot
\underline{\sigma}_{\sigma\sigma'}
\quad \text{for}\quad \eta=-\eta'=+\quad,\quad
\underline{L}^2\,=\,-{1\over 2}\,(\underline{L}^+ \,+\, \underline{L}^-) \quad,
\end{equation}
together with $G_{11'}(E)=-G_{1'1}(E)$ for $\eta=-\eta'=-$. Using this ansatz together with
the form \eqref{eq:L_kondo} for the Liouvillian in the RG equations and omitting the terms
$\sim L^a$ in the RG equation for the vertex (which generate higher orders), one finds after
some straightforward algebra \cite{kondo_E-RTRG} the following RG equations
\begin{align}
\label{eq:kondo_gamma_Delta_rg}
{\partial\over\partial E}\Gamma_\Delta(E)\,&=\,i\,\chi_\Delta(E,\hat{\mu}_{\alpha\alpha'})\,
R(\hat{E}_{\alpha\alpha'})\,J_{\alpha\alpha'}(E)\,J_{\alpha'\alpha}(\hat{E}_{\alpha\alpha'})\quad,\\
\label{eq:kondo_Z'_rg}
{\partial\over\partial E}Z'(E)\,&=\,Z'(E)^2\,
R(\hat{E}_{\alpha\alpha'})\,J_{\alpha\alpha'}(E)\,J_{\alpha'\alpha}(\hat{E}_{\alpha\alpha'})\quad,\\
\label{eq:kondo_J_rg}
{\partial\over\partial E}J_{\alpha\alpha'}(E)\,&=\,-{1\over 2}\,
R(\hat{E}_{\alpha\alpha''})\,J_{\alpha\alpha''}(E)\,J_{\alpha''\alpha'}(\hat{E}_{\alpha\alpha''})
\,-\,{1\over 2}\,
R(\hat{E}_{\alpha''\alpha'})\,J_{\alpha''\alpha'}(E)\,J_{\alpha\alpha''}(\hat{E}_{\alpha''\alpha})
\quad,
\end{align}
where $\hat{E}_{\alpha\alpha'}=E+\hat{\mu}_{\alpha\alpha'}$, 
$\hat{\mu}_{\alpha\alpha'}=\mu_\alpha-\mu_{\alpha'}$, 
$\chi_\Delta(E,\hat{\mu}_{\alpha\alpha'})=\hat{\mu}_{\alpha\alpha'}
+i\tilde{\Gamma}_\Delta(\hat{E}_{\alpha\alpha'})$, and $Z'(E)$ and $R(E)$ have been defined 
in \eqref{eq:gamma_decomposition} and \eqref{eq:kondo_tilde_Gamma_Z'}. The initial conditions
at $E=iD$ are $\Gamma_\Delta=0$, $Z'=1$ and $J_{\alpha\alpha'}=2\sqrt{x_\alpha x_{\alpha'}}J^{(0)}$.\newline

We first start with the analytic solution in the regime of {\bf high energies} $|E|\gg V$. Neglecting 
$\hat{\mu}_{\alpha\alpha'}$ everywhere gives the solution $\Gamma_\Delta(E)\approx 0$ and
$J_{\alpha\alpha'}(E)\approx 2\sqrt{x_\alpha x_{\alpha'}}J(E)$, together with
\begin{align}
\label{eq:Z'_high_energies_rg}
{\partial\over\partial E}Z'(E)\,=\,{4\over E}\,Z'(E)^2\,J(E)^2
\,=\,{4\over E}\,J(E)^2 + O(J^3)\quad,\\
\label{eq:J_high_energies_rg}
{\partial\over\partial E}J(E)\,=\,-{2\over E}\,Z'(E)\,J(E)^2
\,=\,-{2\over E}\,J(E)^2 + O(J^3)\quad,
\end{align}
where we have used $Z'= 1+O(J)$ on the r.h.s. (see Eq.~\eqref{eq:Z'_J_high_energies}). We find that
$-{1\over 2 J(E)}+\ln(-iE)\equiv \ln{T_K}$ is an invariant and ${\partial Z'\over\partial J}= -2$, which
gives the solution
\begin{equation}
\label{eq:Z'_J_high_energies}
Z'(E)\,=\,1\,-\,2\,(J(E)\,-\,J^{(0)})\,\rightarrow 1\,-\,2\,J(E) \quad,\quad
J(E)\,=\,{1\over 2\ln{-iE\over T_K}}\quad,
\end{equation}
where the Kondo temperature $T_K$ is defined in \eqref{eq:kondo_pole} and we used the scaling limit 
in the first equation, defined by $D\rightarrow\infty$, $J^{(0)}\rightarrow 0$, such that
$T_K$ remains a constant. We have chosen $-iE$ in the argument of the logarithm to define a real value
for the Kondo temperature at $E=iD$ and since we want the branch cut of the logarithm to point into the
direction of the negative imaginary axis. In the solution \eqref{eq:Z'_J_high_energies} all
logarithmic terms $\sim (J^{(0)}\ln{D\over -iE})^k$ have been resummed.
$J(E)$ is the poor man scaling solution of the Kondo model,
already introduced in the lecture B3 by T. Costi, but with the difference that $-iE$ plays now the
role of the effective energy scale. Most importantly, the solution would diverge at $E=iT_K$ when
extrapolated to small energies, indicating an increase of antiferromagnetic spin fluctuations at the
scale of the Kondo temperature. However, in this regime the solution can not be used since the 
RG flow becomes very different for $|E|\lesssim V$. The solution at high energies can be used to
evaluate the universal solution \eqref{eq:kondo_short_times} for {\bf short times} $t\ll {1\over V}$
\begin{equation}
\label{eq:kondo_short_times_explicit}
\langle \underline{S} \rangle(t)\,=\,
\left(1\,-\,{1\over|\ln(T_K t)|}\right)\,\langle \underline{S} \rangle(t)_{t=0} \quad.
\end{equation}
In this result all logarithmic terms $\sim (J^{(0)}\ln(Dt))^k$ have been resummed, which can be
seen from $1/\ln(T_K)=-2J^{(0)}/(1-2J^{(0)}\ln(Dt))$. Sub-leading terms are not included but
can be taken into account by truncating the RG equations at $O(G^3)$ \cite{pletyukhov_PRL10}.
For ferromagnetic Kondo models the universal short time behaviour has also been discussed 
in Ref.~\cite{Hackl_09} using flow equation methods.
\newline

To find the solution at {\bf intermediate and small energies} $|E|\lesssim V$ but not exponentially close to the
singularites such that $J(E)\ll 1$ is still fulfilled (we state below what this precisely means),
we first set the initial value by expanding the solution \eqref{eq:Z'_J_high_energies} at high energies
for the case when $E$ starts to approach $V$ such that $|E|\gg V$ is still fulfilled 
but $J_V|\ln{-iE\over V}|\ll 1$, where $J_V=1/(2\ln(V/T_K))$ is the exchange coupling at high energies 
evaluated at the scale $-iE=V$, as introduced in \eqref{eq:kondo_pole}. This gives
\begin{equation}
\label{eq:crossover_high_low}
J_{\alpha\alpha'}(E)\,\approx\,2\sqrt{x_\alpha x_{\alpha'}}\,J_V\,(1\,-\,2\,J_V\,\ln{-iE\over V})\quad,\quad
Z'(E)\,\approx\,1\,-\,2\,J_V\,+\,4\,J_V^2\,\ln{-iE\over V}\quad.
\end{equation}
Using the first term of this expansion in the r.h.s. of the full RG equations
\eqref{eq:kondo_gamma_Delta_rg}, \eqref{eq:kondo_Z'_rg} and \eqref{eq:kondo_J_rg}, we can 
easily integrate the equations up to $O(J_V^2)$ by using
$R(E)={\partial\over\partial E}\ln{-i(E+i\tilde{\Gamma}_\Delta(E))\over V} + O(J_V^2)$
\begin{align}
\label{eq:kondo_gamma_Delta_low}
\tilde{\Gamma}_\Delta(E)\,&=\,i\,4\,x_\alpha x_{\alpha'}\,J_V^2\,
(\hat{\mu}_{\alpha\alpha'}\,+i\tilde{\Gamma}_\Delta(\hat{E}_{\alpha\alpha'}))\,
\ln{-i(\hat{E}_{\alpha\alpha'}+i\tilde{\Gamma}_\Delta(\hat{E}_{\alpha\alpha'}))\over V}\quad,\\
\label{eq:kondo_Z'_low}
Z'(E)\,&=\,1\,-\,2\,J_V\,+\,
4\,x_\alpha x_{\alpha'}\,J_V^2\,
\ln{-i(\hat{E}_{\alpha\alpha'}+i\tilde{\Gamma}_\Delta(\hat{E}_{\alpha\alpha'}))\over V}\quad,\\
\nonumber
J_{\alpha\alpha'}(E)\,&=\,2\sqrt{x_\alpha x_{\alpha'}}\,J_V\,\left\{1\,-\,
x_{\alpha''}\,J_V\,\ln{-i(\hat{E}_{\alpha\alpha''}+i\tilde{\Gamma}_\Delta(\hat{E}_{\alpha\alpha''}))\over V}
\right.\\
\label{eq:kondo_J_low}
&\hspace{4cm}\left.\,-\,
x_{\alpha''}\,J_V\,\ln{-i(\hat{E}_{\alpha''\alpha'}+i\tilde{\Gamma}_\Delta(\hat{E}_{\alpha''\alpha'}))\over V}
\right\} \quad.
\end{align}
The integration constants have been chosen such that for $|E|\gg V$ and $J_V|\ln{-iE\over V}|\ll 1$, 
the result \eqref{eq:crossover_high_low} at high energies is reproduced.
From the solution we can see that branch cuts appear starting at the singularities 
$z_n=-i\Gamma^*+\hat{\mu}_{\alpha\alpha'}$, as stated in \eqref{eq:kondo_branch_cuts}. Furthermore, we
can see that the expansion is well-defined provided that $J_V|\ln{V\over |E-z_n|}|\ll 1$, which is the precise
condition that $E$ should not be exponentially close to the branching points. This is the reason why the
scale $V\sim|z_n|$ has been chosen as reference scale in the logarithm to integrate the RG equations 
perturbatively for intermediate and small energies. In the solution all logarithmic terms
$\sim (J^{(0)}\ln{D\over V})^k$ have been resummed in $J_V$, whereas a perturbative treatment has been
used for the logarithmic terms $J_V|\ln{V\over |E-z_n|}|\ll 1$.\newline

Since $\tilde{\Gamma}_\Delta(E)$ is a weakly varying function for $|E|\lesssim|z_n|$, we can replace 
$\tilde{\Gamma}_\Delta(\hat{E}_{\alpha\alpha'})\rightarrow\Gamma^*$ in the above equations and neglect
the term $\sim J_V^2\Gamma^*~\sim J_V^4$ in \eqref{eq:kondo_gamma_Delta_low}. Inserting 
$E=-i\Gamma^*+\delta$ in \eqref{eq:kondo_gamma_Delta_low} (where $|\delta|\ll\Gamma^*$ is a small 
scale to exclude an exponentially small region around $z_0^p=-i\Gamma^*$), we find 
straightforwardly the result \eqref{eq:kondo_pole} for $\Gamma^*$. Inserting the solution for $Z'(E)$
in \eqref{eq:kondo_F_p}, we can calculate the pre-exponential function for the contribution from the
branching pole at $E=z_0^p=-i\Gamma^*$. For {\bf long times} $t\gg {1\over V}$ we obtain
\begin{equation}
\label{eq:kondo_F_branching_pole}
F^p_0(t)\,=\,1\,-\,2\,J_V\,-\,4\,\sum_\alpha\,x_\alpha^2\,J_V^2\,\ln(Vt)\quad,
\end{equation}
whereas, for intermediate times $t\sim {1\over V}$, the contribution in $O(J_V^2)$ is not 
logarithmic and unimportant (the precise coefficient is also influenced by other sub-leading terms).
Several interesting features appear in this result. The first term is the result from a Markov
approximation, where only the pole without the residuum is considered. We note that the
pole position is also influenced by non-Markovian contributions arising when 
\eqref{eq:kondo_gamma_Delta_low} is solved self-consistently for $\Gamma^*$. Here, this
is a very weak effect occuring in $O(J_V^4)$. For quantum dot models such non-Markovian
contributions have been dicussed perturbatively in Ref.~\cite{contreras_PRB12}.
All other terms of \eqref{eq:kondo_F_branching_pole} are of pure non-Markovian nature
arising from the term linear in $E$ of the effective Liouvillian (leading to the $Z'$-factor). 
The second term linear in $J_V$ can not be obtained from perturbation theory since this term
of the $Z'$-factor involves
the difference $J_V-J^{(0)}$ (see Eq.~\eqref{eq:Z'_J_high_energies}), which reduces to $J_V$
only in the scaling limit. It arises from a resummation of a series of logarithmic terms
$\sim (J^{(0)}\ln{D\over V})^k$ which starts at $k=2$, i.e. the $k=1$ term is absent.
The last term $\sim J_V^2 \ln(Vt)$ is logarithmic in time and becomes
of $O(1)$ for exponentially large times $t\sim {1\over V}e^{1/J_V^2}$. In this regime the 
solution can no longer be used since it corresponds to the regime of energies $E$ exponentially
close to $z_0^p$. In this regime, the full RG equation \eqref{eq:kondo_J_rg} for the vertex
shows that $J(E)$ does not stay small, i.e. a strong coupling problem arises and the truncation
scheme is no longer controlled. As a consequence we see that, concerning the long-time evolution
at exponentially large times, even in the regime $V\gg T_K$, a strong coupling method is needed
to calculate pre-exponential functions. On the other hand, the exponential decay $e^{-\Gamma^*t}$
leads to a very small contribution for exponentially large times, so that it is of no practical use to
know the pre-exponential function in this regime. However, for other problems with quantum critical
points, like e.g. multi-channel Kondo models or the sub-ohmic spin boson model, it happens that
the pole $z_0^p=0$ lies at the origin such that no exponential decay appears. For such models, it
is an interesting subject for the future to calculate the precise form of $F_0^p(t)$ for exponentially
large times. E.g., for a multi-channel Kondo model with many channels $N\gg 1$, which turns out to
be a weak-coupling problem in the whole complex plane, it has been shown
in Ref.~\cite{kondo_E-RTRG} that $F_0^p(t)\sim ({1\over T_K t})^{4/N}$.\newline

Finally, to calculate the branch cut contribution \eqref{eq:kondo_F_b}, we need also the
jump $\delta\Gamma(z_{\alpha\alpha'}^b-ix)$ for $x\sim 1/t\lesssim V$. This is obtained from
the RG equation \eqref{eq:delta_L_rg_explicit}, which reads with 
$J_{\alpha\alpha'}(E)\rightarrow 2\sqrt{x_\alpha x_{\alpha'}}J_V$
\begin{equation}
\label{eq:kondo_delta_gamma_rg}
{\partial\over\partial x}\delta\Gamma(z_{\alpha\alpha'}^b-ix)\,=\,
-\,8\pi i\,x_\alpha x_{\alpha'}\,J_V^2\,\theta(x)\quad.
\end{equation}
This leads to $\delta\Gamma(z_{\alpha\alpha'}^b-ix)=-8\pi i x_\alpha x_{\alpha'} J_V^2 x\theta(x)$.
Inserting this result in \eqref{eq:kondo_F_b} and using $Z'=1+O(J_V^2)$, we obtain for
{\bf long times} $t\gg {1\over|\mu_\alpha-\mu_{\alpha'}|}$ (note that $\alpha\ne\alpha'$
and we assumed that $|\mu_\alpha-\mu_{\alpha'}|\gg\Gamma^*$)
\begin{equation}
\label{eq:kondo_F_branching_point}
F_{\alpha\alpha'}^b(t)\,=\,-\,4 x_\alpha x_{\alpha'}\,J_V^2\,
\left({1\over (\mu_\alpha-\mu_{\alpha'})t}\right)^2\quad.
\end{equation}
Other time regimes $t\sim {1\over|\mu_\alpha-\mu_{\alpha'}|}$ can also be studied leading to
exponential integrals \cite{pletyukhov_PRL10}. Inserting \eqref{eq:kondo_F_branching_point} in 
\eqref{eq:kondo_total_evolution_pb} we get an oscillating term 
$\sim J_V^2({1\over (\mu_\alpha-\mu_{\alpha'})t})^2 e^{-\Gamma^* t}e^{-i(\mu_\alpha-\mu_{\alpha'})t}$
for the time evolution of the local spin. It appears in second order in $J_V$ and is again of
non-Markovian nature. In contrast to the Markov contribution it oscillates with a frequency set
by the differences of chemical potentials and the pre-exponential function decays as a power law $\sim 1/t^2$ 
for long times. This behaviour is quite generic for models with spin or orbital fluctuations. In 
higher orders the oscillation frequencies are set by the renormalized excitation energies of the system
associated with certain processes. E.g. a process where a particle is transferred from reservoir $\alpha$
to reservoir $\alpha'$ involves an energy cost $\mu_{\alpha'}-\mu_{\alpha}$, which gives the oscillation
frequency. In the presence of a local magnetic field $h^{(0)}$, the same process costs the energy
$\mu_{\alpha'}-\mu_{\alpha}\pm h$ if the local spin is flipped, where $h$ is the renormalized magnetic
field. As a consequence, these scales define further oscillation frequencies. In addition, each process
has its own decay rate, setting the scale of the exponential decay. These issues have been discussed in
detail in Ref.~\cite{pletyukhov_PRL10} for the case of the anisotropic Kondo model at finite magnetic field.

\subsection{Ohmic spin boson model}

Here, we consider the ohmic spin boson model at zero bias $\epsilon=0$ and zero temperature $T=0$.
We will follow Ref.~\cite{kashuba_PRB13} where the model has recently been solved for weak damping 
$\alpha\ll 1$ by a systematic RG analysis using the E-RTRG method. In contrast
to the Kondo model it turns out that the effective vertex $G(E)=G_1(E)_{\bar{\omega}=0}$ at zero 
frequency stays small in the whole complex plane allowing for a full solution of the problem on all
time scales. We show here only the solution since the derivation is very similiar to the one for the Kondo
model, except that the algebra is more involved and the solution of the RG equations can also be derived
for $E$ close to the branching points $z_n$. It turns out that the resolvent $R(E)$ has four poles at 
\begin{equation}
\label{eq:sb_poles}
z_{\text{st}}\,=\,0\quad,\quad 
z_0\,=\,-i\Gamma \quad,\quad z_\pm\,=\,\pm\tilde{\Delta}\,-\,i\Gamma/2 \quad,
\end{equation}
where 
\begin{equation}
\label{eq:sb_tilde_Delta}
\tilde{\Delta}\,=\,\Delta\,\left({\tilde{\Delta}\over D}\right)^\alpha\,=\,
\Delta\,\left({\Delta\over D}\right)^{\alpha\over 1-\alpha} \quad,\quad
\Gamma\,=\,\pi\alpha\tilde{\Delta} \quad.
\end{equation}
$\tilde{\Delta}$ is called the renormalized tunneling which is kept fixed in the scaling
limit $D\rightarrow\infty$ and $\alpha\rightarrow 0$. In leading order truncation at $O(\alpha)$ it turns 
out that no branching poles appear, i.e. all poles are isolated. In addition, the eigenvalue
$\lambda_0(E)$ has two branch cuts starting at $z_\pm$ and the eigenvalues $\lambda_\pm(E)$
have a branch cut starting at $z_0$. Therefore, according to the general expression 
\eqref{eq:total_evolution_pb} we get
\begin{equation}
\label{eq:rho_tot}
\rho(t)\,=\,\rho_{\text{st}}\,+\,\sum_{k=0,\pm}\,\left(F_k^p(t) \,+\, F_k^b(t)\right)\,
e^{-i z_k t}\,\rho_{t=0}\quad,
\end{equation}
i.e. all singularities can either act as a pole or as a branch cut.
Ordering the four possible states in Liouville space by $++,--,+-,-+$, where $\pm$ are the
two local states, one can show that the stationary density matrix is given by 
$\rho_{\text{st}}={1\over 2}(1,1,{\tilde{\Delta}\over \Delta},{\tilde{\Delta}\over \Delta})$
and the pre-exponential functions for {\bf long times} $\tilde{\Delta} t\gg 1$ (note that this includes
the important regime $\Gamma t\sim O(1)$ where the exponentials are of $O(1)$) are given by \cite{kashuba_PRB13}
\begin{align}
\label{eq:F_p}
F_0^p(t) &=\, {\tilde{\Delta}\over\Delta}\,
\left(\begin{array}{cc}0 & 0 \\ -1 & \tilde{\Delta}/\Delta  \end{array}\right)
\otimes\tau_+ \quad,\quad
F_\pm^p(t) \,=\, {1\over 2}\,
\left(\begin{array}{cc}1 & \pm\tilde{\Delta}/\Delta \\ 
\pm\tilde{\Delta}/\Delta  & (\tilde{\Delta}/\Delta)^2  \end{array}\right)
\otimes\tau_- \quad,\\
\label{eq:F_b}
F_0^b(t) &\,=\, -\,2\alpha\,{1\over(\tilde{\Delta} t)^2}\,
\left(\begin{array}{cc}1 & 0 \\ 0 & 0  \end{array}\right)
\otimes\tau_- \quad,\quad
F_\pm^b(t) \,=\, -\,\alpha\,{s(t)\over(\Delta t)^2}\,
\left(\begin{array}{cc}0 & 0 \\ 0 & 1  \end{array}\right)
\otimes\tau_+ \quad,
\end{align}
where $\tau_\pm={1\over 2}(1\pm\sigma_x)$, and, for two $2\times 2$-matrices $A$ and $B$, 
we have defined the $4\times 4$-matrix 
\begin{equation}
\label{eq:tensor_def}
A\otimes B\,\equiv\,\left(\begin{array}{cc}A_{11}B & A_{12}B \\ A_{21}B & A_{22}B  \end{array}\right)
\quad.
\end{equation}
Furthermore, the logarithmic function $s(t)$ is defined by 
\begin{equation}
\label{eq:f_t}
s(t)\,=\,\left({1\over [1\,+\,\alpha\ln(\tilde{\Delta} t)]\,
[1\,-\,\ln(1\,+\,\alpha\ln(\tilde{\Delta} t))]}\right)^2
\quad.
\end{equation}
In terms of the expectation values of the Pauli matrices 
$\langle \sigma_i \rangle(t) =\text{Tr}\sigma_i\rho(t)$, these equations can also be written as
\begin{align}
\nonumber
\left(\begin{array}{c} 1 \\ \langle \sigma_x \rangle(t) \end{array}\right)
 &\,=\, 
\left(\begin{array}{c} 1 \\ \tilde{\Delta}/\Delta \end{array}\right)
\,+\,
{\tilde{\Delta}\over\Delta}\,
\left(\begin{array}{cc}0 & 0 \\ -1 & \tilde{\Delta}/\Delta  \end{array}\right)\,
\left(\begin{array}{c}1 \\ \langle \sigma_x \rangle_{t=0} \end{array}\right)
\,e^{-i z_0 t}\\
\label{eq:rho_+}
&\hspace{3cm}
-\,\alpha\,{s(t)\over(\Delta t)^2}\,
\left(\begin{array}{cc}0 & 0 \\ 0 & 1  \end{array}\right)\,
\left(\begin{array}{c}1 \\ \langle \sigma_x \rangle_{t=0} \end{array}\right)
\,\sum_{\sigma=\pm}\,e^{-i z_\sigma t}\quad,\\
\nonumber
\left(\begin{array}{c}\langle \sigma_z \rangle(t) \\ -i\langle \sigma_y \rangle(t) \end{array}\right)
 &\,=\,
{1\over 2}\,\sum_{\sigma=\pm}\,
\left(\begin{array}{cc}1 & \sigma\tilde{\Delta}/\Delta \\ 
\sigma\tilde{\Delta}/\Delta  & (\tilde{\Delta}/\Delta)^2  \end{array}\right)\,
\left(\begin{array}{c} \langle \sigma_z \rangle_{t=0} \\ -i\langle \sigma_y \rangle_{t=0} \end{array}\right)
\,e^{-i z_\sigma t}\quad,\\
\label{eq:rho_-}
&\hspace{2.2cm}
-\,2\alpha\,{1\over(\tilde{\Delta} t)^2}\,
\left(\begin{array}{cc}1 & 0 \\ 0 & 0  \end{array}\right)\,
\left(\begin{array}{c}\langle \sigma_z \rangle_{t=0} \\ -i\langle \sigma_y \rangle_{t=0} \end{array}\right)
\,e^{-i z_0 t} \quad.
\end{align}
In this result all logarithmic terms at high energies $\sim (\alpha\ln{D\over\Delta})^k$ have
been resummed in the renormalized tunneling $\tilde{\Delta}$, and all logarithmic terms at low energies 
(or large times) $\sim(\alpha\ln(\tilde{\Delta} t))^k$ are contained in $s(t)$. For the pre-exponential
function $F_0^b(t)$ it turns out that, in leading order, no logarithmic terms are present at large times.
This has to be contrasted to the solution within the noninteracting blip 
approximation (NIBA) \iffindex{NIBA} \cite{functional_integral}, where, for 
$\langle \sigma_z \rangle_{t=0}=1$ and $\langle \sigma_y \rangle_{t=0}=0$, one obtains
\begin{equation}
\label{eq:niba}
\langle\sigma_z\rangle(t)_{NIBA}\,=\,e^{-{\Gamma\over 2}t}\cos(\tilde{\Delta} t)\,-\,
2\alpha\,{1\over(\tilde{\Delta} t)^{2-2\alpha}}\,,
\end{equation}
whereas the correct result from \eqref{eq:rho_-} reads
\begin{equation}
\label{eq:sigma_z_correct}
\langle\sigma_z\rangle(t)\,=\,e^{-{\Gamma\over 2}t}\cos(\tilde{\Delta} t)\,-\,
2\alpha\,{1\over(\tilde{\Delta} t)^2}\,e^{-\Gamma t}\,.
\end{equation}
Besides the missing exponential part in the second term, which has already been discussed at
the end of Section~\ref{sec:diagrammatic_expansion}, the NIBA predicts a different exponent for
the pre-exponential power law. This shows that power-law exponents of pre-exponential functions
can only be calculated by resumming consistently all logarithmic terms for long times. The E-RTRG
method predicts that no such logarithmic terms are present for $\langle\sigma_{y,z}\rangle(t)$ 
but they appear for $\langle\sigma_x\rangle(t)$ within the logarithmic function $s(t)$.
The leading power-law behaviour $\sim ({1\over t})^2$ of the pre-exponential function is the same 
as for the Kondo model and can also be obtained from perturbative calculations \cite{vincenzo_PRB05}.
There are always two terms with different decay rates $\Gamma$ and $\Gamma/2$ for the time evolution. 
If one transforms to the exact eigenbasis 
$\underline{e}_{1/2}={1\over\sqrt{2}}(|+\rangle\pm |-\rangle)$ of the local system, the 
expectation values $\langle \gamma_{i}\rangle$ of the Pauli matrices in the new basis are related 
to the ones of the original basis by $\langle \gamma_x\rangle=\langle \sigma_z\rangle$,
$\langle \gamma_y\rangle=-\langle \sigma_y\rangle$ and $\langle \gamma_z\rangle=\langle \sigma_x\rangle$.
Thus the Markovian term $\sim e^{-\Gamma t}$ from the pole contribution describes the decay of
the diagonal matrix elements of the density matrix in the new basis, whereas the one
$\sim e^{-(\Gamma/2)t}e^{\pm\tilde{\Delta} t}$ corresponds to the decay of the nondiagonal matrix elements.
Therefore, $\Gamma$ is called the relaxation rate, whereas $\Gamma/2$ is the decoherence rate, in
accordance with the general rule that, in the absence of pure dephasing, the relaxation rate is 
always twice as large as the decoherence rate. 
\newline

For large energies $|E|\gg\tilde{\Delta}$, one needs the function $Z'(E)$ to determine the regime 
of {\bf short times} $t\ll {1\over \tilde{\Delta}}$ from \eqref{eq:short_times} (the contribution from the 
exponential is a small correction and can be neglected). One obtains the result
\begin{equation}
\label{eq:sb_Z'_high_energies}
Z'(E)\,=\,\sum_{\sigma=\pm}\,
\left(\begin{array}{cc}1 & 0 \\ 0 & Z_\sigma(E) \end{array}\right)\otimes\tau_\sigma
\quad,\quad Z_\pm(E)\,\approx\,\left({-iE\over D}\right)^{2\alpha}
\end{equation}
This gives rise to the universal short time behavior 
\begin{equation}
\label{eq:sb_short_times}
\rho(t)\,=\,\left(\begin{array}{cc}1 & 0 \\ 0 & ({1\over Dt})^{2\alpha}\end{array}\right)
\otimes 1\,\rho_{t=0}\quad,
\end{equation}
or
\begin{equation}
\label{eq:sb_short_times_sigma}
\langle\sigma_{x,y}\rangle(t)\,=\,({1\over Dt})^{2\alpha}\,\langle\sigma_{x,y}\rangle_{t=0}
\quad,\quad
\langle\sigma_z\rangle(t)\,=\,\langle\sigma_z\rangle_{t=0}\quad.
\end{equation}
This agrees with previous predictions and can also be obtained from the exact solution \eqref{eq:sb_exact}
at $\Delta=T=0$ in the universal regime $t\gg {1\over D}$. Again we can see that all logarithmic terms
$\sim(\alpha\ln(Dt))^k$ have been resummed in this result.

\subsection{Interacting resonant level model}

Finally we discuss the IRLM for the special case of a single reservoir with chemical potential $\mu=0$ 
and zero level position $\epsilon=0$ (i.e. in resonance with the reservoir). As discussed in 
Section~\ref{sec:models}, this model can be mapped to the ohmic spin boson model close to the
exactly solvable point $\alpha={1\over 2}$. In particular, we want to understand where the crossover
from coherent to incoherent time evolution by changing the sign of $U=1-\sqrt{2\alpha}$ comes from.
We follow Ref.~\cite{kennes_PRL13,kashuba_kennes_PRB13}, where the IRLM has been studied by using
E-RTRG and functional RG. \newline

We concentrate on the time evolution of the occupation $\langle n\rangle(t)$ of the local level 
which is related via Eq.~\eqref{eq:irlm_sb_occupation} to the expectation 
$\langle\sigma_z\rangle(t)$ within the spin boson 
model by $2\langle n \rangle(t)-1=\langle \sigma_z\rangle(t)$. For $\langle \sigma_z\rangle(t)$, one can
show that the result can be written in the form
$\langle \sigma_z \rangle(t)=P(t)\langle\sigma_z\rangle_{t=0}$, with
\begin{equation}
\label{eq:Prelax}
P(t)\,=\,{i\over 2\pi}\,\int_{-\infty+i0^+}^{\infty+i0^+} \,dE\,e^{-iEt}\,{1\over E\,+\,i\Gamma_1(E)}
\quad.
\end{equation}
For this special case there is no $Z'$-factor and $\Gamma_1(E)$ is a slowly varying logarithmic
function describing the energy dependent charge relaxation rate. It is determined from the RG equations
\begin{equation}
\label{eq:irlm_rg}
{\partial\over\partial E}\Gamma_1(E)\,=\,-g\,R_2(E)\,\Gamma_1(E)
\quad,\quad
{\partial\over\partial E}\Gamma_2(E)\,=\,-g\,R_1(E)\,\Gamma_1(E)
\quad,
\end{equation}
where $g=2U-U^2=1-2\alpha$ and the resolvents $R_{1/2}(E)$ are defined by
\begin{equation}
\label{eq:R_1_2}
R_1(E) \,=\, {1\over E\,+\,i\Gamma_1(E)}\quad,\quad
R_2(E) \,=\, {1\over E\,+\,i\Gamma_2(E)/2}\quad.
\end{equation}
The initial conditions are given by $\Gamma_{1/2}(E=iD)=\Gamma^{(0)}$. 
$\Gamma_2(E)/2$ is also a slowly varying logarithmic function and describes the energy dependent
broadening of the local level corresponding to the decoherence mode for nondiagonal matrix elements
of the local density matrix w.r.t. the charge states (note, however, that such elements can not be prepared).
As we will see below the subtle coupling of the two RG equations for $\Gamma_{1/2}(E)$ leads to the 
interesting effect that, for $g>0$, the resolvent $R_1(E)$ can have poles with a finite real part 
although the local system has no finite excitation energy. \newline

We start by solving the RG equations at {\bf high energies} $E\gg\Gamma_{1/2}(E)$. Neglecting $\Gamma_{1/2}(E)$
on the r.h.s. of the RG equations, we find the solution
\begin{equation}
\label{eq:solution_high_energies}
\Gamma_{1/2}(E)\,=\,\Gamma^{(0)}\,\left({D\over -iE}\right)^g\,=\,
\tilde{\Delta}\,\left({\tilde{\Delta}\over -iE}\right)^g ,
\end{equation}
where 
\begin{equation}
\label{eq:irlm_tilde_Delta}
\tilde{\Delta}\,=\,\Gamma^{(0)}\,\left({D\over \tilde{\Delta}}\right)^g\,=\,
\Gamma^{(0)}\,\left({D\over \Gamma^{(0)}}\right)^{g/(1+g)}
\end{equation}
is the renormalized tunneling which is kept fixed in the scaling limit $D\rightarrow\infty$
and $\alpha,\Gamma^{(0)}\rightarrow 0$. Using the relation $g=1-2\alpha$ and 
$\Gamma^{(0)}={\Delta^2\over D}$ to the spin boson model, one can see that it is identical 
to the definition \eqref{eq:sb_tilde_Delta} of the renormalized tunneling for the spin boson model.
As discussed in detail in Refs.~\cite{RTRG_irlm,kashuba_kennes_PRB13}, the solution at
high energies contains all leading logarithmic terms $\sim (U\ln{D\over -iE})^k$ and 
all subleading ones $\sim U(U\ln{D\over -iE})^k$. From the solution \eqref{eq:irlm_tilde_Delta} 
at high energies we can calculate with \eqref{eq:short_times} the time evolution 
for {\bf short times} $t\ll 1/\tilde{\Delta}$ as
\begin{equation}
\label{eq:irlm_short_times}
P(t)\,\approx\,e^{-\Gamma_1(1/t)t}\,\approx\,e^{-(\tilde{\Delta} t)^{1+g}},
\end{equation}
i.e. the relaxation rate in the exponent is cut off at the energy scale $1/t$. In contrast
to the spin boson model at small $\alpha$ and the Kondo model, there is no $Z'$-factor and
therefore the exponential provides the leading order. Expanding the exponential we find
$P(t)=1-(\tilde{\Delta} t)^{1+g}$ in agreement with previous results \cite{functional_integral}.
Since $(\tilde{\Delta} t)^{1+g}=(Dt)^g \Gamma^{(0)}t$, we see again that all logarithmic
terms $\sim (g\ln(Dt))^k$ have been resummed for small times.
\newline

Next we study the analytic structure of the resolvent $R_1(E)$ to find the time evolution for
intermediate and long times. As we will show below, for positive $g>0$, $R_1(E)$ has two poles 
at $z_\pm$ (followed by a branch cut with jump of $O(g^2)$ which can be neglected) and one
branch cut starting at $z_0$ (with jump of $O(g)$), where the singularities $z_n$, $n=0,\pm$, are given by
\begin{equation}
\label{eq:z_i}
z_0\,=\,-\,i\,{\tilde{\Delta}\over 2} \quad,\quad
z_\pm\,=\,\pm\,\Omega\,-\,i\,\tilde{\Delta} \quad,\quad 
\Omega\,=\,\pi g \tilde{\Delta} \quad.
\end{equation}
For $g<0$, there is only a branch cut starting at $z_0$. Thereby, $z_0$ is the position
of the pole of the resolvent $R_2(E)$, i.e. $z_0$ and $z_\pm$ can be determined from the
equations
\begin{equation}
\label{eq:poles_R12}
z_\pm\,+\,i\Gamma_1(z_\pm)\,=\,0 \quad,\quad z_0\,+\,i\Gamma_2(z_0)/2\,=\,0 \quad.
\end{equation}
Note that, in contrast to the singularities \eqref{eq:sb_poles} for the spin boson model at
small $\alpha$, for the IRLM (or the spin boson model at $\alpha={1\over 2}$) the renormalized 
tunneling determines the rate and not the oscillation frequency. Furthermore, we note that
the pole of $R_1(E)$ describes the charge relaxation mode, whereas for the spin boson model
at small $\alpha$ it corresponds to the decoherence mode w.r.t. the exact eigenstates of the
local system. Therefore, $z_0$ corresponds to the decoherence mode for the IRLM and its 
imaginary part is half of the one of the relaxation poles $z_\pm$. To derive the result for the 
positions of the singularities we solve the RG equations for intermediate and small energies
$\,|E|\lesssim\tilde{\Delta}\,$ but $\,g\ln{\tilde{\Delta}\over |E-z_n|}\ll 1\,$, i.e. $E$ should not
be exponentially close to the singularities. Expanding in the small parameter 
$g\ln{\tilde{\Delta}\over |E-z_n|}\ll 1$ and fixing the integration constants by comparing
with the solution \eqref{eq:solution_high_energies} at high energies in the usual way, we find
\begin{equation}
\label{eq:gam12_intermediate}
\Gamma_1(E)/\tilde{\Delta}\,\approx\,1\,-\,g\,\ln{-iE\,+\,\Gamma_2(E)/2\over \tilde{\Delta}} 
\,\,,\,\,
\Gamma_2(E)/\tilde{\Delta}\,\approx\,1\,-\,g\,\ln{-iE\,+\,\Gamma_1(E)\over \tilde{\Delta}} .
\end{equation}
In contrast to the corresponding equation \eqref{eq:kondo_gamma_Delta_low} for the Kondo model,
there is a subtle coupling of the singularites of $\Gamma_1(E)$ and $\Gamma_2(E)$, which leads
to the new feature that $z_\pm$ obtains a finite real part for $g>0$. We note that although
the equations can not be used for $E$ exponentially close to the singularities, they can be
used for $|E-z_n|\sim g^2$ since $g\ln(g)\ll 1$ for $g\ll 1$. Therefore, the equations can be 
used to determine the positions of the branching points of $\Gamma_{1/2}(E)$ up to $O(g)$.
From the equations we can see that $\Gamma_1(E)$ ($\Gamma_2(E)$) have a branch cut with jump 
of $O(g)$ starting at the branching point of the logarithmic function where \eqref{eq:poles_R12}
is fulfilled, i.e. at $z_0$ ($z_\pm$). Thereby, the branch cut of $\Gamma_2(E)$ starting at
$z_\pm$ leads also to a branch cut for $\Gamma_1(E)$ at the same position but this branch cut
has a jump of $O(g^2)$ and can be neglected. Inserting the leading order results
$\Gamma_{1/2}(E)\approx\tilde{\Delta}$, $z_0\approx-i\tilde{\Delta}/2$ and $z_\pm\approx-i\tilde{\Delta}$
on the r.h.s. of \eqref{eq:gam12_intermediate}, we find for the position of the singularities
the result \eqref{eq:z_i}
\begin{align}
\label{eq:z_0_determination}
2i z_0/ \tilde{\Delta} &= \Gamma_2(z_0)/\tilde{\Delta} 
\approx  1-g\ln(-iz_0/\tilde{\Delta}+1) 
\approx  1-g\ln \left( -{1\over 2}+1 \right)  \approx 1 \quad,\\
\nonumber
i z_\pm/\tilde{\Delta} &= \Gamma_1(z_\pm)/\tilde{\Delta} 
\approx  1-g\ln \left( -iz_\pm/\tilde{\Delta}+{1\over 2} \right) 
\approx  1-g\ln \left( -1\mp i\Omega/\tilde{\Delta}+{1\over 2} \right) \\
\label{eq:z_+-_determination}
&\approx 1\pm i\pi g \quad.
\end{align} 
\newline

Due to the analytic structure of the resolvent $R_1(E)$ the time evolution can be written as
\begin{equation}
\label{eq:irlm_intermediate_times}
P(t)\,=\,\theta(g)\,\sum_{\sigma=\pm}\,e^{-iz_\sigma t}
\,+\,F_0^b(t)\,e^{-iz_0t}\,=\,\theta(g)\,2\,\cos(\Omega t)\,e^{-\tilde{\Delta}t}
\,+\,F_0^b(t)\,e^{-(\tilde{\Delta}/2)t}\,,
\end{equation}
where the first term involves the contribution from the isolated poles (we have neglected
corrections of $O(g)$ to the residuum) and the second term involves the analog of the branch cut integral
\eqref{eq:F_branching_point}, which can be written as
\begin{equation}
\label{eq:branch_cut_integral}
F_0^b(t)\,=\,{1\over \pi}\,\text{Im}\,\int_0^\infty\,dx\,e^{-xt}\,
{1\over -i(z_0\,+\,i\Gamma_1(z_0-i/t+0^+))\,-\,x}\quad.
\end{equation}
For {\bf intermediate and long times} $t\gtrsim {1\over\tilde{\Delta}}$ but 
$g\ln(\tilde{\Delta}t)\ll 1$, $F_0^b(t)$ can be evaluated by using the result 
\eqref{eq:gam12_intermediate}, where we obtain 
$-i\Gamma_1(z_0-i/t+ 0^+)=z_+(1 + O(g\ln(\tilde{\Delta}t))$.
In particular one has to consider the fact that $x\sim {1\over t}$
can not be neglected compared to the difference $|z_0-z_\pm|\sim\tilde{\Delta}$ for 
intermediate times $t\sim {1\over\tilde{\Delta}}$. This time regime is of particular interest here 
since the exponentials of the time evolution \eqref{eq:irlm_intermediate_times}
decay on the time scale ${1\over\tilde{\Delta}}$. Therefore, the
integral \eqref{eq:branch_cut_integral} has to be calculated more carefully in terms of
the exponential integral $E_1(z)$
\begin{equation}
\label{eq:bc_intermediate_times}
F_0^b(t)\,=\,-\,{1\over\pi}\,\text{Im}\,\left\{e^{-i(z_0-z_+)t}\,E_1(-i(z_0-z_+)t)\right\}\quad.
\end{equation}
This result has been used in Refs.~\cite{kennes_PRL13,kashuba_kennes_PRB13} to discuss the
competition between the oscillating (i.e. coherent) and the purely decaying (i.e. incoherent) term of
the time evolution in Eq.~\eqref{eq:irlm_intermediate_times}. Since the incoherent term decays on 
a longer time scale it turns out that it wins very rapidly such that the coherent term leads only
to a few number of oscillations, in contrast to the physics of a classical damped harmonic oscillator.
For {\bf long times} $t\gg{1\over\tilde{\Delta}}$ but still $g\ln(\tilde{\Delta}t)\ll 1$, the incoherent 
term dominates and, using the asymptotic expansion $E_1(z)=e^z/z$ of the exponential integral, one obtains
\begin{equation}
\label{eq:bc_long_times}
F_0^b(t)\,\approx\,-4g\,{1\over \tilde{\Delta}t}\quad,
\end{equation}
i.e. a power law $\sim 1/t$ typical for models with charge fluctuations.\newline

Finally, for {\bf exponentially large times} $g\ln(\tilde{\Delta}t)\sim O(1)$, we need the solution for $\Gamma_1(E)$
for energies $E$ exponentially close to the branching point $z_0$. In this regime, we can replace
$\Gamma_2(E)/2\rightarrow iz_0$ on the r.h.s. of the RG equation \eqref{eq:irlm_rg} for $\Gamma_1(E)$,
which gives the solution
\begin{equation}
\label{eq:gam1_exponentially_small_energies}
\Gamma_1(E)\,=\,\tilde{\Delta}\,\left({\tilde{\Delta}\over -i(E-z_0)}\right)^g\quad,
\end{equation}
where the integration constant has been fixed by comparison with the solution \eqref{eq:gam12_intermediate}
at intermediate and small energies. Using this solution for the evaluation of the branch cut
integral \eqref{eq:branch_cut_integral} for exponentially large times, we can neglect $x$  
in the denominator and find with 
$\Gamma_1(z_0-i/t+0^+)=\tilde{\Delta}(\tilde{\Delta}t)^g+2\pi i g\tilde{\Delta}(\tilde{\Delta}t)^g$ 
the result (neglecting terms of $O(g^2)$ in the denominator)
\begin{equation}
\label{eq:bc_exponentially_large_times}
F_0^b(t)\,\approx\,-g\,
{1\over (1/2\,-\,(\tilde{\Delta}t)^g)^2}\,
{1\over (\tilde{\Delta}t)^{1-g}}\quad.
\end{equation}
This result holds for all times $t\gg{1\over\tilde{\Delta}}$. For long times with $g\ln(\tilde{\Delta}t)\ll 1$ it 
reduces to the result \eqref{eq:bc_long_times}. However, for exponentially large times where
$(\tilde{\Delta}t)^g$ is some number of $O(1)$, the result changes. In the extreme regime
$(\tilde{\Delta}t)^{|g|}\gg 1$, it reduces to
\begin{equation}
\label{eq:bc_extrem_long_times}
F_0^b(t)\,\approx\,-g\,[1\,+\,3\theta(-g)]\,{1\over (\tilde{\Delta}t)^{1+|g|}}\quad.
\end{equation}
This result agrees with the prediction of the NIBA \cite{functional_integral} and its improved version 
\cite{egger_PRE97} (where the exponential term $e^{-(\tilde{\Delta}/2)t}$ has also been obtained, 
see Eq.~\eqref{eq:irlm_intermediate_times}). However, as we have seen, it holds only for extremely 
long times and, for $g>0$, the prefactor is different from the
result \eqref{eq:bc_long_times} for more realistically long times $\tilde{\Delta}t\gg 1$ with 
$g\ln(\tilde{\Delta}t)\ll 1$. Therefore, we see that the regime of long times is very subtle and
the result can change significantly by entering the regime of exponentially large times.\newline

Finally, as already mentioned in Section~\ref{sec:RG}, it has not yet been studied to a full extent
how the RG equation \eqref{eq:irlm_rg} looks like in higher orders in the tunneling. There is some
evidence that all higher order terms in $\Gamma$ are of the form
\begin{equation}
\label{eq:higher_orders}
U^n\,\left({\Gamma_i\over\tilde{\Delta}}\right)^k\,(\ln{E-z_n\over \tilde{\Delta}})^l \quad
\text{for}\quad n=1,2
\quad,\quad
U^n\,{\Gamma_i\over E-z_n}\,\left({\Gamma_i\over\tilde{\Delta}}\right)^k
\text{for}\quad n>2\quad,
\end{equation}
i.e., after integration, either vanish in the limit $E\rightarrow z_n$ or contribute to higher 
orders in $U$, but this is still under investigation. Furthermore, the results have been compared to 
functional RG in Refs.~\cite{kennes_PRL13,kashuba_kennes_PRB13}, where all orders in the tunneling have been 
resummed keeping only the lowest order term in the Coulomb interaction. The numerical results of functional
RG agree quite nicely with the analytical result \eqref{eq:bc_intermediate_times} 
for intermediate and long times and, 
in particular for extremely long times, the result \eqref{eq:bc_extrem_long_times} has been confirmed 
analytically by functional RG. Therefore, there is good evidence that also within E-RTRG 
higher orders in the tunneling will not change the results at least in leading order in $U$.\newline

{\bf Acknowledgments.} I am particularly thankful to M. Wegewijs for a thorough reading of the manuscript.

%%%%%%%%%%%%%%% Appendices %%%%%%%%%%%%%%%%%
%\newpage
%\section*{Appendices}

\appendix

%\section{Evaluation of the time evolution}

\newpage

%%%%%%%%%%%%%%% References %%%%%%%%%%%%%%%%%

\end{document}